\documentclass[a4paper,11pt]{article}
\usepackage{silence}
\usepackage{jheppub}
\usepackage{physics} 
\usepackage{bm, bbm} 
\usepackage{algorithm, algpseudocode} 
\usepackage{subcaption}
\usepackage{placeins}
\usepackage{dsfont}
\usepackage{xurl}
\usepackage{tikz}

\allowdisplaybreaks 

\usepackage{xcolor}

\newcommand{\customparentket}{\mathclose{)}}
\newcommand{\customparentbra}{\mathopen{(}}

\newcommand{\ketc}[1]{|#1\customparentket}

\newcommand{\braketc}[2]{\customparentbra #1|#2\customparentket}
\newcommand{\melc}[3]{\customparentbra #1|#2|#3\customparentket}

\title{\boldmath From phase space to Krylov space, one shell at a time}

\author[a]{N. De Ro, }
\affiliation[a]{Department of Theoretical Physics, University of Geneva, \\ 24 quai Ernest-Ansermet, 1205 Geneva, Switzerland\\}

\author[b]{A. Sánchez-Garrido }
\affiliation[b]{School of Mathematical Sciences \& STAG Research Centre,\\ University of Southampton, SO17 1BJ, Southampton, United Kingdom}

\author[a]{and J. Sonner, }

\emailAdd{nicolas.dero@unige.ch}
\emailAdd{A.Sanchez-Garrido@soton.ac.uk}
\emailAdd{Julian.Sonner@unige.ch}

\abstract{In this work, we develop and study the classical Lanczos algorithm allowing us to define Krylov complexity using the symplectic structure of phase space: Poisson brackets take on the role of the quantum commutators and phase-space integrals furnish the inner product needed to define the Lanczos recursion. We show, using general methods of quantum mechanics in phase space, that the $\hbar \to 0$ limit of the usual quantum mechanical Krylov framework smoothly goes over into the classical one. In theories with well-defined semiclassical limits, we show that classical Krylov complexity accurately approximates quantum complexity at early enough times, and thus is a useful characteristic of early-time chaotic dynamics. We define a Krylov-Ehrenfest time, which quantifies the eventual divergence of classical and quantum complexities, corresponding to a characteristic depth of the Krylov chain, $n\sim n_*(\hbar)$, which in the time domain translates to the well-known scale, $t_*\sim\lambda_K^{-1}\log(1/\hbar)$, in generic chaotic systems. We additionally define microcanonical Krylov complexities, both in the classical and quantum setting, which allows one a fine-grained study of complexity, energy shell by energy shell. We apply this framework to the Lipkin-Meshkov-Glick (LMG) and Feingold-Peres (FP) models, which are collective spin systems known to classicalize in the thermodynamic limit. In particular, while the FP model features spectral chaos for some range of coupling values, the LMG model is known to exhibit early-time saddle-dominated scrambling. Our analysis shows that the instability in LMG is resolved by the microcanonical Krylov complexity, which is controlled by the integrable structure of the Hamiltonian in spectral windows away from the instability, both at early and late times.}

\begin{document}

\maketitle

\section{Introduction}
Krylov complexity (K-complexity) and the structure of its underpinning Lanczos sequence have emerged as a compelling probe of operator growth and other aspects of chaotic quantum dynamics. These developments were triggered by the universal operator growth hypothesis \cite{Universal_Growth_Hypothesis}, linking maximally chaotic dynamics to asymptotic linear growth of the Lanczos $b$-sequence, $\{b_n \}$, in the thermodynamic limit. Focusing attention on finite-size effects, the late-time saturation of K-complexity has been proposed as a sensitive diagnostic distinguishing chaotic from integrable dynamics \cite{Adrian_1,Adrian_2,Adrian_3}. For a comprehensive review of the state of the art in K-complexity, we refer the reader to \cite{Nandy:2024evd,Rabinovici:2025otw,Baiguera:2025dkc}. For the most part, the formalism is quantum mechanical, operating on a Hilbert space and operators acting on it, utilizing the Frobenius inner product or a thermal variant thereof at every step of the construction of the Lanczos sequences $\{a_n, b_n \}$. A crucial aspect of this framework has received much less attention: is there a meaningful classical limit of the Krylov construction, and what regimes of Krylov dynamics would it usefully govern? Indeed, this raises the natural question of what becomes of the Lanczos algorithm in the classical limit, and on how to translate the associated physics into the arena of phase space. Early discussions of this little-explored subject can be found in \cite{Universal_Growth_Hypothesis,Recursion_method_book}. 

In this work, we expand on this area and propose a formal study of the classical counterpart of the Krylov dynamics framework, demonstrating the usefulness of the resulting formalism on example semiclassical systems that have been considered in the recent literature \cite{Universal_Growth_Hypothesis,saddle_Krylov}. Given a classical Hamiltonian system on a compact symplectic manifold, one may run a Lanczos algorithm on the algebra of phase space observables, with the Poisson bracket furnishing the Liouvillian and the phase space integral providing the inner product. We concretely analyze this construction for systems whose phase space is $S^2$ or a Cartesian product thereof. In this case, it is judicious to use spherical harmonics as an orthonormal basis \cite{Universal_Growth_Hypothesis}, which forms the natural classical analogue of the operator basis in the quantum theory. As in the original quantum case, the algorithm is not restricted to a single choice of inner product. We formulate it with microcanonical inner products that restrict the Lanczos dynamics to fixed energy shells, turning K-complexity from a single number into a function of energy. On the quantum side, the corresponding construction replaces the Frobenius inner product with one built from a microcanonical density matrix. We are able to show that the formulations agree in the classical limit. While we work out the details for collective spin systems, the framework itself is general. It applies to any system admitting a classical limit, such as large-$S$ spin models, large-$N$ matrix and gauge theories, and WKB regimes of quantum mechanics, to name but a selection of examples. Only the choice of phase space geometry and basis would change from case to case.

Introducing an energy-resolved version of the Krylov apparatus has a number of inherent advantages, especially when approaching a classical phase space limit. An infinite-temperature inner product will average democratically over the entire spectrum, while interesting phenomena associated with chaotic dynamics often emerge only once one is able to resolve energy scales separately. This includes issues such as localized instabilities and quantum phase transitions, as well as regions in which chaotic and integrable behavior coexist. At the same time, passing from what is essentially a canonical formalism -- as has hitherto been largely the case for Krylov techniques -- to a microcanonical one is a natural step to take that has the potential to reveal a host of new phenomena and clarify existing ones, while not sacrificing any generality of the existing framework, similar to the relationship between canonical and microcanonical ensembles in standard statistical mechanics. In the context of K-complexity, this means that phenomena tied to specific energies, such as excited-state quantum phase transitions or unstable fixed points of the classical dynamics, can be isolated and studied, rather than being obscured by a global spectral average.

In a first demonstration of the reach and power of the formalism, we apply it to two different collective spin models, both of which possess a semiclassical limit. We consider the exactly integrable Lipkin-Meshkov-Glick (LMG) model as well as the Feingold-Peres (FP) model, which  admits a tunable transition from integrability to chaos. Both feature saddle points in their classical phase space, but only the latter exhibits genuine chaos. The presence of saddle points implies that  nearby trajectories diverge exponentially, mimicking a hallmark of chaos, without giving rise to the remaining phenomenology of chaotic dynamics. Previous Krylov analyses of the LMG model have reported seemingly chaotic signatures, such as linear Lanczos growth and large K-complexity saturation, despite being integrable. Our microcanonical framework resolves this tension by showing that saddle points enhance the K-complexity saturation value $\overline{C_K}$ locally at their energy shell in the LMG model, but leave the rest of the spectrum consistent with integrable behavior, exhibiting strong and persistent complexity oscillations. By contrast, genuine chaos produces uniform saturation across all energies.

Along the way, we derive a number of new and potentially useful features of the Krylov dynamics in semiclassical systems. We rigorously establish that, for systems admitting a well-defined classical limit in some $\hbar\to 0$ limit, standard quantum Krylov/Lanczos-related quantities, including the Lanczos coefficients and K-complexity, converge to their classical counterparts obtained from the phase-space Lanczos construction. We refer to this correspondence as a \emph{Krylov-Ehrenfest theorem}, and show that it holds independently of the choice of inner product, whether infinite-temperature or microcanonical. 

Using the classical version of the Lanczos algorithm, we study the infinite-temperature Lanczos coefficients and K-complexity of the LMG and FP models directly in their classical limit, thereby realizing their Krylov dynamics as a lattice hopping problem generated by the classical Liouvillian. Our analysis shows that the Lanczos coefficients exhibit enhanced growth in the presence of dynamical instabilities, namely when the LMG phase space contains a saddle point or when the FP model becomes chaotic, driving the classical K-complexity to grow exponentially at early times. Extending the analysis to the microcanonical formulation of the classical LMG model reveals a more nuanced picture: while the enhanced growth persists on the energy shell containing the saddle point, away from this shell the Lanczos coefficients develop large fluctuations around their linear trend, resulting in a suppression of the early-time growth of K-complexity.

We then return to the quantum mechanical framework and investigate the LMG and FP models at finite size. Exploiting the discrete symmetries of these systems, we derive symmetry-resolved upper bounds on the Krylov dimension, that are stronger than the general bounds of Ref.~\cite{Adrian_1} and depend explicitly on the charge of the initial operator. Since physically relevant initial operators are typically dense in the energy basis, these bounds are in fact saturated, yielding exact symmetry-determined Krylov dimensions that are independent of the model parameters -- and importantly of the underlying dynamical regime.

Finally, we revisit the crucial question of the role played by saddle points in the late-time Krylov dynamics of the LMG and FP models. We first demonstrate that the conventional infinite-temperature inner product is not well suited to address this issue, especially in the LMG model. This is because many \textit{a priori} reasonable choices of initial operator turn out to be ``non-generic'', in the sense that they commute with the Hamiltonian in some parameter limit, and therefore the late-time saturation value of their K-complexity is strongly constrained by their autocorrelation function. Even for ``generic'' operators in the LMG model, the Krylov dynamics exhibits no clear signature distinguishing saddle-point instabilities (or lack thereof) from genuine chaos. In this context, we further show that the approach of K-complexity to the predicted late-time saturation value is controlled by quasi-degeneracies in the spectrum, which are related to an excited-state quantum phase transition. In the case of the FP model, choosing sufficiently ``generic'' operators allows us to observe a reduction in the infinite-temperature K-complexity saturation value in the integrable regime of the model as compared to the chaotic one.

The microcanonical framework provides a natural resolution of the aforementioned limitations in the LMG analysis by strictly isolating the instability to its corresponding energy shell while unfolding the Krylov dynamics at other energies to be studied independently. We discuss the freedom in choosing a microcanonical inner product (and argue that our conclusions are robust with respect to this choice). Within this framework, K-complexity is maximized on the saddle-point energy shell, where its time evolution displays features closely resembling chaotic dynamics. Away from this shell, however, the expected signatures of integrability are recovered, including a suppressed K-complexity and persistent coherent oscillations in its time evolution. Taken together, these results provide new insight into the interplay between saddle-point instabilities in classical phase space and quantum chaotic dynamics in collective spin systems.

The paper is organized as follows. Section~\ref{sec:LMG_FP_models} introduces the LMG and FP models and reviews the properties required for our subsequent study of their classical and quantum Krylov dynamics. Specifically, we discuss the structure of their spectra, their discrete symmetries, and their integrable and chaotic regimes. We conclude the section with a brief toolbox listing the basic definition of the Lanczos algorithm and K-complexity. In Section~\ref{sec:ClassicalKC}, we develop the classical counterpart of the Krylov dynamics framework. After introducing the general formalism of K-complexity in phase space, we derive a \emph{Krylov-Ehrenfest theorem} based on the Stratonovich--Weyl correspondence. We then construct the classical Lanczos algorithm, compute the corresponding Lanczos coefficients and K-complexities for the LMG and FP models, and finally extend the construction to a microcanonical inner product. Section~\ref{sec:quantum_K_complexity} is devoted to the quantum analysis of the LMG and FP models. We first derive symmetry-resolved upper bounds on the Krylov dimension before investigating the infinite-temperature Krylov dynamics and its microcanonical extension. Finally, in Section~\ref{sec:discussion}, we summarize our results and discuss possible directions for future work.

\section{Models and tools}\label{sec:LMG_FP_models}

This section provides a self-contained overview of the main ingredients of this work. Sections~\ref{sec:LMG} and \ref{sec:FP} introduce the collective spin LMG and FP models, emphasizing the aspects most relevant for the subsequent analysis of their Krylov dynamics. In particular, we discuss their discrete symmetries and the resulting effects on the structure of the spectrum. Section~\ref{sec:quantum_integrability_chaos} analyzes the integrable and chaotic regimes of both models. Finally, Section~\ref{sec:Krylov_toolbox} provides a concise review of the general framework of Krylov dynamics for operators, which is extensively used throughout this work.

\subsection{Lipkin-Meshkov-Glick model}\label{sec:LMG}

We begin by introducing the LMG model, a prototypical example of a fully connected spin system that is integrable across all parameter regimes. In the sense that will be made precise below, the model becomes classical in the large-spin limit, and its phase space features a tunable saddle point. This makes it a particularly instructive setting for exploring the connection between quantum and classical K-complexity, both in the presence and absence of phase-space instabilities.

\subsubsection{Presentation of the model}

Consider a system of $N$ spin-$1/2$ particles, and let $\hat{\sigma}_i^{\mu}$ ($\mu=1,2,3$) denote the usual Pauli matrices acting on site $i$. The collective spin operators are defined as $\hat S_\mu=\sum_{i=1}^N \hat \sigma^\mu_i/2$ and describe the collective spin degrees of freedom of the system. They satisfy the Lie algebra $\mathfrak{su}(2)$, $[\hat S_\mu,\hat S_\nu]= i\sum_{\sigma=1}^3 \varepsilon_{\mu \nu \sigma} \hat S_\sigma$. Throughout this work, we use units in which Planck's constant is set to $\hbar=1$.

The collective spin operators can be rescaled as $\hat s_\mu\equiv \hat S_\mu/S$, where $S=(N\ \text{mod}\ 2)/2,\cdots\\,N/2-1,N/2$ is the quantum number associated with the eigenvalue $S(S+1)$ of the collective Casimir operator $\hat{\bm{S}}^2=\sum_{\mu=1}^3 \hat S_\mu^2$. Accordingly, $\hat{s}_\mu$ denote the components of a collective spin with unit magnitude. The rescaled collective spin operators satisfy the same $\mathfrak{su}(2)$ algebra but with an effective Planck constant $\hbar_{\mathrm{eff}}\equiv 1 / S$, which vanishes in the limit $S\to \infty$. The large-spin limit can therefore be interpreted as a classical limit.

In terms of the rescaled collective spin operators $\hat{s}_\mu$ and restricted to the sector of maximal collective spin $S=N/2$, the LMG Hamiltonian \cite{LMG_1, LMG_2, LMG_3} is given by
\begin{equation}\label{LMG_rescaled_TSS} 
    \frac{\hat{H}_{\text{LMG}}}{N} = -\frac{J}{2} \, \hat s_z^2- h\, \hat s_x\, .
\end{equation}
Within this maximal-spin sector, the classical limit ($S\to\infty$) coincides with the thermodynamic limit ($N\to \infty$). Here, $J$ is the coupling strength between interacting spin pairs and $h$ is the transverse magnetic field. In the following, we consider $J>0$, corresponding to a ferromagnetic coupling, and $h\geq 0$ since the spectrum of the LMG Hamiltonian is even under the transformation $h\to -h$ \cite{LMG_entaglement}. The Hamiltonian commutes with the Casimir operator $\hat{\bm{S}}^2$, and its eigenstates can therefore be labeled by the quantum number $S$. Furthermore, the LMG model is integrable and can be solved exactly via the Bethe ansatz \cite{LMG_integrable1, LMG_integrable2, LMG_integrable3}.

We note that the Hamiltonian in Eq.~\eqref{LMG_rescaled_TSS} takes the form $\hat H = H( \hat{s}_\mu)$, where $\hat H\equiv \hat{H}_{\text{LMG}}/N$\footnote{\label{fn:convention}In the context of the LMG or the FP model (see below), $\hat{H}$ denotes the corresponding intensive Hamiltonian in the rest of this work. This convention will also be used when defining the classical limit.} is a polynomial function of the components of the rescaled collective spin. Unless stated otherwise, all calculations in this work are performed using the Hamiltonian in the form given in Eq.~\eqref{LMG_rescaled_TSS}.

\subsubsection{Symmetries and spectrum}\label{sec:symmetries_spectrum_LMG}

The LMG Hamiltonian \eqref{LMG_rescaled_TSS} is invariant under a global rotation of the collective spin by an angle $\pi$ around the $x$-axis, which transforms $\hat S_x \to \hat S_x$ and $\hat S_z \to -\hat S_z$. This symmetry operation is implemented by the unitary operator $e^{- i  \pi \hat S_x}$, which commutes with the Hamiltonian. Consequently, the spectrum decomposes into two symmetry sectors, with eigenvalues $\pm 1$ if $S\in\mathbb{N}$ and $\pm i$ if $S\in \mathbb{N}/2$\footnote{With a slight abuse of notation, we use in this work the notation $\mathbb{N}/2$ to denote the set of positive half-integers, i.e. $\mathbb{N}/2=\{1/2, 3/2, 5/2, \cdots\}$.}, see Appendix~\ref{app:LMG_symmetries}. 

At $h=0$, all energy levels except the one at zero energy are doubly degenerate\footnote{For $h=0$, one can prove that the eigenstates corresponding to degenerate pairs belong to different symmetry sectors.}, while at finite $h$, this degeneracy is lifted and the spectrum becomes nondegenerate, as can be verified numerically. 

In the thermodynamic limit $N\to\infty$, the LMG model undergoes a second-order ground-state quantum phase transition (QPT) at the critical point $h=J$ \cite{QPT_LMG_1, QPT_LMG_2}, accompanied by the spontaneous breaking of the $\mathbb{Z}_2$ spin-inversion symmetry ($\hat S_z \to -\hat S_z$). In the disordered phase ($h>J$), the ground-state expectation value $\langle \hat{S}_z \rangle$ vanishes and the transverse field dominates. In the ordered phase ($h<J$), the ground state becomes doubly degenerate, with one state in each symmetry sector, and $\langle \hat{s}_z \rangle$ increases continuously to $\pm 1$ when $h=0$. For finite $N$, no true transition exists. Instead, observables display nontrivial finite-size scaling behavior \cite{QPT_LMG_3, QPT_LMG_4}. 

\begin{figure}[t!]
    \centering
    \includegraphics[width=.45\textwidth]{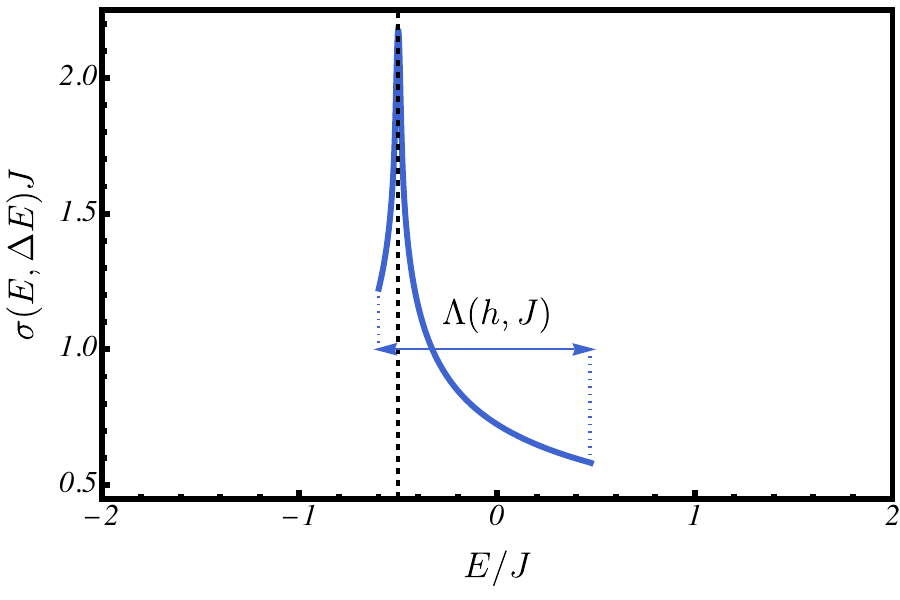}\qquad
    \includegraphics[width=.45\textwidth]{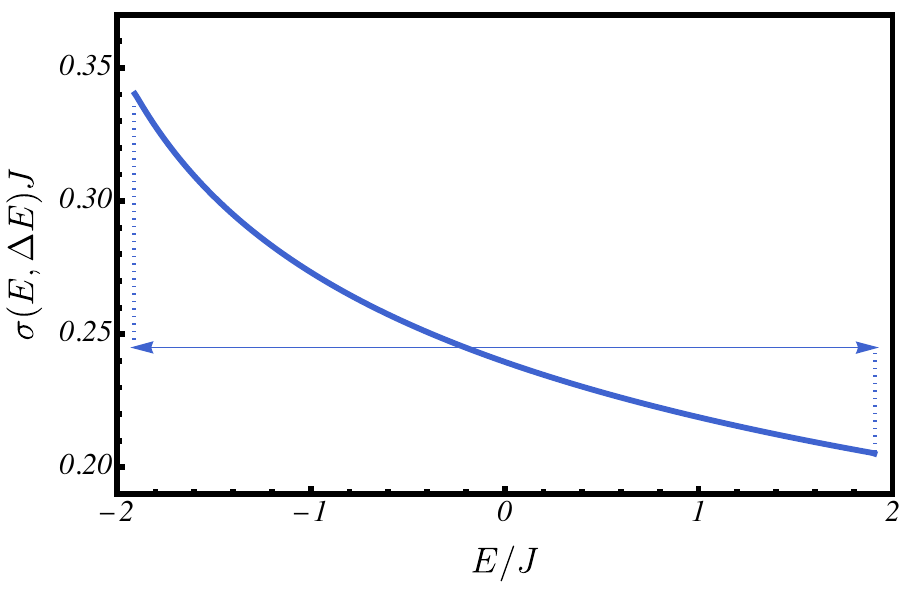}
    \includegraphics[width=.45\textwidth]{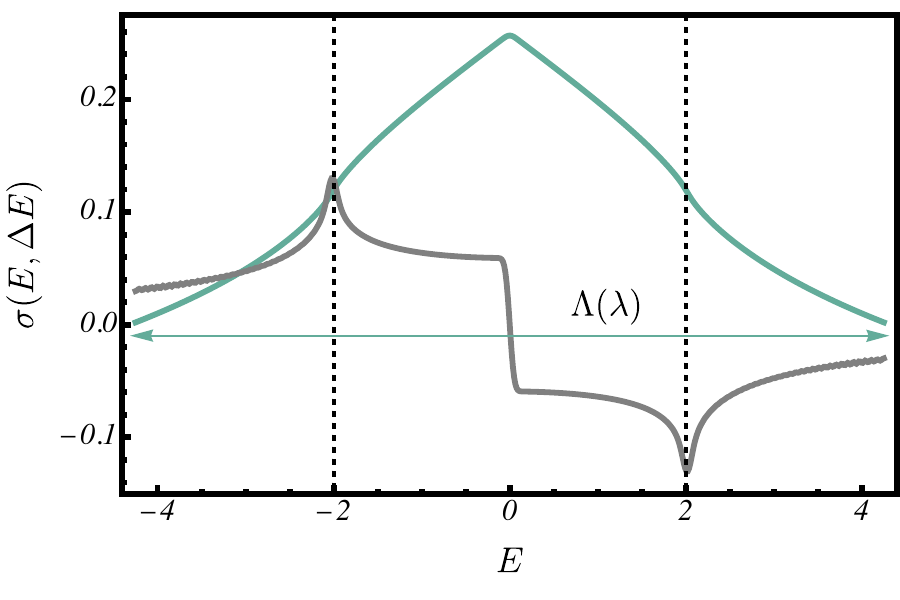}\qquad
    \includegraphics[width=.45\textwidth]{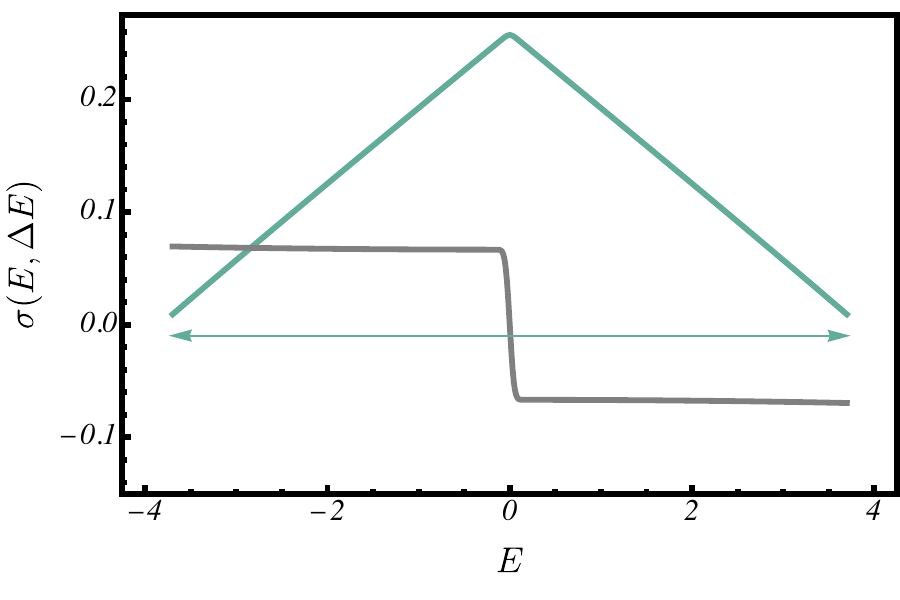}
    \caption{Normalized DOS for the full spectrum of the LMG and FP Hamiltonians, defined in Eqs.~\eqref{LMG_rescaled_TSS} and  \eqref{FP_rescaled}, respectively, obtained by exact diagonalization and smoothed using kernel density estimation. The DOS within individual symmetry sectors is qualitatively identical to that of the full spectrum. Varying the model parameters modifies the spectral bandwidth $\Lambda$ of the corresponding Hamiltonian. (\textbf{Upper row}) LMG model ($S=7500$) with \textbf{(left)} $h=1/2$, $J=1$ and \textbf{(right)} $h=2$, $J=1$. The vertical dashed line indicates the critical energy at $E=-h$. \textbf{(Lower row)} FP model ($L=60$) with \textbf{(left)} $\lambda=0$ and \textbf{(right)} $\lambda=0.9$. The gray curves denote the derivative of the DOS with respect to the energy $E$, and the vertical dashed lines indicate the critical energies $E=\pm 2(1+\lambda)$.}
    \label{fig:DOS_LMG_FP}
\end{figure}

The LMG spectrum can be analyzed through the density of states (DOS) $\sigma(E, \Delta E)$, such that $N(E,\Delta E)\equiv\sigma(E, \Delta E) \Delta E$ is the number of eigenstates in the energy window $[E-\Delta E/2, E+\Delta E/2]$. For $h<J$, the DOS exhibits a singularity at the critical energy $E=-h$, which becomes a true singularity in the thermodynamic limit (see upper row of Figure~\ref{fig:DOS_LMG_FP}). This singularity corresponds to an excited-state quantum phase transition (ESQPT) -- a QPT occurring at excited energies rather than in the ground state (see Ref.~\cite{ESQPT_review} for a recent review). For $h>J$, the ESQPT disappears and the DOS is a smooth function \cite{ESQPT_LMG}.

\begin{figure}
    \centering
    \includegraphics[width=0.45\linewidth]{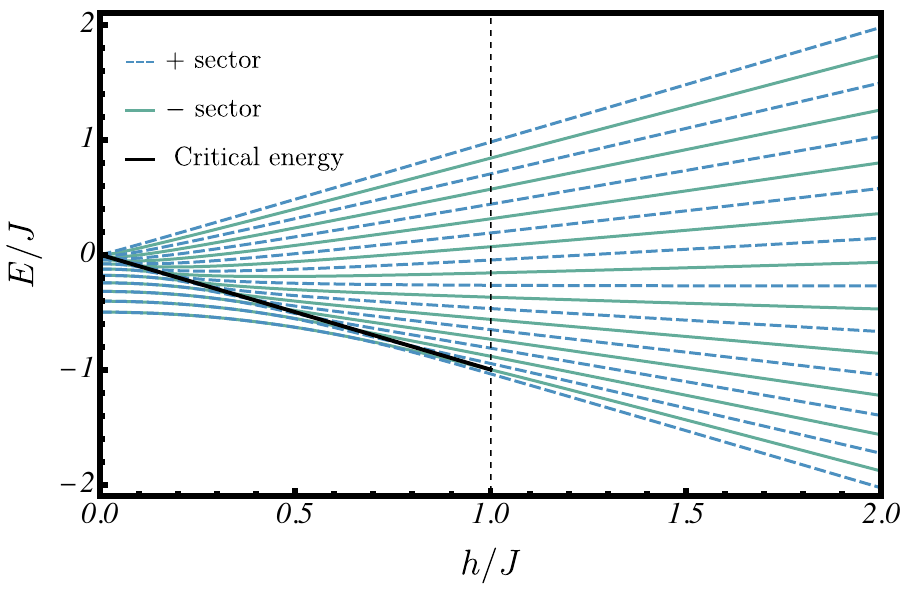}\qquad 
    \includegraphics[width=0.45\linewidth]{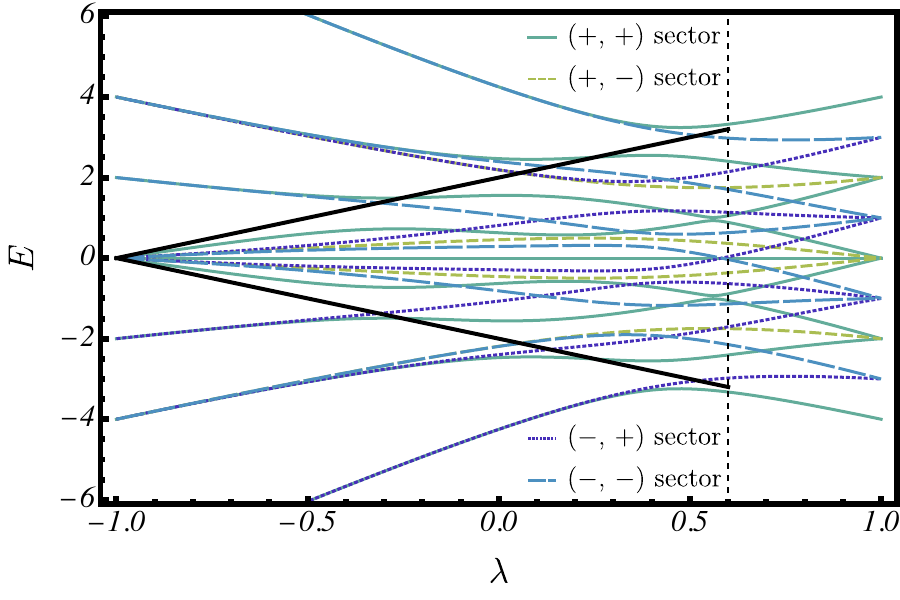}
    \caption{Energy spectra of the LMG and FP models. Colors indicate the symmetry sector of each eigenstate. Solid black lines mark the ESQPT critical energies, while vertical dashed lines indicate the ground-state QPT. \textbf{(Left)} LMG Hamiltonian \eqref{LMG_rescaled_TSS} ($S=10$) as a function of $h/J$ ($J=1$). For $h<J$, the spectrum exhibits quasi-degenerate pairs below the critical energy $E=-h$ (not resolvable at this scale). These quasi-degeneracies are lifted for $h>J$. \textbf{(Right)} FP Hamiltonian \eqref{FP_rescaled} ($L=3$) as a function of $\lambda$. The spectrum is shown in all four symmetry sectors, although only those associated with the $\hat{U}_1$ symmetry are relevant for detecting quasi-degeneracies, which are found within the region $|E|\geq 2(1+\lambda)$ for $-1<\lambda<3/5$.}
    \label{fig:LMG_FP_sectors}
\end{figure}

The QPT and the ESQPT can be understood within a unified framework. In the thermodynamic limit, the classical LMG Hamiltonian maps onto that of a one-dimensional particle in an effective potential \cite{ESQPT_general}. For $h<J$, the potential has two symmetric minima. At the critical point $h=J$, it is purely quartic, and for $h>J$ it has a single quadratic minimum. The QPT is thus associated with a structural change of the potential. For $h<J$, the spectrum consists of pairs of degenerate levels in alternating symmetry sectors, corresponding to wavefunctions localized in the two wells. For finite $N$, these degeneracies are lifted into quasi-degeneracies with a splitting of order $e^{-N}$ (see for example Ref.~\cite{Alessio}). This behavior is illustrated in the left panel of Figure~\ref{fig:LMG_FP_sectors}, which shows the lifting of these quasi-degeneracies upon crossing the critical energy. As discussed in Section~\ref{sec:late-time_K-complexity}, these quasi-degeneracies leave imprints in the late-time dynamics of K-complexity in LMG.

The dynamics generated by the classical LMG Hamiltonian undergoes a bifurcation at the critical point $h=J$, see Appendix~\ref{app:classical_LMG}. For $h<J$, where the quantum LMG model displays an ESQPT, the classical phase space contains an unstable saddle point. Such saddles are known to result in an initial exponential separation of nearby trajectories and Lyapunov dynamics, reminiscent of chaotic systems, that have recently been linked to quantum scrambling dynamics, fast operator growth in out-of-time-ordered correlation functions (OTOCs) and K-complexity \cite{Scrambling_chaos, saddle_Krylov, Krylov_SU2, Surprise_LMG}. In Sections \ref{sec:ClassicalKC} and \ref{sec:quantum_K_complexity}, we analyze to what extent the imprints of the model's integrability can be traced back despite the presence of the unstable saddle point, both in the classical and in the quantum regimes. 

\subsection{Feingold-Peres model}\label{sec:FP}

We now introduce the FP model, which can be viewed as a generalization of the LMG model involving two collective spins. It exhibits a tunable transition from integrability to chaos. Like the LMG model, it admits a classical description in the large-spin limit, whose phase space also features saddle points. This makes it a particularly compelling setting for investigating the competing effects of phase-space instabilities and integrability or chaos on both quantum and classical K-complexity.

\subsubsection{Presentation of the model}

The FP model is defined on the tensor product of two irreducible representations of $\mathrm{SU}(2)$ with Casimir quantum numbers $L$ and $M$. By a slight abuse of notation, we write $\hat{L}_\mu\equiv \hat{L}_\mu\otimes \mathbbm{1}$ and $\hat{M}_\mu \equiv \mathbbm{1}\otimes \hat{M}_\mu$ for the collective spin operators acting on the corresponding factor. They satisfy independent Lie algebras $\mathfrak{su}(2)$ and commute with one another, $[\hat L_\mu,\hat M_\nu]=0$. We take both spins to have the same magnitude, $L=M$, while retaining distinct notations for the two spins to distinguish between them. The Hilbert space has dimension $(2L+1)^2$ and is spanned by the basis $\left\{ \ket{m_1,m_2}\hspace{.1cm}|\hspace{.1cm}-L\leq m_1, m_2\leq L\right\}$ consisting of simultaneous eigenstates of $\hat L_z \otimes\hat M_z $. 

In terms of the rescaled collective spin operators, the FP Hamiltonian \cite{FP_1, FP_2} is given by
\begin{equation}\label{FP_rescaled}
    \frac{\hat{H}_{\text{FP}}}{L}=-\left(1+\lambda\right) (\hat l_z+\hat m_z)-4(1-\lambda)\hat l_x \hat m_x\, ,
\end{equation}
where energy units are implicitly set to $1$, matching the convention in Refs.~\cite{Universal_Growth_Hypothesis, FP_modern}. The FP model has been extensively studied \cite{Coupled_tops_1,Coupled_tops_2,Coupled_tops_QPT,Coupled_tops_classical_1,Coupled_tops_classical_2} and is known to exhibit a transition from integrability to chaos, both from the perspective of level spacing statistics and from that of operator matrix elements and eigenstate thermalization, see references~\cite{PhysRevA.30.504,FP_2,PhysRevA.34.591,FP_modern,Coupled_tops_2,Coupled_tops_classical_1}.

We note that the Hamiltonian in Eq.~\eqref{FP_rescaled} takes the form $\hat{H}=H( \hat{l}_\mu,\hat{m}_\nu)$, where $\hat H\equiv \hat{H}_{\text{FP}}/L$\footnote{See footnote \ref{fn:convention}.} is a polynomial function of the components of the two rescaled spins.

\subsubsection{Symmetries and spectrum}\label{sec:symmetries_spectrum_FP}

The FP Hamiltonian \eqref{FP_rescaled} is invariant under a simultaneous global rotation of both spins by an angle $\pi$ around the $z$-axis, which transforms $(\hat L_x, \hat M_x)\to -(\hat L_x, \hat M_x)$ and $(\hat L_z, \hat M_z)\to (\hat L_z, \hat M_z)$. This symmetry is implemented by the unitary operator $\hat{U}_1\equiv e^{- i\pi \left(\hat L_z + \hat M_z \right)}$, which commutes with the Hamiltonian and acts on the basis states as $\hat{U}_1\ket{m_1,m_2}=(- i)^{2 \left(m_1+m_2 \right)}\ket{m_1,m_2}$.

Within each $(L,L)$ sector, the FP Hamiltonian is also invariant under exchange of the two collective spins. The corresponding unitary operator $\hat U_2$ acts as $\hat{U}_2\ket{m_1,m_2}=\ket{m_2,m_1}$, and its (unnormalized) eigenstates are of the form $\ket{m_1,m_2}\pm \ket{m_2,m_1}$. Moreover, we have that $\hat{U}_2 (\hat{L}_\mu \otimes \hat{M}_\nu) \hat{U}_2^\dagger = \hat{M}_\nu \otimes  \hat{L}_\mu$, which immediately implies that $\hat{U}_2$ commutes with the FP Hamiltonian. 

Additionally, in the $(L, L)$ sector of the Hilbert space $\mathcal{H}$, the symmetry operators $\hat{U}_1$ and $\hat{U}_2$ commute, yielding the decomposition into invariant subspaces \cite{FP_modern, Haake} $\mathcal{H}_{(L,L)}=\mathcal{H}_{++}\oplus\mathcal{H}_{+-}\oplus\mathcal{H}_{-+}\oplus\mathcal{H}_{--}$ for $L\in\mathbb{N}$. The explicit form and dimensions of these subspaces are given in Appendix~\ref{app:FP_symmetries}. The restriction of the FP Hamiltonian \eqref{FP_rescaled} to the $\mathcal{H}_{-+}$ and $\mathcal{H}_{--}$ subspaces belongs to the orthogonal symmetry class, whereas its restriction to the $\mathcal{H}_{+-}$ and $\mathcal{H}_{++}$ subspaces belongs to the chiral orthogonal class \cite{FP_modern}\footnote{As long as $\lambda\neq \pm 1$, which is a special point in parameter space where the spectrum of the Hamiltonian contains many degeneracies. For $\lambda \in (-1,1)$, it can be numerically verified that the spectrum is nondegenerate in each symmetry sector.}. These nonstandard symmetry classes \cite{PhysRevLett.70.3852,PhysRevLett.72.2531,PhysRevLett.76.3420,Altland_nonstandard_symmetry,Zirnbauer:1996zz} are characterized by operators that anticommute with the Hamiltonian \cite{Altland_nonstandard_symmetry}. Defining $\hat V\equiv e^{- i \pi \hat L_x} e^{- i \pi \hat M_y}$, it can be verified that it anticommutes with the Hamiltonian \eqref{FP_rescaled}. Consequently, every eigenstate with energy $E$ has a partner with energy $-E$, so the spectrum is symmetric about $E=0$. It was shown in \cite{FP_modern} that the subspaces $\mathcal{H}_{++}$ and $\mathcal{H}_{+-}$ are invariant under the action of $\hat V$, while $\mathcal{H}_{--}$ and $\mathcal{H}_{-+}$ are exchanged by $\hat V$. The eigenstate associated with zero energy belongs to $\mathcal{H}_{++}$ if $L$ is even, and to $\mathcal{H}_{+-}$ if $L$ is odd \cite{FP_modern}.

In the limit $L\to \infty$, the FP Hamiltonian undergoes a second-order ground-state QPT at the critical point $\lambda=3/5$, accompanied by the spontaneous breaking of the $\mathbb{Z}_2$ spin-inversion symmetry $(\hat L_x, \hat M_x)\to -(\hat L_x, \hat M_x)$. The transition separates a disordered phase ($\lambda>3/5$) from an ordered one ($\lambda<3/5$), where $\langle \hat{L}_x \otimes \hat{M}_x\rangle$ serves as the order parameter.

Owing to the spectral symmetry under which energies appear in pairs $(-E,E)$, the DOS is symmetric about $E=0$, as shown in the lower row of Figure~\ref{fig:DOS_LMG_FP}. For $\lambda<3/5$, the FP model exhibits ESQPTs at critical energies $E=\pm 2(1+\lambda)$, appearing as singularities in the first derivative of the DOS \cite{Coupled_tops_2, ESQPT_FP}, as illustrated in Figure~\ref{fig:DOS_LMG_FP}. Consequently, they do not force any clustering of states.

The QPT is associated with a bifurcation of the classical phase space, see Appendix~\ref{sec:classical_FP}, where for $-1<\lambda<3/5$, the classical Hamiltonian has two symmetric minima that coalesce into a single minimum for $3/5<\lambda<1$. As in the LMG model, in the regime $-1<\lambda<3/5$, the spectrum consists of pairs of degenerate levels in alternating $\mathbb{Z}_2$ symmetry sectors. At finite $L$, these degeneracies are lifted into quasi-degeneracies. Figure~\ref{fig:LMG_FP_sectors} shows the spectrum of the FP model \eqref{FP_rescaled} as a function of $\lambda$. In particular, quasi-degeneracies occur only in the range $-1<\lambda<3/5$ and are lifted upon crossing one of the critical energies.

\subsection{Characterization of quantum integrability and chaos}\label{sec:quantum_integrability_chaos}

As mentioned earlier, the LMG model is integrable, whereas the FP model exhibits a transition from quantum integrability to chaos. To make these statements more precise, we review the standard spectral diagnostics of quantum chaos based on the statistics of consecutive energy-level spacings. These are used to identify the relevant dynamical regimes of both models in which the K-complexity analyses of sections \ref{sec:ClassicalKC} and \ref{sec:quantum_K_complexity} are carried out.

\subsubsection{General considerations}

Let $\hat H$ be a Hamiltonian with a nondegenerate spectrum $\{E_k\}$ with $k=1,2,\cdots$, ordered such that $E_k<E_{k+1}$. The energy-level spacings are defined as $s_k \equiv (E_{k+1}-E_k)/\Delta $, where $\Delta$ is the mean level spacing such that the mean value of $\{s_k\}$ is one. For quantum integrable systems, the distribution of the set $\{s_k\}$ follows a Poisson law $P(s)=e^{-s}$ \cite{BT_conjecture}, whereas for quantum chaotic systems in the Gaussian Orthogonal Ensemble (GOE), the distribution follows the Wigner-Dyson distribution, $P(s)=\pi s \,e^{-\pi s^2/4}/2$ \cite{BGS_conjecture}. In order to probe the integrability of a quantum system, one can compute the ratio of consecutive level spacings $r_k$ \cite{r_ratio_Huse}, defined as
\begin{equation}
\label{rstat_def}
    r_k\equiv \frac{\text{min}(s_k,\, s_{k+1})}{\text{max}(s_k,\, s_{k+1})}=\text{min}\left(\frac{s_{k+1}}{s_k},\frac{s_k}{s_{k+1}}\right) ,
\end{equation}
where by construction $0\leq r_k \leq 1$. For a Poisson distribution of energy-level spacings, the corresponding probability distribution is $P(r)=(1+r)^{-2}$ with mean value $\langle r \rangle=2\ln{2}-1\simeq 0.386$. For sufficiently large GOE matrices, the probability distribution for $r$ is accurately described by $P(r)=27(r+r^2)(1+r+r^2)^{-5/2}/8$ and $\langle r\rangle=4-2\sqrt{3}\simeq0.536$ \cite{RMT_math}. The mean ratio $\langle r \rangle$ thus provides a convenient diagnostic for distinguishing integrable from chaotic quantum systems.

\subsubsection{Lipkin-Meshkov-Glick model}

\begin{figure}[t!]
    \centering
    \includegraphics[width=.45\textwidth]{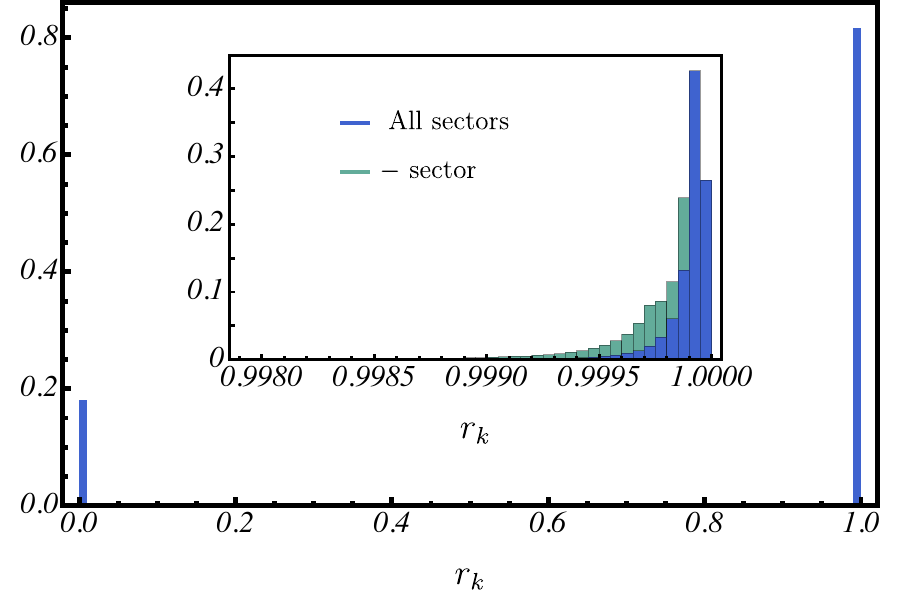}\qquad 
    \includegraphics[width=.45\textwidth]{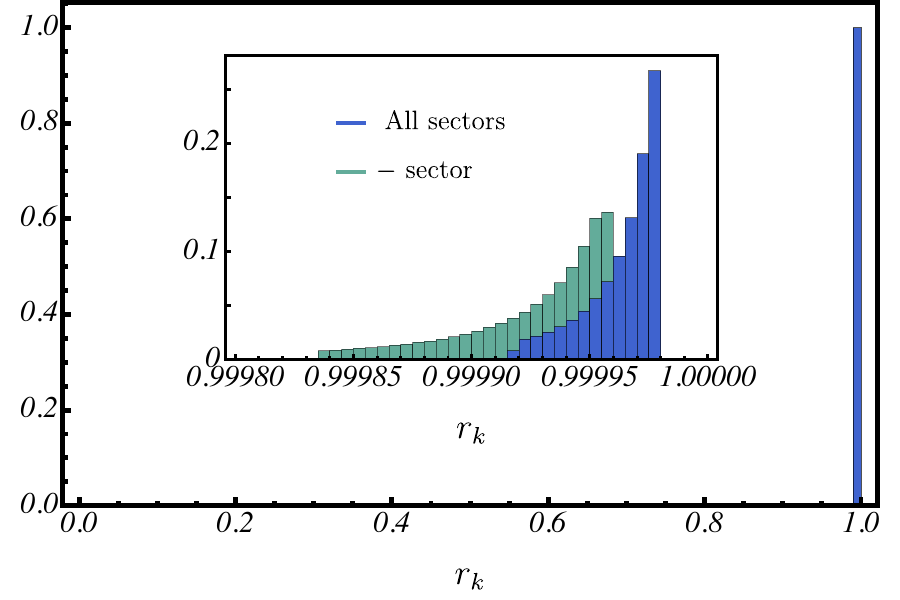}
    \caption{Distribution of the ratio of consecutive level spacings $r_k$ for the full spectrum of the LMG Hamiltonian \eqref{LMG_rescaled_TSS} ($S=7500$) with parameters \textbf{(left)} $h=1/2$, $J=1$ and \textbf{(right)} $h=2$, $J=1$. For $h<J$, quasi-degenerate pairs from different symmetry sectors produce a peak near $r_k=0$ in the full-spectrum statistics. For $h>J$, quasi-degeneracies are absent and the values of $r_k$ cluster near $1$. The insets show a magnified view of the distribution near $r_k=1$ for the full spectrum and for a single symmetry sector (the two sectors overlap). Note the different horizontal scales between the two panels. In all cases, the mean ratio $\langle r \rangle$ remains close to $1$.}
\label{fig:quantum_integrability_chaos_1}
\end{figure}

The low- and high-energy eigenstates of the LMG Hamiltonian are highly correlated and can be described by bosonic excitations obtained from a Holstein–Primakoff expansion \cite{Holstein_Primakoff} around the classical ground state \cite{Alessio, frequency_Heiss}. The low-energy spectrum of the Hamiltonian in Eq.~\eqref{LMG_rescaled_TSS} scales as $E_k\sim \sqrt{J^2-h^2}\,k/2$ for $h<J$ and $E_k\sim \sqrt{h(h-J)}\,k$ for $h>J$, where $k=0,1,2,\cdots$. The high-energy spectrum has the same form in both regimes, $E_k\sim \sqrt{h(h+J)}\,k$. Such scalings also hold within each symmetry sector (up to an irrelevant numerical prefactor). As one can check numerically, see Figure~\ref{fig:LMG_FP_sectors}, the bulk of the spectrum is similarly rigid (except for energies near the critical energy). Consequently, the level-spacing ratios $r_k$ are concentrated around $1$, see Figure~\ref{fig:quantum_integrability_chaos_1}.

The situation for the LMG model is qualitatively similar to that of a single harmonic oscillator, whose equally spaced spectrum is not Poissonian: in the context of the Berry-Tabor (BT) conjecture \cite{BT_conjecture}, this non-generic behavior of the harmonic oscillator is known to be fragile, in the sense that a large number of degrees of freedom combined with any weak perturbation (such as a small nonlinear coupling term, or a small amount of disorder) eventually yields a Poissonian spectrum \cite{Level_spacings_harmonic}. The LMG model has only one collective spin degree of freedom, which explains why the BT conjecture is not immediately applicable. In this sense, the LMG model is not a ``generic'' many-body system.

\begin{figure}[t!]
    \centering
    \includegraphics[width=.45\textwidth]{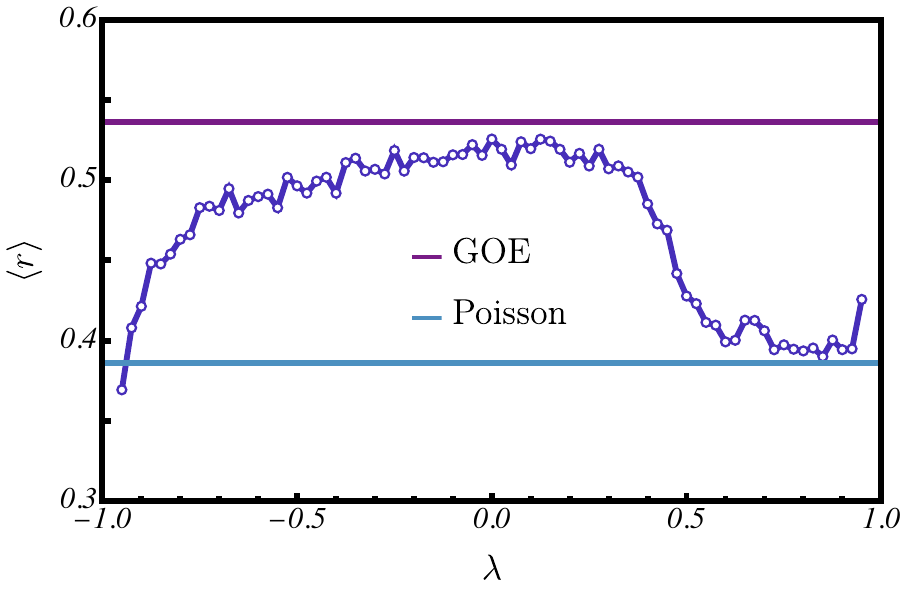}
    \caption{Mean ratio $\langle r \rangle$ of the FP Hamiltonian \eqref{FP_rescaled} ($L=60$) in the $\mathcal{H}_{-+}$ subspace, shown as a function of the parameter $\lambda$. A regular-to-chaotic transition is observed as $\lambda$ decreases from $1$. The solid horizontal lines correspond to the Poisson and GOE predictions. In the limit $\lambda\to 1$, the Hamiltonian becomes proportional to $\hat{l}_z+ \hat{m}_z$, yielding a highly regular spectrum that departs from Poisson statistics.}
    \label{fig:quantum_integrability_chaos_FP}
\end{figure}

The integrability of the LMG model nevertheless remains manifest and is often investigated from the perspective of algebraic integrability, in which the spectrum is obtained by solving algebraic equations for the roots, thereby highlighting its highly structured nature. This includes approaches based on the Bethe ansatz (see references above), Richardson-Gaudin integrable models \cite{LMG_integrable4, Richardson_Gaudin}, the zeros of Husimi function \cite{Integrability_Husimi}, and deformed polynomial algebras \cite{LMG_deformed_polynomials}, among others.

\subsubsection{Feingold-Peres model}

The FP model has two degrees of freedom and thus provides a minimal setting in which the BT and Bohigas-Giannoni-Schmit (BGS) \cite{BGS_conjecture} conjectures apply. The FP Hamiltonian in the $\mathcal{H}_{-+}$ and $\mathcal{H}_{--}$ subspaces interpolates between GOE and Poisson statistics as a function of the coupling parameter $\lambda$. This crossover is captured by the mean level-spacing ratio $\langle r\rangle$, signaling a transition between quantum integrability and chaos. Accordingly, the classical limit of the FP Hamiltonian exhibits a corresponding transition from regular to chaotic dynamics \cite{FP_1}. In Figure~\ref{fig:quantum_integrability_chaos_FP}, we compute the mean level-spacing ratio for various values of $\lambda$ over the interval $\lambda\in [-1,1]$. Whenever $\langle r\rangle$ approaches its Poisson or GOE value, the full level-spacing distribution agrees with the corresponding theoretical prediction.

\subsection{Krylov complexity}\label{sec:Krylov_toolbox}

In this section, we briefly review the Lanczos algorithm and the notion of K-complexity for operators, emphasizing the concepts used throughout this work. For further details, we refer the reader to the references~\cite{Rabinovici:2025otw,Baiguera:2025dkc,Nandy:2024evd}.

Consider a quantum system with a $D$-dimensional Hilbert space $\mathcal{H}$. Observables are Hermitian operators in $\mathrm{End}(\mathcal{H})$, the $D^2$-dimensional operator space. Since $\mathrm{End}(\mathcal{H})$ is itself a vector space, operators $\hat{\mathcal{O}}$ can be represented as states $\ketc{\mathcal{O}}$. In this notation, the time evolution of operators in the Heisenberg picture is given by $|\mathcal{O}(t))=e^{it\mathcal{L}}|\mathcal{O})$ where the Liouvillian is defined by $\mathcal{L}\, \cdot \equiv [\hat{H},\cdot]$ and serves as the generator of time evolution. The Krylov space generated by $\hat{\mathcal{O}}$ is
\begin{equation}
    \label{eq:Krylov_space_def}
    \mathcal{H}_{\hat{\mathcal{O}}}\equiv\mathrm{span}\left\{\mathcal{L}^n|\mathcal{O})\right\}_{n\geq 0}\,.
\end{equation}
Given an inner product on operator space, the Lanczos algorithm \cite{Lanczos_algorithm_historical,Krylov:1931} builds an orthonormal Krylov basis $\{|\mathcal{O}_n)\}_{n=0}^{K-1}$, where $K$ is the Krylov dimension. Assuming $\hat{\mathcal{O}}$ is normalized, the first basis element is $\ketc{\mathcal{O}_0}=\ketc{\mathcal{O}}$, and each vector $\mathcal{L}^n|\mathcal{O})$ is a linear combination of the first $n$ Krylov basis elements. The algorithm also produces the Lanczos coefficients $\{a_n\}_{n=0}^{K-1}$ and $\{b_n\}_{n=1}^{K-1}$\footnote{The coefficients $a_n$ vanish if and only if the autocorrelation function of the initial operator $(\mathcal{O} |\mathcal{O}(t))$ is even in $t$, see for example \cite{Universal_Growth_Hypothesis}. This holds for Hermitian operators when using the regularized inner product \eqref{eq:Operator_Inner_product_sym} with a density matrix commuting with the Hamiltonian, $[\hat H,\hat \rho]=0$, and, in particular, at infinite temperature, where $\hat \rho=\mathbbm{1}$. The coefficients $b_n$ are strictly positive for $n=1,\dots,K-1$, as they arise as norms in the Lanczos algorithm, see Appendix~\ref{appx:Lanczos_Lanczos} for further details.
}. The Krylov basis elements satisfy the following recursion relation
\begin{equation}
    \label{eq:Krylov_elements_recursion}
    \mathcal{L}|\mathcal{O}_n) = b_{n+1}|\mathcal{O}_{n+1}) + a_n|\mathcal{O}_n) + b_n|\mathcal{O}_{n-1})\,.
\end{equation}

Additionally, it is useful to note that the Lanczos coefficients $\{a_n,b_n\}$ are in one-to-one correspondence with the moments $\mu_{n}$ of the operator autocorrelation function, defined through $\left(\mathcal{O}|\mathcal{O}(t)\right)=\sum_{n\geq 0}(it)^n \mu_n/n!$. We refer the reader to Refs.~\cite{Rabinovici:2025otw,Universal_Growth_Hypothesis} for a detailed discussion of this so-called recursion method. For our purposes, it suffices to note that the moments $\mu_{2k}$ with $k=0,\dots,n$ determine the coefficients $a_k$ and $b_k$, and vice versa\footnote{Note that $a_n=0$ for all $n$ if and only if $\mu_{2n+1}=0$ for all $n$, i.e. if and only if the autocorrelation function is an even function of $t$.}.

The Krylov dimension associated with an operator $\hat{\mathcal{O}}$ equals the number of Liouvillian eigenspaces onto which $|\mathcal{O})$ has a nonzero projection. Let $|\omega_{ab})\equiv |E_a\rangle\langle E_b|$ be the Liouvillian eigenstates with eigenvalues $\omega_{ab}\equiv E_a-E_b$, with $a,b=1,\cdots, D$. The Krylov space is spanned by the so-called eigenspace representatives, defined as the projections of $|\mathcal{O})$ onto each Liouvillian eigenspace. The representative associated with a given phase $\omega$ is denoted by $\ketc{\mathcal{K}_\omega}$ and defined as
\begin{equation}
    \ketc{\mathcal{K}_\omega}=\sum_{(a,b)\,\in\, I} \mathcal{O}_{ab} \, \ketc{\omega_{ab}}\, ,
\end{equation}
where the set $I$ contains all index pairs $(a,b)$ satisfying $\omega_{ab}=\omega$, and $\mathcal{O}_{ab}\equiv \mel{E_a}{\mathcal{O}}{E_b}$ are the matrix elements of the initial operator in the energy eigenbasis. The Krylov dimension is then simply the number of nonzero eigenspace representatives,
\begin{equation}\label{Krylov_dimension_K_omega}
    K=\text{card}\left\{\ketc{\mathcal{K}_\omega}\neq 0 \hspace{.1cm}|\hspace{.1cm} \omega \in \text{spec}(\mathcal{L}) \right\} .
\end{equation}
Operationally, $K$ is obtained by counting the number of distinct phases $\omega_{ab}$ for which $\mathcal{O}_{ab} \neq 0$. In practice, however, finite-precision arithmetic makes it difficult to determine whether a matrix element vanishes exactly. Symmetries provide a way around this issue by imposing selection rules that force certain matrix elements to be identically zero. This allows one to derive analytical upper bounds on the Krylov dimension, which are typically saturated in strongly interacting integrable systems once all symmetries have been resolved, since matrix elements not forbidden by symmetry are typically nonzero \cite{PhysRevB.102.075127}. In Section~\ref{sec:K_dim}, we exploit this approach to derive analytical upper bounds for the Krylov dimension of the LMG and FP models.

The choice of inner product on operator space plays a central role, as it determines how different regions of the spectrum contribute to Krylov dynamics. Throughout this work, we shall mainly consider inner products of the form
\begin{equation}
    \label{eq:Operator_Inner_product}
    \left(\mathcal{A}|\mathcal{B}\right)=\frac{1}{\text{Tr}[\hat{\rho}]}\text{Tr}\left[\hat{\rho} \hat{\mathcal{A}}^\dagger \hat{\mathcal{B}}\right],\quad \text{for any }\hat{\mathcal{A}},\hat{\mathcal{B}}\in\mathrm{End}(\mathcal{H})\, ,
\end{equation}
where $\hat \rho$ is the density matrix specifying the state in which correlation functions are taken. Choosing $\hat \rho=\mathbbm{1}$ yields the infinite-temperature inner product, while $\hat \rho = e^{-\beta \hat H}$ and a projector onto an energy shell define the canonical and microcanonical inner products, respectively. In Section~\ref{sec:quantum_K_complexity}, we consider both infinite-temperature and microcanonical inner products to probe different regions of the LMG and FP spectra. 

A commonly used alternative is the regularized inner product, in which the density matrix is distributed between the operator insertions \cite{Recursion_method_book,Universal_Growth_Hypothesis}. In the thermal case, it turns the inner product into a thermal correlator on some Schwinger-Keldysh contour where operators are arranged along the thermal circle\footnote{These choices of inner products are often convenient in quantum field theory contexts \cite{Universal_Growth_Hypothesis,Dymarsky:2021bjq,Avdoshkin:2022xuw,Camargo:2022rnt}. See \cite{Kundu:2023hbk} for a different choice of inner product, based on the operator-state correspondence, in the context of conformal field theory.}. An example is the inner product with symmetric arrangement
\begin{equation}
    \label{eq:Operator_Inner_product_sym}
    \left(\mathcal{A}|\mathcal{B}\right)^{(w)}=\frac{1}{\text{Tr}[\hat{\rho}]}\text{Tr}\left[\sqrt{\hat{\rho}} \hat{\mathcal{A}}^\dagger \sqrt{\hat{\rho}} \hat{\mathcal{B}}\right] ,
\end{equation}
often referred to as the Wightman inner product, hence the superscript ``$^{(w)}$''.
Note that both inner products in Eqs.~\eqref{eq:Operator_Inner_product} and \eqref{eq:Operator_Inner_product_sym} coincide in the infinite-temperature limit. 

Since the basis elements $|\mathcal{O}_n)$ form an orthonormalized version of the vectors $\mathcal{L}^n|\mathcal{O})$, they retain a natural ordering associated with time evolution and are progressively explored in time by $ |\mathcal{O}(t))$. This motivates the definition of K-complexity as the mean position of $ |\mathcal{O}(t))$ on the Krylov chain \cite{Universal_Growth_Hypothesis},
\begin{equation}
    \label{eq:KC_from_wavefn}
    C_K(t)=\sum_{n=0}^{K-1}n|\varphi_n(t)|^2 ,
\end{equation}
where the Krylov wavefunctions are defined by $\varphi_n(t)=(-i)^n\braketc{\mathcal{O}_n}{\mathcal{O}(t)}$. For a normalized operator, unitarity implies that $\sum_{n=0}^{K-1}|\varphi_n(t)|^2=1$ at all times, and $\varphi_n(0)=\delta_{n,0}$. If $a_n=0$ for all $n$, then $\varphi_n(t)\in\mathbb{R}$ at all (real) times\footnote{This follows from the fact that, when all $a_n$ vanish, the operators $i^n\hat{\mathcal{O}}_n$ are Hermitian.}. Following \cite{Adrian_2}, the late-time saturation value of K-complexity is given by
\begin{equation}
    \label{eq:KC_LongTimeAvg}
    \overline{C_K}\equiv\lim_{T\to \infty}\frac{1}{T}\int_0^T \dd t\,C_K(t) = \sum_{n,i=0}^{K-1}n\left|(\mathcal{O}_0|\omega_i)\right|^2\left|(\omega_i|\mathcal{O}_n)\right|^2,
\end{equation}
where the $\ketc{\omega_i}$ are the eigenvectors of the Liouvillian in the Krylov basis. $\overline{C_K}$ can be compactly expressed as $\overline{C_K}= \sum_{n=0}^{K-1} n Q_{0n}$ where the $Q_{0n}$ denote the late-time averages of the Krylov wavefunctions,
\begin{equation}\label{Q_0n}
    Q_{0n}=\lim_{T\to \infty}\frac{1}{T}\int_0^T\dd t\,|\varphi_n(t)|^2=\sum_{i=0}^{K-1}\left|(\mathcal{O}_0|\omega_i)\right|^2\left|(\omega_i|\mathcal{O}_n)\right|^2 ,
\end{equation}
which are normalized such that $\sum_{n=0}^{K-1} Q_{0n}=1$.

In later sections, we characterize integrable and chaotic dynamics in the LMG and FP models through the K-complexity of collective spin operators, both at infinite temperature and within microcanonical energy shells, with a focus on the early-time growth and the late-time saturation value of K-complexity. The former is the subject of the universal operator growth hypothesis \cite{Universal_Growth_Hypothesis}, according to which maximally chaotic systems exhibit linearly growing Lanczos coefficients, $b_n\sim \alpha n$, in the thermodynamic limit. This implies exponential growth of K-complexity with a so-called Krylov exponent $\lambda_K=2\alpha$, conjectured to bound the Lyapunov exponent $\lambda_L$ of the four-point OTOCs\footnote{This bound was proven in \cite{Universal_Growth_Hypothesis} at infinite temperature, and remains conjectural at finite temperature due to the ambiguity in the regularization of the inner product. For further details, see \cite{Rabinovici:2025otw} and references therein.}. For finite systems, the saturation value of K-complexity has been argued to be sensitive to chaos \cite{Adrian_2,Adrian_3}. This sensitivity can be understood in terms of localization effects in the Krylov chain that get enhanced in integrable systems, lowering the saturation value -- see \cite{Rabinovici:2025otw} and references therein for details.

Finally, we discuss the normalization of Lanczos coefficients. To compare different Hamiltonians, it is convenient to use common energy units. Since the Lanczos coefficients are matrix elements of the Liouvillian, rescaling the Hamiltonian by a constant rescales all $b_n$ by the same factor. A sensible option is the spectral bandwidth $\Lambda$, such that the width of the spectrum of the (dimensionless) normalized Hamiltonian is one. This, in turn, sets the time unit equal to the inverse bandwidth, which can be understood as the typical time step in a quantum system at infinite temperature. From a more mathematical perspective, for finite-bandwidth systems, the Lanczos coefficients approach a plateau $b_n\to b_\infty\propto \Lambda$ as $n\to\infty$ \cite{Recursion_method_book,barbon2019}, making this normalization particularly convenient. Alternatively, one may normalize by the mean level spacing $\Delta = \Lambda/(D-1)$, so that the natural time unit becomes the Heisenberg time $t_H\equiv 1/\Delta$ \cite{Altland:2020ccq}. We adopt this convention in Section~\ref{sec:ClassicalKC} to compare the quantum and classical Lanczos coefficients of the collective spin models studied here.

\section{Classical Krylov complexity}\label{sec:ClassicalKC}

As discussed in Section~\ref{sec:Krylov_toolbox}, the Lanczos algorithm provides a completely general framework for defining K-complexity: it only requires a Hilbert space structure, a generator of time evolution, and an initial condition. In particular, it is known \cite{Recursion_method_book} that the Lanczos algorithm can be applied to classical phase space dynamics generated by a Hamiltonian flow, equipped with the standard inner product of classical statistical mechanics. In this section, we use this formalism to define a classical notion of K-complexity that is generally applicable to dynamics on symplectic manifolds and to study its relation to the quantum Lanczos algorithm and the associated K-complexity in quantum systems with a classical limit. In addition, we propose the use of a classical microcanonical inner product as a tool to scan across energy shells in phase space and to define the corresponding classical microcanonical K-complexity within an energy window.

Collective spin models such as LMG and FP are examples of quantum systems that become effectively classical in the thermodynamic limit. Furthermore, the phase space of their classical limit features interesting properties, such as the existence of unstable points that are related to excited-state quantum phase transitions of their quantum counterparts \cite{ESQPT_review,ESQPT_general,PhysRevE.66.016217,ESQPT_FP}. For instance, in the LMG model, the phase space saddle point has been qualitatively associated with the early-time exponential growth of operator K-complexity observed in the corresponding quantum Hamiltonian, see Refs.~\cite{Nandy:2024evd,Krylov_SU2}. In order to systematically analyze these properties, it appears natural to develop a classical K-complexity formalism, as it provides a direct way to analyze the influence of unstable phase space saddles on the classical Krylov space dynamics and to understand how this relates to the corresponding quantum computation. 

In Section~\ref{subsec:Classical_Krylov_Generic_setup}, we present the general framework of classical K-complexity, and in Section~\ref{subsec:Classical_Limit_Krylov}, we use the phase space formulation of quantum mechanics to argue that this definition arises naturally as the classical limit of the quantum Lanczos algorithm. As a byproduct, we define a Krylov-Ehrenfest scale which characterizes the departure between classical and quantum K-complexities on the Krylov chain and aligns with the standard definition of the Ehrenfest time, once translated into the time domain. Subsequently, in Section~\ref{sec:general_prescription}, we specialize this framework to classical collective spin models, whose phase space manifold is the sphere (or Cartesian product of spheres), and, in particular, we explain  how the Lanczos algorithm can be implemented on such manifolds. Sections~\ref{sec:classical_LMG} and \ref{sec:classical_FP} are devoted to the study of the classical Lanczos sequences for the LMG and FP models at infinite temperature. Finally, in Section~\ref{sec:classical_microcanonical_Lanczos}, we compute the classical microcanonical Lanczos sequence and K-complexity profile for the LMG model. The results show that the influence of the unstable saddle point in phase space is maximal in the energy shell containing the saddle and becomes suppressed in shells away from it. 

\subsection{Krylov complexity in phase space}\label{subsec:Classical_Krylov_Generic_setup}

In this section, we describe the generic framework for computing K-complexity in the phase space of a classical system. Earlier discussions of the classical Lanczos algorithm and K-complexity can be found in \cite{Recursion_method_book,Universal_Growth_Hypothesis}. Here, we expand on this construction and lay out the necessary details for computing classical K-complexity in (compact) symplectic manifolds.

In classical mechanics, observables are described by functions $f$ defined on a phase space manifold $\mathcal{M}$, which is equipped with a symplectic two-form $\omega^{(2)}$ \cite{Arnold}. To each phase space function $f$, one can associate a Hamiltonian vector field $X_f$ defined implicitly by $\iota (X_f)\,\omega^{(2)}=df$ , where $\iota(X_f)$ is the interior product. The Poisson bracket of two functions $f$ and $g$ is then given by $\left\{f,g\right\}=\omega^{(2)}(X_f,X_g)$, and one may write $X_f\equiv \{\cdot,f\}$. 

With these definitions, the time evolution of a given observable $f$ along classical trajectories is generated by the Hamiltonian vector field associated with the Hamiltonian function $H$, namely $X_H\equiv\{\cdot,H\}$. Assuming that $f$ has no explicit time dependence, its time evolution is given by
\begin{equation}\label{eq:classical_Liouvillian_operator}
    \dv{f}{t}=\left\{f,\,  H \right\}=i\, \mathcal{L}_c f\, , \qquad \mathcal{L}_c f \equiv i \left\{H, f \right\},
\end{equation}
where, for later convenience, we have defined the classical Liouvillian as $\mathcal{L}_c \equiv-iX_H\equiv i\{H,\cdot\}$\footnote{We have implicitly promoted the space of phase space functions to be a space of complex-valued functions.}. The solution of \eqref{eq:classical_Liouvillian_operator} is obtained by exponentiating the time evolution generator,
\begin{equation}
\label{eq:Classical_evolution_generic}
    f(t)= e^{tX_H}f=e^{it\mathcal{L}_c}f\,.
\end{equation}

To formulate a well-defined Krylov problem, the remaining ingredient is to equip the space of phase space functions $\mathcal{F}(\mathcal{M})$ with an inner product, thereby promoting it to a Hilbert space. As anticipated in the introduction to Section \ref{sec:ClassicalKC}, we adopt the standard inner product in classical statistical mechanics, 
\begin{equation}
    \label{eq:Classical_Inner_product_generic}
    (f|g)=\frac{1}{Z}\int_{\mathcal{M}}f^{*}g\,p\,\omega\,,\quad \text{for any }f,g\in\mathcal{F}(\mathcal{M})\,,
\end{equation}
where $\omega$ is the volume form in the symplectic manifold\footnote{For a $2N$-dimensional symplectic manifold, the volume form is given by $\omega_N = \left(\omega^{(2)}\right)^{\wedge N}/N!$. In the discussion above, we denote it as $\omega$ for simplicity.} and we use the notation $|\cdot)$ to explicitly denote elements of the Hilbert space obtained by endowing $\mathcal{F}(\mathcal{M})$ with this inner product\footnote{We use the same notation as for the elements of the operator Hilbert space $\text{End}(\mathcal{H})$. This slight abuse of notation was judged convenient given that, as we shall discuss later, there is a sense in which $\mathcal{F}(\mathcal{M})$ is the classical limit of $\text{End}(\mathcal{H})$.}. The function $p$ defines the statistical ensemble, and the normalization constant is $Z=\int_{\mathcal{M}} p\,\omega$. In the canonical ensemble, $p=e^{-\beta H}$ for some inverse temperature $\beta$, and $Z\equiv Z(\beta)$ becomes the canonical partition function. In Section~\ref{sec:classical_microcanonical_Lanczos}, we will instead consider $p$ to be a microcanonical indicator function in order to scan over energy shells in phase space. With respect to the inner product in Eq.~\eqref{eq:Classical_Inner_product_generic}, the classical Liouvillian $\mathcal{L}_c$ is Hermitian\footnote{\label{footnote:boundary_terms}This holds if no boundary terms are picked up in the application of Stokes' theorem, which is guaranteed in compact manifolds. More generally, for hermiticity of the classical Liouvillian to hold, one might need to restrict $\mathcal{F}(\mathcal{M})$ to sufficiently fast-decaying functions.}. As a result, the time evolution operator $e^{it\mathcal{L}_c}$ is unitary, and the norm of $f(t)$ is preserved under the Liouvillian flow in Eq.~\eqref{eq:Classical_evolution_generic}.

Up to this point, we have intentionally worked in coordinate-free notation. This makes explicit the close analogy between the present setup and the formalism of quantum operators reviewed in Section \ref{sec:Krylov_toolbox}, and shows that it is readily amenable to the framework of Krylov methods. From Eq.~\eqref{eq:Classical_evolution_generic}, we define the classical Krylov space associated with an observable $f$ as the following subspace of $\mathcal{F}(\mathcal{M})$,
\begin{equation}
    \label{eq:Classical_Krylov_space_def}
    \mathcal{H}^{(c)}_f\equiv\mathrm{ span}\left\{\mathcal{L}_c^n|f)\right\}_{n\geq 0}\,.
\end{equation}
Unless the phase space function $f$ happens to overlap with a finite subset of the eigenfunctions of $\mathcal{L}_c$, the corresponding classical Krylov space will typically be infinite-dimensional\footnote{The recent work \cite{Das:2026hbw} proposes a notion of classical K-complexity based on the application of the Lanczos algorithm in a specific finite-dimensional truncation of phase space, dubbed the Koopman-Krylov space. We do not make use of this approximation and, as we will show in Section \ref{subsec:Classical_Limit_Krylov}, our construction arises naturally as the classical limit of quantum K-complexity.}. An orthonormal Krylov basis $|f_n)$ can then be constructed by applying the Lanczos algorithm in this classical setting. The classical Lanczos algorithm is entirely analogous to the quantum version for operators described in Appendix \ref{appx:Lanczos_Lanczos}, and we therefore do not repeat it here. Specifically, the algorithm is entirely determined once a Hilbert space with an inner product, an element of that space serving as initial condition, and a time evolution generator acting on it are given. The classical Lanczos algorithm is obtained from taking the inner product to be that of Eq.~\eqref{eq:Classical_Inner_product_generic}, replacing $\hat{\mathcal{O}}$ by $f$, and $\mathcal{L}$ by $\mathcal{L}_c$. The output consists of a (typically infinite) orthonormal Krylov basis $\{|f_n)\}_{n\geq 0}$ and a sequence of classical Lanczos coefficients $b^{(c)}_n$, as the $a^{(c)}_n$ coefficients vanish identically, as we will momentarily prove. As for the quantum case, the classical Liouvillian is tridiagonal in the Krylov basis, see Eq.~\eqref{eq:Krylov_elements_recursion}, and one may define the classical K-complexity as
\begin{equation}
    \label{eq:Classical_KC}
    C_K^{(c)}(t)=\sum_{n\geq 0}n\left|\braketc{f_n}{f(t)}\right|^2.
\end{equation}

We are not aware of any analytical upper bounds for the growth of the classical Lanczos coefficients, but in generic nonlinear systems they typically grow linearly\footnote{In the context of fluctuation-dissipation theorems, $b_n\sim n^\delta$ with $\delta>1$ is not compatible with an entire dissipation function, see Section 5.4 of \cite{Recursion_method_book}.} $b^{(c)}_n\sim \alpha n$, which yields \cite{Universal_Growth_Hypothesis} an exponentially growing K-complexity with Krylov exponent $\lambda_K=2\alpha$. In the classical setting, the bound $\lambda_L\leq \lambda_K$, where $\lambda_L$ is the classical Lyapunov exponent, is not proven but has been conjectured \cite{Universal_Growth_Hypothesis}\footnote{In the context of spin models, the authors of \cite{Universal_Growth_Hypothesis} proved a weaker bound, $\lambda_L\leq 2\lambda_K$. They additionally performed numerical checks in the FP model which suggested compatibility with the conjectured version $\lambda_L\leq \lambda_K.$}.
Finally, the classical Lanczos coefficients satisfy a classical version of the recursion method, as they are in one-to-one correspondence with the Taylor series coefficients (i.e. moments) of the classical autocorrelation function $C(t)=(f|f(t))$. The aforementioned vanishing of the $a_n^{(c)}$ coefficients implies that $C(t)$ is an even function of $t$, and therefore that all odd moments vanish -- this is a known property in the context of classical fluctuation-dissipation theorems \cite{Recursion_method_book}. 

Due to its importance, we conclude this section by providing a self-contained proof of the statement that $a_n^{(c)}=0$. We make two assumptions: (i) the (non-normalized) phase space distribution $p$ defining the inner product in Eq.~\eqref{eq:Classical_Inner_product_generic} satisfies $\{p,H\}=0$, which is particularly true if $p$ is a function of the Hamiltonian, $p\equiv p(H)$, and (ii) boundary terms arising from Stokes' theorem can generically be dropped\footnote{See footnote~\ref{footnote:boundary_terms}.}. The coefficients $a_n^{(c)}$ are given by diagonal matrix elements of the Liouvillian in the Krylov basis, i.e. $a_n^{(c)}=(f_n|\mathcal{L}_c|f_n)$. Notice that any diagonal element of the form $\melc{f}{\mathcal{L}_c}{f}$, where $\ketc{f}\in \mathcal{F}(\mathcal{M})$ satisfies $f^*=\pm f$\footnote{Since we take $\mathcal{F}(\mathcal{M})$ to be the space of complex-valued functions over a real symplectic manifold $\mathcal{M}$, complex conjugation ``$^*$'' can be unambiguously denoted without reference to a coordinate system.}, is given by
\begin{equation}
\label{eq:Nico_magic_proof_1}
 (f|\mathcal{L}_c|f)= i\int_{\mathcal{M}}f^*\{H,f\}\,p\,\omega= \pm\frac{i}{2} \int_{\mathcal{M}} \{H,f^2 p\} \,\omega=\mp\frac{i}{2} \int_{\mathcal{M}} X_H(f^2 p)\,\omega\, ,
\end{equation}
where assumption (i) was used to move the phase space distribution $p$ inside the Poisson bracket. In addition, it is a general result of Hamiltonian mechanics that for any smooth scalar field $g$ on $\mathcal{M}$,
\begin{equation}
\label{eq:Nico_magic_proof_2}
    \int_\mathcal{M} X_H(g)\, \omega =\int_\mathcal{M} L_{X_H}(g)\, \omega=\int_\mathcal{M} L_{X_H}(g\, \omega)=0\, ,
\end{equation}
where $L_{X_H}$ is the Lie derivative along the Hamiltonian vector field $X_H$, and we used Cartan's magic formula together with assumption (ii) in the last step. Setting $g=f^2p$ immediately yields $\melc{f}{\mathcal{L}_c}{f}=0$. With this in hand, one can now proceed inductively and, by repeated application of identities \eqref{eq:Nico_magic_proof_1}-\eqref{eq:Nico_magic_proof_2}, show that the Lanczos algorithm implies that the classical Krylov basis elements $f_n$ satisfy $i^n f_n\in\mathbb{R}$, and consequently $a_n^{(c)}=0$ for all $n\geq 0$. This concludes the proof. 

In conclusion, for quantum systems whose classical limit satisfies assumptions (i) and (ii), the diagonal Lanczos coefficients $a_n$ can be viewed as a purely quantum effect. We will return to this point later in this section.

\subsection{Classical limit and Krylov-Ehrenfest time}\label{subsec:Classical_Limit_Krylov}

In this section, we show that the classical Lanczos algorithm in phase space, described in Section~\ref{subsec:Classical_Krylov_Generic_setup}, can be obtained as the classical limit ($\hbar\to 0$) of its counterpart for quantum operators evolving in the Heisenberg picture, which we discussed in Section~\ref{sec:Krylov_toolbox} and Appendix~\ref{appx:Lanczos_Lanczos}. 
In spirit, the correspondence between the quantum and classical descriptions relies on the Dirac correspondence, $[\cdot,\cdot]\mapsto i\hbar\{\cdot,\cdot\}+\mathit{O}(\hbar^2)$, but another key ingredient of the Lanczos algorithm that will be addressed in this discussion is the inner product. In particular, we are here interested in systems whose phase space is some compact symplectic manifold rather than the standard $\mathbb{R}^{2N}$ ($N$ being the number of degrees of freedom). To facilitate a more precise semiclassical analysis, it is therefore convenient to use the phase space formulation of quantum mechanics\footnote{See \cite{Marino:2021lne} for a pedagogical introduction to the subject, and \cite{Rundle:2021sku} for a recent review discussing the case of compact symplectic manifolds. Both include references to relevant and seminal papers in the field.}. For the purposes of the present discussion, we will keep the $\hbar$ factors explicit in all expressions.

The phase space formulation of quantum mechanics describes a quantum system in terms of functions over a phase space equipped with a twisted algebra (defined by a non-commutative Moyal product), which is designed to be isomorphic to the corresponding algebra of quantum operators. It provides a systematic approach to quantizing a classical system and, in particular, to studying quantum corrections order by order in $\hbar$. Additionally, it is a powerful tool for making quantitative statements about the classical limit of a quantum system. Specifically, consider a quantum system with some Hamiltonian $\hat{H}$, whose time evolution operator is $e^{-it\hat{H}/\hbar}$ (note the explicit $\hbar$), where the dynamics is controlled by an underlying operator algebra with (quantum) structure constants given by $f^{(q)}_{ijk}=i\hbar f^{(c)}_{ijk}$, where $f_{ijk}^{(c)}$ are the structure constants of the Poisson algebra on the group manifold that describes the phase space of the classical limit of the system. The discussion that follows is kept general, though it is useful to keep the following particular cases in view: the Heisenberg-Weyl group, corresponding to systems of position and momentum operators $(\hat{q},\hat{p})$, for which the classical structure constants are given by the flat symplectic form, and the $\mathrm{SU}(2)$ group, relevant for collective spin systems, where $f_{ijk}^{(c)}=\varepsilon_{ijk}$. As discussed in Section~\ref{sec:LMG_FP_models}, the $\mathrm{SU}(2)$ algebra of collective spins introduces an effective spin-dependent Planck constant, $\hbar \mapsto \hbar_\mathrm{eff} \equiv \hbar/S$. In both cases, the Moyal product has been computed explicitly, as well as for other compact groups such as $\mathrm{SU}(N)$, and even for toroidal lattices representing the classical limit of quantum systems with a discrete Hilbert space\footnote{Refs.~\cite{Basu:2024tgg,Basu:2025mmm} apply the phase space formalism of discrete Hilbert spaces to study the Wigner negativity of Krylov basis elements as a measure of their non-classicality. The classical limit, however, was not considered in these works. See \cite{Pal:2026otf} for related work.}. A recent concise review can be found in \cite{Rundle:2021sku}.

Without delving into the technical details of its construction, which can be found in the references above, we briefly summarize some properties of the phase space formulation and its associated Moyal product, as they will be useful for analyzing the classical limit of the Lanczos algorithm. The bijective correspondence between quantum operators and phase space functions on generic group manifolds is implemented via the Stratonovich-Weyl kernel \cite{Stratonovich1957}, denoted here by $\hat{\Delta}(\Omega)$, which parametrically depends on the set of canonical phase space coordinates symbolically denoted by $\Omega$,
\begin{equation}
    \label{eq:Stratonovich_Weyl_kernel}
    \hat{\mathcal{O}}\longmapsto\mathcal{O}_W(\Omega)=\mathrm{Tr}[\hat{\Delta}(\Omega)\hat{\mathcal{O}}]\,,
\end{equation}
where the subscript $W$ indicates the Wigner transform of $\hat{\mathcal{O}}$. This expression can be interpreted as the (non-normalized) projection of $\hat{\mathcal{O}}$ along $\hat{\Delta}(\Omega)$. The Stratonovich axioms ensuring that Eq.~\eqref{eq:Stratonovich_Weyl_kernel} defines a sensible isomorphism between the operator algebra and the twisted phase space algebra are as follows,
\begin{enumerate}
    \item Linearity, which is already guaranteed by \eqref{eq:Stratonovich_Weyl_kernel}.
    \item Reality, $\mathcal{O}_W^{*} (\Omega)=(\mathcal{O}^\dagger)_W(\Omega)$.
    \item Standardization,
    \begin{equation}
        \label{eq:Stratonovich_axiom_Standardization}
        \frac{1}{\mathrm{Vol}(\mathcal{M})}\int_{\mathcal{M}} \mathcal{O}_W\,\omega = \frac{1}{\mathcal{N}}\,\mathrm{Tr}[\hat{\mathcal{O}}]\,,
    \end{equation}
    where $\mathcal{N}$ is a normalization factor associated with the trace in the operator Hilbert space.
    \item Traciality, 
    \begin{equation}
        \label{eq:Stratonovich_axiom_Traciality}
        \frac{1}{\mathrm{Vol}(\mathcal{M})}\int_{\mathcal{M}}\mathcal{A}_W^{*}\mathcal{B}_W\, \omega=\frac{1}{\mathcal{N}}\,\mathrm{Tr}[\hat{\mathcal{A}}^\dagger\hat{\mathcal{B}}]\,.
    \end{equation}
\end{enumerate}
A fifth axiom, namely covariance of the kernel under the symmetry group defining the group manifold $\mathcal{M}$, is often also imposed \cite{VARILLY1989107}. These axioms ensure the invertibility of the kernel, with inverse transform
\begin{equation}
    \label{eq:Weyl_Inverse_of_Wigner}
    f\longmapsto \widehat{W}_f\equiv\frac{\mathcal{N}}{\mathrm{Vol}(\mathcal{M})} \int_{\mathcal{M}}f(\Omega) \hat{\Delta}(\Omega)\, \omega(\Omega)\,.
\end{equation}
In general, the Stratonovich-Weyl kernel satisfying the above requirements is not unique, reflecting the freedom in the choice of a quantization scheme for a given classical system. This will not be important in our discussion, where we shall discuss the classical limit of a given quantum system: in this sense, a choice of quantization scheme, and hence of kernel $\hat{\Delta}(\Omega)$, has already been made. Correspondingly, we will not need the explicit form of the kernel, provided it satisfies the properties listed above.

The explicit construction of the kernel involves harmonic analysis and generalized coherent states in the corresponding group manifold. As an illustrative example, consider the Heisenberg-Weyl group, for which $\Omega\equiv (q,p)$. Although this group is not compact, the normalization inherited from flat-space Fourier analysis effectively replaces the factor $\mathcal{N}/\mathrm{Vol}(\mathcal{M})$ by $1/(2\pi \hbar)$, see for example \cite{Marino:2021lne} for details. On the other hand, the Stratonovich-Weyl kernel for $\mathrm{SU}(2)$ was first constructed in \cite{VARILLY1989107} (see also \cite{Amiet_1991,Amiet:2000rju,PhysRevA.63.012105}). In this case, $\mathrm{Vol}(\mathcal{M}=S^2)=4\pi$, and the trace normalization is $\mathcal{N}=2S+1$, ensuring the correct normalization of the identity in a given irreducible representation of total spin $S$. In the large-spin limit, this behaves as $\mathcal{N}\sim 2/\hbar_\mathrm{eff}$.

In any case, we have reviewed the Stratonovich axioms to emphasize that they are designed to ensure compatibility between the (Moyal-twisted) phase space inner product and the operator inner product. This compatibility is in particular guaranteed by axioms 3 and 4 above. As announced, this property is crucial for the semiclassical analysis of the Lanczos algorithm, which is, at its core, an orthogonalization procedure. This correspondence can be made more explicit by introducing the Moyal product $\times$, a non-commutative deformation of the standard pointwise product of functions. It can be derived from the Stratonovich-Weyl kernel and is constructed so as to mirror the product of operators,
\begin{equation}
    \label{eq:Moyal_product_compatibility}
    \mathcal{A}_W(\Omega)\times \mathcal{B}_W(\Omega) = (\mathcal{AB})_W(\Omega)\,,
\end{equation}
for any two operators $\hat{\mathcal{A}}$ and $\hat{\mathcal{B}}$. Let us now highlight three properties of the Moyal product that will be relevant for our analysis:
\begin{enumerate}
    \item Classical limit:
    \begin{equation}
        \label{eq:Moyal_classical}
        \lim_{\hbar\to 0}f\times g = fg\,,
    \end{equation}
    for any phase space functions $f,g$.
    \item Compatibility with the cyclical property of the trace:
    \begin{equation}
        \label{eq:Moyal_product_trace_cyclicity}
        \int_{\mathcal{M}}f(\Omega)\times g(\Omega)\,\omega(\Omega) = \int_{\mathcal{M}}f(\Omega) g(\Omega)\, \omega(\Omega)\,,
    \end{equation}
    for any phase space functions $f,g$. This property is consistent with the cyclicity of the trace, compare with Eq.~\eqref{eq:Stratonovich_axiom_Traciality}.
    \item Connection to the Poisson bracket: we can define the Moyal commutator of two phase space functions $f,g$ as $[f,g]_{\times}\equiv f\times g - g\times f$. It satisfies the following property,
    \begin{equation}
        \label{eq:Moyal_commutator_to_PB}
        [f,g]_\times = i\hbar \left\{f,g\right\}+\mathit{O}(\hbar^2)\,.
    \end{equation}
    This relation between the Moyal commutator and the Poisson bracket is a central result of the phase space formulation of quantum mechanics for the Heisenberg-Weyl group\footnote{For this group, the first nonzero correction to the Poisson bracket appears at order $\hbar^3$ and vanishes when the functions are at most quadratic in the canonical coordinates $q,p$, see for example \cite{Marino:2021lne}.}. It has also been shown to hold for $\mathrm{SU}(2)$ (upon replacing $\hbar$ by $\hbar_\mathrm{eff}$) in \cite{Klimov_2002}, using the differential form of the Moyal product constructed in \cite{VARILLY1989107}. Analogous results extend to other compact groups such as $\mathrm{SU}(N)$ \cite{Rundle:2021sku}.
\end{enumerate}
We now have all the necessary ingredients to address the semiclassical limit of the Lanczos algorithm. To this end, let us first introduce two twisted versions of the phase space inner product, in direct analogy with the operator inner products in Eqs.~\eqref{eq:Operator_Inner_product} and \eqref{eq:Operator_Inner_product_sym},
\begin{subequations}
\begin{align}
    & \big(f|g\big)_\times \equiv \frac{1}{Z}\int_\mathcal{M} p\times f^{*}\times g\,\omega\,, \label{eq:Classical_inner_product_twisted}\\
    &\big(f|g\big)^{(w)}_\times \equiv \frac{1}{Z}\int_\mathcal{M} p^{\times \frac{1}{2}}\times f^{*}\times p^{\times \frac{1}{2}}\times g\,\omega\,, \label{eq:Classical_inner_product_twisted_wightman}
\end{align}
\end{subequations}
for some phase space function $p$. Here, $p^{\times \frac{1}{2}}$ denotes a function whose square with respect to the Moyal product equals $p$\footnote{The \textit{deformed} square root need not be unique. For the purposes of this discussion, we assume the existence of a solution continuously connected to $\sqrt{p}$ parametrically in $\hbar$, and denote it by $p^{\times \frac{1}{2}}$.}. The normalization factor $Z$ is defined as the phase space integral of $p$. Given two operators $\hat{\mathcal{A}}$ and $\hat{\mathcal{B}}$, it then follows from Eqs.~\eqref{eq:Stratonovich_axiom_Standardization} and \eqref{eq:Moyal_product_compatibility} that
\begin{subequations}
\begin{align}
    & (\mathcal{A}|\mathcal{B}) = \big( \mathcal{A}_W|\mathcal{B}_W\big)_\times\,, \label{eq:Standard_op_inner_both_formulations}\\
    & (\mathcal{A}|\mathcal{B})^{(w)} = \big( \mathcal{A}_W|\mathcal{B}_W\big)^{(w)}_\times\,,\label{eq:Wightman_op_inner_both_formulations}
\end{align}
\end{subequations}
where, in both cases, the relation between the phase space function $p$ and the operator $\hat \rho$ entering the definition of the inner product is given by the Wigner transform, $p = \rho_W$. These relations show that Eqs.~\eqref{eq:Standard_op_inner_both_formulations}–\eqref{eq:Wightman_op_inner_both_formulations} provide equivalent formulations of the operator inner product in phase space, without invoking any semiclassical approximation.

An important observation follows: in the limit $\hbar\to0$, the Moyal product becomes commutative, see Eq.~\eqref{eq:Moyal_classical}. As a consequence, both twisted inner products \eqref{eq:Classical_inner_product_twisted}-\eqref{eq:Classical_inner_product_twisted_wightman} coincide in this limit and reduce to the classical phase space inner product \eqref{eq:Classical_Inner_product_generic}. Combining this observation with Eqs.~\eqref{eq:Standard_op_inner_both_formulations}-\eqref{eq:Wightman_op_inner_both_formulations}, we obtain 
\begin{equation}
    \label{eq:Op_products_equal_classically}
    \lim_{\hbar\to 0}(\mathcal{A}|\mathcal{B})=\lim_{\hbar\to 0}(\mathcal{A}|\mathcal{B})^{(w)}=(\mathcal{A}_W|\mathcal{B}_W)\,,
\end{equation}
i.e. both operator inner products coincide in the classical limit. 

The same conclusion holds for any regularized operator inner product defined by a different arrangement of the density matrix within the trace, since the argument ultimately relies on the fact that the Moyal product reduces to the standard pointwise product of functions as $\hbar\to 0$. Consequently, differently regularized correlation functions also coincide in the classical limit. For instance, this provides a justification for the uniqueness of the classical Lyapunov exponent, in contrast to the various definitions of quantum Lyapunov exponent \cite{Maldacena:2015waa}, all of which converge to the former when $\hbar\to 0$.

With these ingredients in place, one can formulate the Lanczos algorithm for operators presented in Appendix~\ref{appx:Lanczos_Lanczos} in phase space. Specifically, the Lanczos recursion for an operator $\hat{\mathcal{O}}$, with time evolution generator $\mathcal{L}/\hbar\equiv [\hat{H}/\hbar,\cdot]$, and defined with respect to an operator inner product such as \eqref{eq:Operator_Inner_product} or \eqref{eq:Operator_Inner_product_sym}, can be mapped to a Lanczos recursion in (twisted) phase space. In this formulation, the initial condition is $\mathcal{O}_W(\Omega)$, the time evolution generator is $\mathcal{L}_W/\hbar\equiv[H_W,\cdot]_\times/\hbar$, and the inner product is replaced by the corresponding twisted versions of the phase space inner product \eqref{eq:Classical_inner_product_twisted} or \eqref{eq:Classical_inner_product_twisted_wightman}. In particular, the Krylov basis elements $\hat{\mathcal{O}}_n$ are mapped to their Wigner transforms $(\mathcal{O}_n)_W(\Omega)$. 

To compute the classical limit of the recursion, it suffices to combine Eq.~\eqref{eq:Op_products_equal_classically} with the relation
\begin{equation}
    \label{eq:Classical_Liouvillian_as_limit}
    \lim_{\hbar\to 0}\frac{\mathcal{L}_W}{\hbar} = i\left\{H_W,\cdot\right\}=\mathcal{L}_c\,,
\end{equation}
which follows from Eqs.~\eqref{eq:Moyal_commutator_to_PB} and \eqref{eq:classical_Liouvillian_operator}\footnote{With a slight abuse of notation, we identify the canonical Hamiltonian function $H(\Omega)$ with the Wigner transform of the quantum Hamiltonian $H_W(\Omega)$. These coincide whenever $\hat{H}$ is given by the Weyl quantization of a function of the canonical phase space variables \cite{Polkovnikov_phase_space_QM}, as is the case for the Hamiltonians considered in this work.}. In conclusion, in the classical limit, the Lanczos recursion for the operator $\hat{\mathcal{O}}$ reduces to the classical recursion (as described in Section \ref{subsec:Classical_Krylov_Generic_setup}) of its Wigner transform $\mathcal{O}_W$. This, in turn, implies the following limits for the Lanczos coefficients,
\begin{equation}
    \label{eq:Classical_limit_Lanczos_coefficients}
    \lim_{\hbar\to 0}b_n^{(q)}=b_n^{(c)}\,,\qquad \lim_{\hbar\to 0}a_n^{(q)}=a_n^{(c)}\,,
\end{equation}
where $a_n^{(q)}$ and $b_n^{(q)}$ are computed with the usual Lanczos algorithm for operators, see Appendix~\ref{appx:Lanczos_Lanczos}, for any choice within the class of regularized operator inner products discussed above, while $a_n^{(c)}$ and $b_n^{(c)}$ are obtained from the classical recursion defined in Section~\ref{subsec:Classical_Krylov_Generic_setup}, using the standard inner product on phase space. 

We close this section by noting an interesting corollary of the above discussion. At the quantum level, the Lanczos algorithm for a Hermitian operator $\hat{\mathcal{O}}=\hat{\mathcal{O}}^\dagger$ can be implemented using various choices of inner products, such as in Eqs.~\eqref{eq:Operator_Inner_product} and \eqref{eq:Operator_Inner_product_sym}. For $[\hat{\rho},\hat{H}]=0$, the choice \eqref{eq:Operator_Inner_product} generically yields nonzero coefficients $a^{(q)}_n$, whereas \eqref{eq:Operator_Inner_product_sym} leads to $a_n^{(q)}=0$, as can be shown inductively. In the classical limit, however, all such quantum inner products coincide and reduce to the classical inner product in phase space, see Eq.~\eqref{eq:Op_products_equal_classically}, for which we showed in Section~\ref{subsec:Classical_Krylov_Generic_setup} that $a_n^{(c)}=0$. It follows that
\begin{equation}
    \label{eq:an_coeffs_vanish_classically}
    \lim_{\hbar\to 0}a_n^{(q)}=0\, ,
\end{equation}
for any choice of regularized operator inner product defined by splitting the density matrix $\hat{\rho}$ among the operator insertions in the trace. Our numerical results for the LMG model are in agreement with this, see Figure~\ref{fig:microcanonical_classical_LMG}.

\subsubsection{A Krylov-Ehrenfest theorem}

We have shown that, in the classical limit $\hbar\to 0$, the quantum Lanczos coefficients reduce to their classical counterparts, see Eq.~\eqref{eq:Classical_limit_Lanczos_coefficients}. Given that K-complexity is entirely determined by the Lanczos coefficients\footnote{This follows by reformulating the recurrence relation \eqref{eq:Krylov_elements_recursion} as a differential recursion for the Krylov space wavefunction, see Ref.~\cite{Rabinovici:2025otw}.}, it follows that
\begin{equation}
    \label{eq:Classical_Limit_KC}
    \lim_{\hbar\to 0}C_K(t)=C_K^{(c)}(t)\,,\quad \text{for all }t\in\mathbb{R}\,.
\end{equation}
For $\hbar>0$, one may further expect that $C_K^{(c)}(t)$ provides a good approximation to $C_K(t)$ for sufficiently early times. This can be viewed as an analogue of Ehrenfest's theorem in the context of K-complexity, and is the focus of the present discussion. 

The moment method provides a simple estimate of the critical value of $n$ at which the quantum Lanczos coefficients begin to deviate from their classical counterparts, since they are in one-to-one correspondence with the Liouvillian moments,
\begin{align}
    m_n &= \left(\mathcal{O}\left|\left(\frac{\mathcal{L}}{\hbar}\right)^n\right|\mathcal{O}\right)=\left(\mathcal{O}^{(\hbar)}_W\left|\left(\frac{\mathcal{L}_W}{\hbar}\right)^n\right|\mathcal{O}^{(\hbar)}_W\right)_\times \nonumber \\
    &=  \left(\mathcal{O}^{(\hbar)}_W\left|\bigg(\mathcal{L}_c+\mathit{O}(\hbar)\right)^n\right|\mathcal{O}^{(\hbar)}_W\bigg) \approx m_n^{(c)}(1+n\hbar~\#)\,, \label{eq:Moments_Moyal_line2}
\end{align}
where the second equality is merely a rewriting in terms of the Moyal product\footnote{The Wigner transform of an operator $\hat{\mathcal{O}}$, as defined in Eq.~\eqref{eq:Stratonovich_Weyl_kernel}, can depend on $\hbar$. In order to keep track of this, we have denoted it as $\mathcal{O}_W^{(\hbar)}$.}, while Eq.~\eqref{eq:Classical_Liouvillian_as_limit} has been used in the third equality. In the last step, we used the approximation $(1+\varepsilon)^n\approx 1+\varepsilon n$ valid for $\varepsilon\ll 1$.  The symbol $\#$ denotes a dimensionful quantity that does not scale parametrically with $\hbar$, and $m_n^{(c)}\equiv (\mathcal{O}^{(\hbar\to 0)}_W |\mathcal{L}^n_c|\mathcal{O}^{(\hbar\to 0)}_W)$ are the moments of the classical two-point function. Consequently, the classical and quantum moments remain quantitatively similar up to a critical value $n_{*}\sim 1/\hbar$\footnote{More precisely, the relevant dimensionless parameter is $\#/\hbar$, where $\#$ has units of an action and can be interpreted as a characteristic classical action scale. Because we are primarily interested in the parametric dependence of $n_*$, we omit this factor.}. Since the Lanczos coefficients are in one-to-one correspondence with the moments, it follows that $b_n^{(q)}\approx b_n^{(c)}$ and $a_n^{(q)}\approx a_n^{(c)}=0$ for the range $n\lesssim 1/\hbar$. Furthermore, given that K-complexity is an expectation value of the position on a chain whose hopping amplitudes are given by the Lanczos coefficients, it also follows that $C_K(t)\approx C_K^{(c)}(t)$ up to a time $t_{*}$ defined implicitly by $C_K(t_*)\approx n_{*}\approx 1/\hbar$. 

Using typical profiles for the Lanczos coefficients $b_n$ \cite{Universal_Growth_Hypothesis}, one can extract explicit expressions for this \textit{Krylov-Ehrenfest time scale}:
\begin{subequations}
\begin{align}
    & b_n\sim \alpha n \quad \Longrightarrow\quad C_K(t)\sim e^{\lambda_K t},\,\lambda_K=2\alpha\quad \Longrightarrow\quad t_{*}\sim \frac{1}{\lambda_K}\log \left(\frac{1}{\hbar}\right),\label{eq:Ehrenfest_time_line1} \\
    & b_n\sim \alpha n^\delta,\, \delta<1\quad \Longrightarrow\quad C_K(t)\sim t^{\frac{1}{1-\delta}}\quad \Longrightarrow\quad t_{*}\sim \left(\frac{1}{\hbar}\right)^{1-\delta}, \label{eq:Ehrenfest_time_line2}
\end{align}
\end{subequations}
where the first line is expected to describe chaotic systems, and the second to describe integrable systems, in accordance with the universal operator growth hypothesis \cite{Universal_Growth_Hypothesis}. These complexity profiles can be obtained from the Lanczos sequence using a ballistic approximation for the propagation of the Krylov wavepacket, which is a reasonable approximation at early times \cite{Universal_Growth_Hypothesis} (see also Section III.H.2 of \cite{Rabinovici:2025otw} for a detailed discussion). The estimates in Eqs.~\eqref{eq:Ehrenfest_time_line1}-\eqref{eq:Ehrenfest_time_line2} are in qualitative agreement with the standard expression for the Ehrenfest time in chaotic and integrable systems (see Refs.~\cite{Shepelyansky:2020,Haake}), with the Krylov exponent $\lambda_K$ playing a role analogous to that of the Lyapunov exponent $\lambda_L$.

In the collective spin models studied in later sections, the effective Planck constant is $\hbar_{\mathrm{eff}} = \hbar/S$, where $S$ denotes the total spin (which, in the maximal Hilbert space sector, scales proportionally with $N$, as discussed in Sections~\ref{sec:LMG} and \ref{sec:FP}). If the Lanczos coefficients exhibit an initial linear growth, as occurs in the LMG model when the classical dynamics is dominated by an unstable saddle point in phase space, it implies that the Krylov–Ehrenfest time scales as $t_{*}\sim (\log N)/\lambda_K$, thereby explicitly connecting the Ehrenfest and scrambling time scales.

\subsubsection{Classical limit in the Schrödinger picture}

The analysis presented so far has focused on the classical limit of the K-complexity for quantum operators evolving in the Heisenberg picture. Accordingly, it relied on the Lanczos algorithm in its Liouvillian formulation. For completeness, we now briefly comment on the extension of these arguments to the K-complexity of quantum states evolving in the Schrödinger picture.

The first issue that needs to be addressed in that case is the existence of two inequivalent formulations \cite{Caputa:2024vrn}. On the one hand, the state may be treated as a density matrix, viewed as an element of the operator Hilbert space, whose time evolution is generated by the Liouvillian via the von Neumann equation. In this case, the operator formalism described in Section~\ref{sec:Krylov_toolbox} applies directly and can be used to define the K-complexity of the density matrix. On the other hand, pure states can be treated as elements of the Hilbert space, evolving under the Hamiltonian operator $\hat{H}$ (or, more precisely, by $\hat{H}/\hbar$, as dictated by the Schrödinger equation). In this framework, their K-complexity -- often referred to as spread complexity -- can be defined by implementing the Lanczos algorithm in the Hamiltonian representation \cite{Balasubramanian:2022tpr}. 

The inequivalence of these two approaches can be clearly illustrated by the following example~\cite{Caputa:2024vrn}: for a pure state $|\psi\rangle$ with associated density matrix $\rho_\psi=|\psi\rangle\langle\psi|$, the first Lanczos coefficient satisfies $b_1^{(\rho_\psi)}=\sqrt{2}\,b_1^{(|\psi\rangle)}$. We refer to the aforementioned work for a detailed discussion on the early- and late-time differences between the K-complexities of $\rho_\psi(t)$ and $|\psi(t)\rangle$.

Let us now discuss separately the classical limit of both constructions, building on the tools developed in this section:

\begin{itemize}
    \item The classical limit of the K-complexity of density matrices $\hat{\rho}(t)$ in the Schrödinger picture\footnote{In this discussion, $\hat{\rho}$ denotes the density matrix of the evolving state and should not be confused with the density matrix entering the definition of the operator inner product, see Eqs.~\eqref{eq:Operator_Inner_product}-\eqref{eq:Operator_Inner_product_sym}.} can be studied using the same phase space formalism employed throughout Section \ref{subsec:Classical_Limit_Krylov}, since it naturally falls within the operator framework. In the limit $\hbar\to 0$, the K-complexity of $\hat{\rho}(t)$ reduces to the classical K-complexity of its Wigner function $\rho_W(\Omega;t)$, which evolves according to the classical Liouville equation and can be interpreted as the phase space probability distribution describing the classical configuration of the system (see, for instance, Ref.~\cite{Marino:2021lne}).
    
    \item The K-complexity of a pure state $|\psi(t)\rangle$ (often referred to as spread complexity) \cite{Balasubramanian:2022tpr} is defined using the Hamiltonian representation of the Lanczos algorithm, which involves iterated applications of $\hat{H}/\hbar$ over the vector $|\psi\rangle$, viewed as a vector in the Hilbert space. Naively, this object does not admit a natural description within Wigner’s phase space formalism, which is specifically designed to map operators to phase space functions. In fact, considering the operator $|\psi\rangle\langle\psi|$ and its associated Wigner function would bring us back to the density matrix approach described above, which is inequivalent to the Hamiltonian formulation. A direct approach to the classical limit of the K-complexity of $|\psi(t)\rangle$ in the Hamiltonian representation might require the use of other semiclassical techniques, such as saddle-point approximations to path integrals or a WKB analysis of the Schrödinger equation. We leave this as an interesting open problem\footnote{Although Wigner’s formalism does not provide a direct phase space representation of the vector $|\psi(t)\rangle$, the associated moments $\mu_n=\langle \psi|(\hat{H}/\hbar)^n|\psi\rangle$, to which its Lanczos coefficients are in correspondence, admit a description in terms of a phase space calculation, to the extent that they can be written as $\mu_n = \text{Tr}[(\hat{H}/\hbar)^n\hat{\rho}_\psi]$, and the trace expressed as a phase space integral according to Eq.~\eqref{eq:Stratonovich_axiom_Traciality}. It would be interesting to identify a classical phase space construction reproducing these moments.}. 
\end{itemize}

\subsection{Classical phase space of collective spin models}\label{sec:general_prescription}

We now apply the generic framework of K-complexity in phase space developed in Section~\ref{subsec:Classical_Krylov_Generic_setup} to classical collective spin models. We begin by introducing the general setup for systems with $N$ collective spins ($N=1$ for the LMG model and $N=2$ for the FP model), and then explain how the classical Lanczos algorithm can in practice be implemented.

In the classical limit, the rescaled collective spin operator $\hat{\bm{s}}=(\hat{s}_x,\hat{s}_y,\hat{s}_z)$ is represented by classical spin vectors $\bm{s}_i=(x_i,y_i,z_i)$, with $i=1,\cdots,N$, each having conserved\footnote{This follows from the fact that the magnitude of the corresponding rescaled collective spin operators is fixed within a given Hilbert space sector, see Sections~\ref{sec:LMG} and~\ref{sec:FP}.} unit magnitude, $\bm{s}_i\cdot \bm{s}_i=1$, where $\cdot$ denotes the standard Euclidean scalar product on $\mathbb{R}^3$.

For a single classical spin, the phase space is compact and given by the unit sphere $S^2$, which is $2$-dimensional. For a system of $N$ classical spins, the phase space is the Cartesian product of the individual spin phase spaces, $S^2_N\equiv S^2\times \cdots \times S^2$, and is therefore $2N$-dimensional. On each sphere, we introduce spherical coordinates parametrized by the angles $(\theta_i,\varphi_i)$, with $\theta_i\in[0, \pi]$ and $\varphi_i\in [0,2\pi)$. The Cartesian coordinates on each sphere are expressed in terms of these angles as
\begin{equation}
    x_i= \sin \theta_i \cos \varphi_i\, , \qquad y_i=\sin \theta_i \sin \varphi_i\, ,\qquad z_i=\cos \theta_i\, .
\end{equation}
The resulting phase space is equipped with the symplectic form 
\begin{equation}
    \omega_N^{(2)}=\sum_{i=1}^N \dd \varphi_i \wedge \dd \cos \theta_i\, .
\end{equation}
Accordingly, a convenient choice of canonical variables is $q_i=\varphi_i$ and $p_i=\cos \theta_i$. With this identification, the Poisson bracket between two functions $f$ and $g$ defined on $S^2_N$ takes the form
\begin{equation}\label{Poisson_bracket}
    \{f,\, g\}=\omega^{(2)}_N(X_f,X_g)=\sum_{i=1}^N\left(\pdv{f}{\varphi_i}\pdv{g}{\cos \theta_i}-\pdv{f}{\cos \theta_i}\pdv{g}{\varphi_i}\right) ,
\end{equation}
where $X_f$ and $X_g$ are the corresponding Hamiltonian vector fields associated with $f$ and $g$. In particular, one finds $\{x_i,y_j\}=\delta_{ij}\, z_i$, $\{x_i,z_j\}=-\delta_{ij}\, y_i$, $\{y_i,z_j\}=\delta_{ij}\, x_i$, and similarly for cyclic permutations.

Classically, the role of operator space is played by the Poisson algebra on $S^2_N$, which we denote by $\mathcal{F}(S^2_N)$, and is the space of all smooth complex-valued functions on $S^2_N$, equipped with the commutative pointwise product and antisymmetric Poisson bracket. The classical operator space is endowed with the inner product\footnote{Here, $\omega_N$ is understood as the volume form on $S^2_N$ as in Eq.~\eqref{eq:Classical_Inner_product_generic}, where the subscript $N$ is included for clarity in the following discussion.},
\begin{equation}\label{sphere_inner_product}
\braketc{f}{g}=\frac{1}{(4\pi)^N}\int_{S^2_N} f^* g\, \omega_N\, \quad \text{for any }f,g\in\mathcal{F}(S^2_N)\,,
\end{equation}
which corresponds to the infinite-temperature limit of Eq.~\eqref{eq:Classical_Inner_product_generic}, obtained by setting $p=1$ (that is, $\beta=0$), where $f$, $g\in \mathcal{F}(S^2_N)$.

For $N=1$, spherical harmonics $Y_l^m(\theta, \varphi)$ form a convenient orthonormal basis of $\mathcal{F}(S^2)$ with respect to the inner product \eqref{sphere_inner_product}. In particular,
\begin{equation}\label{classical_operator_space}
    \mathcal{F}(S^2)=\text{span}\left\{Y_l^m(\theta, \varphi),\hspace{.2cm} l=0,\, 1,\, 2,\,  \cdots \hspace{.2cm}\text{and}\hspace{.2cm}-l\leq m \leq l \right\},
\end{equation}
with orthonormality condition $\braketc{Y_l^m}{Y_{l'}^{m'}}=\delta_{l,l'}\,\delta_{m,m'}$\footnote{Note that this condition, with the inner product defined in Eq.~\eqref{sphere_inner_product}, implies that the spherical harmonics are defined without the overall factor of $1/\sqrt{4\pi}$ that usually enters their definition \cite{spherical_harmonics_relations}.}. Any function $f\in \mathcal{F}(S^2)$ can be expanded in this basis as a linear combination of these spherical harmonics,
\begin{equation}
    f(\theta,\varphi)=\sum_{l=0}^\infty \sum_{m=-l}^l f^m_l Y_l^m(\theta, \varphi)\, ,
\end{equation}
where the coefficients are given by $f^m_l=\braketc{Y_l^m}{f}$. This construction extends naturally to the product space, $\mathcal{F}(S^2_N)=\bigotimes_{i=1}^N \mathcal{F}(S^2)$, whose basis elements are given by pointwise products of spherical harmonics on each sphere,
\begin{equation}
    \mathcal{F}(S^2_N)=\text{span}\left\{\prod_{i=1}^N (Y_i)_{l_i}^{m_i}(\theta_i, \varphi_i),\hspace{.2cm} l_i=0,\, 1,\, 2,\,  \cdots \hspace{.2cm}\text{and}\hspace{.2cm}-l_i\leq m_i \leq l_i \right\},
\end{equation}
and $(Y_i)_{l_i}^{m_i}(\theta_i, \varphi_i)$ is understood as a spherical harmonic on the $i$th sphere. These basis elements satisfy the orthonormality condition $\braketc{\prod_{i=1}^N (Y_i)_{l_i}^{m_i}}{\prod_{i=1}^N (Y_i)_{l_i'}^{m_i'}}=\prod_{i=1}^N \delta_{l_i,l_i'} \,\delta_{m_i,m_i'}$. Accordingly, any function $f\in \mathcal{F}(S^2_N)$ can be expanded as a linear combination of the pointwise products of spherical harmonics.

Let us now return to the case $N=1$. The components of a single classical spin $\bm{s}$ can be expressed in terms of spherical harmonics as 
\begin{equation}\label{x_to_spherical_harmonics}
    x =-\frac{1}{\sqrt{6}}\left(Y_1^1-Y_1^{-1} \right), \qquad y = \frac{ i}{\sqrt{6}} \left(Y_1^1+Y_1^{-1} \right), \qquad z =\frac{1}{\sqrt{3}}\,Y_1^0\, .
\end{equation}
Given a classical Liouvillian operator \eqref{eq:classical_Liouvillian_operator}, one is interested in computing its action on an arbitrary spherical harmonic. The following relations will play a central role,
\begin{subequations}
\begin{align}
    & \left\{Y_1^0,\, Y_l^m\right\}=- i\,m  \sqrt{3}\, Y_l^m\, ,\label{spherical_harmonics_relation_1}\\
    & \left\{Y_1^{\pm 1},\, Y_l^m\right\}=\pm  i\, \sqrt{\frac{3}{2}}\sqrt{(l\pm m+1)(l\mp m)} \, Y_l^{m\pm 1}\, ,\label{spherical_harmonics_relation_2}
\end{align}
\end{subequations}
and a proof is provided in Appendix~\ref{app:spherical_harmonics_poisson_brackets}. We note that general expressions for Poisson brackets of arbitrary spherical harmonics, $\{Y_{l}^{m},\, Y_{l'}^{m'}\}$, are available in the literature (see for example \cite{spherical_harmonics_Poisson}) and could be useful for extending the analysis presented below.

\subsubsection{Practical implementation of the classical Lanczos algorithm}\label{sec:classical_Lanczos_algo_practical}

We now turn to the practical implementation of the Lanczos algorithm. For simplicity, we describe our strategy in detail for the case of a single classical spin, $N=1$. The generalization to multiple spins is straightforward.

\begin{figure}[t!]
    \centering
    \includegraphics[width=.75\textwidth]{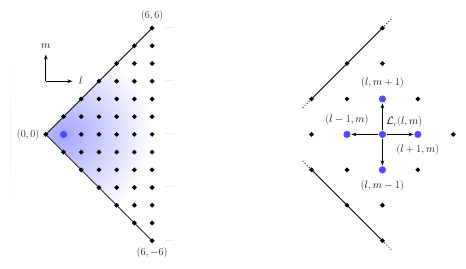}
    \caption{\textbf{(Left)} Grid of nodes $(l,m)$ with cutoff $l_\star=6$. The initial function is $\ketc{Y_1^0}$, corresponding (up to normalization) to the spin component $z$, and having support on the node $(1,0)$ (more general initial conditions may have support on multiple nodes). Black nodes correspond to zero coefficients. As the Lanczos algorithm is iterated, the support spreads across the lattice, generating increasingly complicated functions. \textbf{(Right)} Action of the classical Liouvillian operator associated with the LMG Hamiltonian (see Eq.~\eqref{classical_Liouvillian_LMG} below) on a spherical harmonic $Y_l^m$ illustrating the induced transitions across the grid.}
    \label{fig:classical_Lanczos_algorithm}
\end{figure}

The classical operator space, defined in Eq.~\eqref{classical_operator_space}, can be visualized as a semi-infinite ($l\geq 0$) triangular grid ($-l\leq m\leq l$), where each node corresponds to a spherical harmonic, and is denoted by $(l,m)$. To each node is also associated a complex coefficient, which represents the projection of a given function $f$ onto that spherical harmonic, namely $f_l^m$. Initially, all coefficients are set to zero, except for the nodes over which the initial function $\ketc{f_0}$ has support, see the left panel of Figure~\ref{fig:classical_Lanczos_algorithm}.

 In the Lanczos algorithm, each application of the classical Liouvillian operator acts on the nodes associated with a nonzero coefficient $f_l^m$, producing an updated grid with a new set of coefficients. The corresponding Lanczos coefficient is obtained by computing the norm of the resulting function with respect to the inner product in Eq.~\eqref{sphere_inner_product}. In terms of the projections, this norm reads $\sum_{l=0}^{\infty}\sum_{m=-l}^{l}|f_l^m|^2$.  In practice, however, one must introduce a cutoff $l=l_\star$ to truncate the basis \eqref{classical_operator_space}. The algorithm is then terminated once nonzero projections reach the highest retained spherical harmonic $Y_{l_\star}^m$.

Naturally, the classical Liouvillian and its action on the spherical harmonics basis -- which must be known in advance -- are model-dependent. As we will see below for the classical LMG and FP models, its action on a given spherical harmonic $Y_l^m$ generates a linear combination of neighboring spherical harmonics. Here, the notion of ``neighboring'' refers to proximity on the $(l,m)$ grid. This degree of locality depends on the polynomial degree of the classical spin components entering the Hamiltonian. From this perspective, the action of the Liouvillian can be interpreted as a set of jumps across the lattice, with each allowed transition associated with a specific coefficient, see the right panel of Figure~\ref{fig:classical_Lanczos_algorithm}.

Let us close this discussion with a comment on the case of two classical spins: the grid structure must be extended to a higher-dimensional grid, where each node is labeled by four integers $(l,m,k,n)$ corresponding to the product of two spherical harmonics $(Y_1)_l^m (Y_2)_k^n$. In this case, the classical operator space is spanned by such pointwise products of basis elements, and the Lanczos algorithm evolves coefficients defined over this enlarged lattice. This construction generalizes straightforwardly to the case of multiple spins.

\subsection{Classical Lanczos sequence for Lipkin-Meshkov-Glick model}\label{sec:classical_LMG}

The LMG Hamiltonian in Eq.~\eqref{LMG_rescaled_TSS} reduces to a classical Hamiltonian on the phase space $S^2$ ($N=1$) in the limit $S\to \infty$\footnote{More rigorously, we note that the classical limit $S\to\infty$ corresponds to the convergence of the ground-state expectation values of the rescaled collective spin operators, $\langle \hat{s}_\mu \rangle$, toward the minimum of the corresponding classical Hamiltonian on the unit sphere \cite{Duminil_Copin_semiclassical}.}. Using the coordinates $(\theta,\varphi)$ and the relations in Eqs.~\eqref{x_to_spherical_harmonics}, the classical Hamiltonian is given by 
\begin{equation}\label{LMG_classical}
    H=-\frac{J}{2}z^2-h x=-\frac{J}{6}\left(Y_1^0 \right)^2+\frac{h}{\sqrt{6}}\left(Y_1^1-Y_1^{-1} \right).
\end{equation}
The expression on the right is particularly useful for computing the classical Liouvillian operator associated with that Hamiltonian. Hamilton's equations and their corresponding linear stability analysis are detailed in Appendix~\ref{app:classical_LMG}. A key feature of this system is that for $h<J$, the phase space contains a single saddle point associated with a one-dimensional unstable manifold. At $h=J$, the system experiences a bifurcation and for $h>J$, this saddle is replaced by a center (see Table~\ref{tab:stability_fixed_points}). 

As anticipated in Section~\ref{sec:ClassicalKC}, we aim to analyze how the presence or absence of a saddle point in phase space influences the early-time growth of the classical K-complexity. To achieve this, the classical Lanczos algorithm must be implemented following the prescription described in Section~\ref{sec:classical_Lanczos_algo_practical}. This requires calculating the action of the classical Liouvillian operator on spherical harmonics. Using the relations in Eqs.~\eqref{spherical_harmonics_relation_1} and \eqref{spherical_harmonics_relation_2}, we show in Appendix~\ref{app:classical_Liouvillian_operator} that 
\begin{align}\label{classical_Liouvillian_LMG}
     \mathcal{L}_cY_l^m&=-\frac{J m}{\sqrt{2l+1}}\left(\sqrt{\frac{(l+m+1)(l-m+1)}{2l+3}}Y_{l+1}^m+\sqrt{\frac{(l-m)(l+m)}{2l-1}}Y_{l-1}^m \right)\nonumber \\
     &\hspace{.5cm}-\frac{h}{2}\left(\sqrt{(l+m+1)(l-m)}Y_l^{m+1}+\sqrt{(l-m+1)(l+m)}Y_l^{m-1} \right).
\end{align}
Therefore, the action of the classical Liouvillian operator on the spherical harmonic $Y_l^m$ generates a linear combination of four neighboring spherical harmonics, each associated with a specific transition coefficient (see the right panel of Figure~\ref{fig:classical_Lanczos_algorithm}). Boundary terms, such as $\mathcal{L}_c Y_l^l$ or $\mathcal{L}_c Y_0^0$, may formally generate unphysical terms, such as $Y_{-1}^0$ or $Y_l^{l+1}$, which lie outside the light-cone-shaped grid depicted in the left panel of Figure~\ref{fig:classical_Lanczos_algorithm}. By construction, these terms vanish and thus do not affect the Lanczos algorithm.

\begin{figure}[t!]
    \centering
    \includegraphics[width=.45\textwidth]{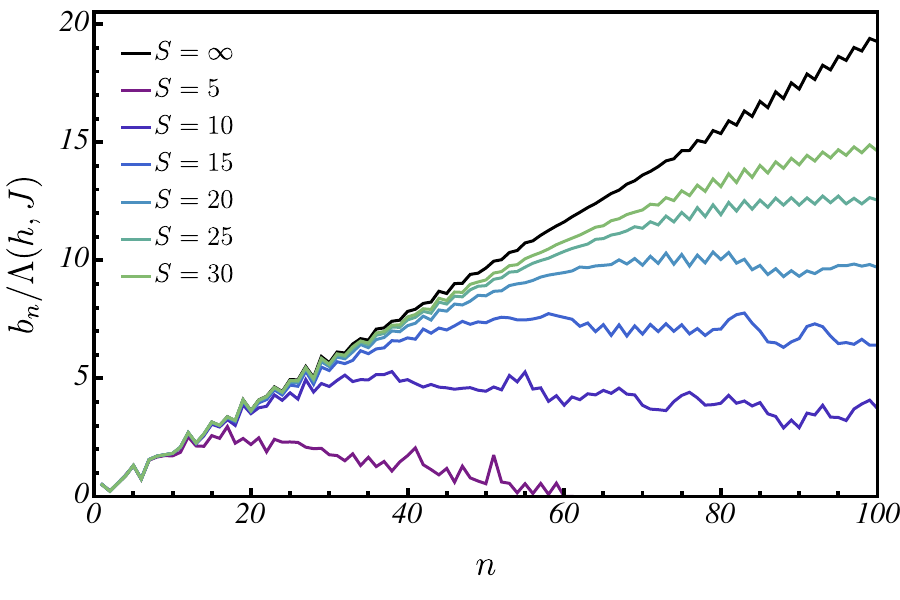}\qquad 
    \includegraphics[width=.45\textwidth]{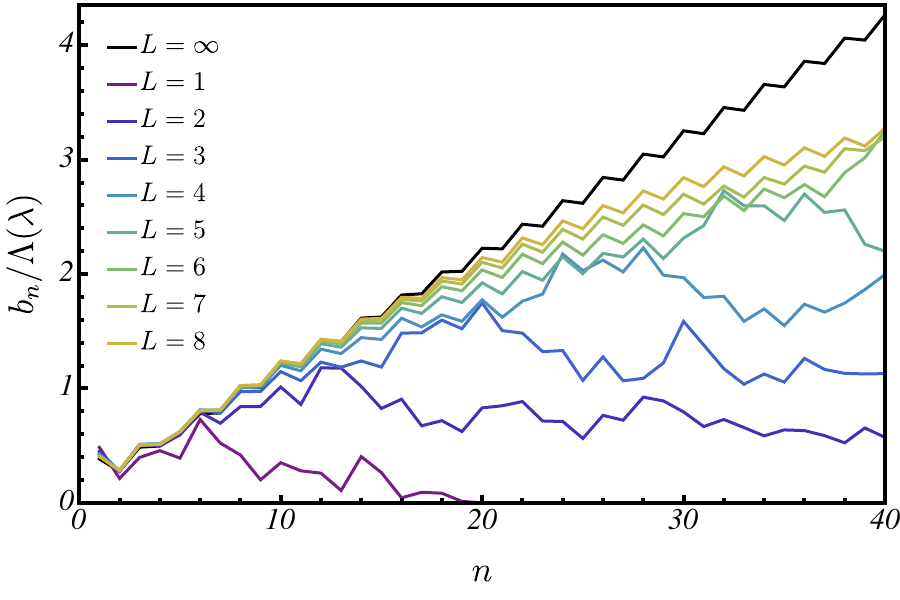}
     \caption{Classical Lanczos coefficients (black curves) shown versus quantum Lanczos coefficients (colored curves) for various values of collective spin size. \textbf{(Left)} For the classical LMG model \eqref{LMG_classical} and its quantum counterpart in Eq.~\eqref{LMG_rescaled_TSS}, with $h=J=1$. In the classical case, the initial function is $\ketc{z}$, and in the quantum case, the initial operator is $\hat{s}_z$. \textbf{(Right)} For the classical FP Hamiltonian \eqref{FP_classical} and its quantum counterpart in Eq.~\eqref{FP_rescaled}, with $\lambda=0$. In the classical case, the initial (normalized) function is $\ketc{x_1 z_2}$, and in the quantum case, the initial operator is (the normalized version of) $\hat{l}_x\otimes\hat{m}_z$.}
    \label{fig:classical_quantum_Lanczos_LMG_FP}
\end{figure}

\begin{figure}[ht!]
    \centering
    \includegraphics[scale=0.295]{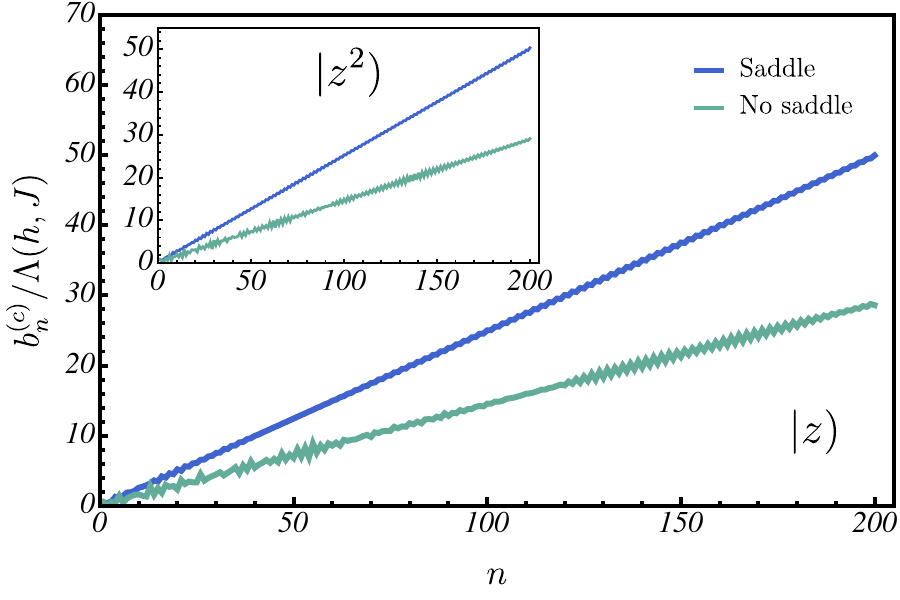}\qquad
    \includegraphics[scale=0.295]{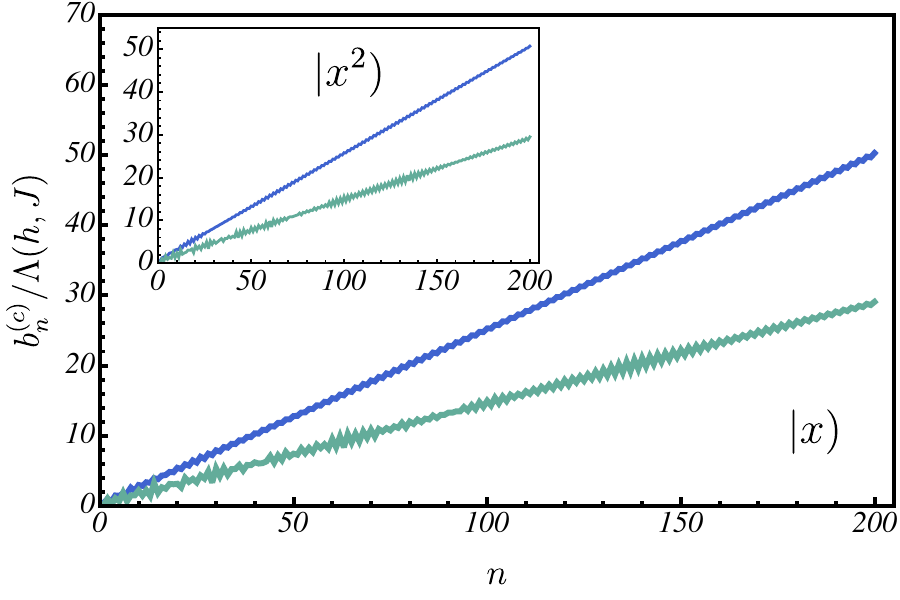}\qquad
    \includegraphics[scale=0.295]{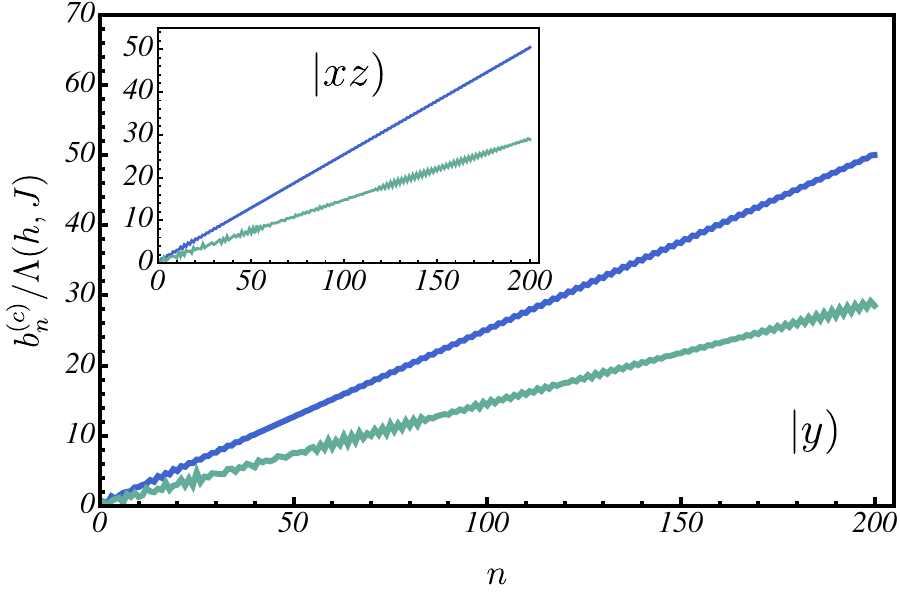}
    \caption{Classical Lanczos coefficients for the LMG model \eqref{LMG_classical} in both saddle (blue lines with $h=1/2$, $J=1$) and no-saddle (green lines with $h=2$, $J=1$) regimes for multiple initial functions up to $n=200$. The Lanczos coefficients are normalized by the spectral bandwidth $\Lambda(h,J)$, see Eq.~\eqref{spectral_bandwidth_LMG}. For all cases illustrated, the Lanczos coefficients grow faster in the presence of a saddle point in the phase space.}
    \label{fig:classical_Lanczos_LMG_multiple}
\end{figure}

The left panel of Figure~\ref{fig:classical_quantum_Lanczos_LMG_FP} displays the classical Lanczos coefficients obtained from the numerical execution of the algorithm using the Liouvillian operator given above for the initial function $\ketc{z}$. These are plotted alongside their quantum counterpart computed from the LMG model in Eq.~\eqref{LMG_rescaled_TSS} with initial operator $\hat{s}_z$ for increasing values of $S$. As derived in Section~\ref{subsec:Classical_Limit_Krylov}, the quantum and classical coefficients effectively align up to $n\simeq S$.

\begin{figure}[t!]
    \centering
    \includegraphics[width=.45\textwidth]{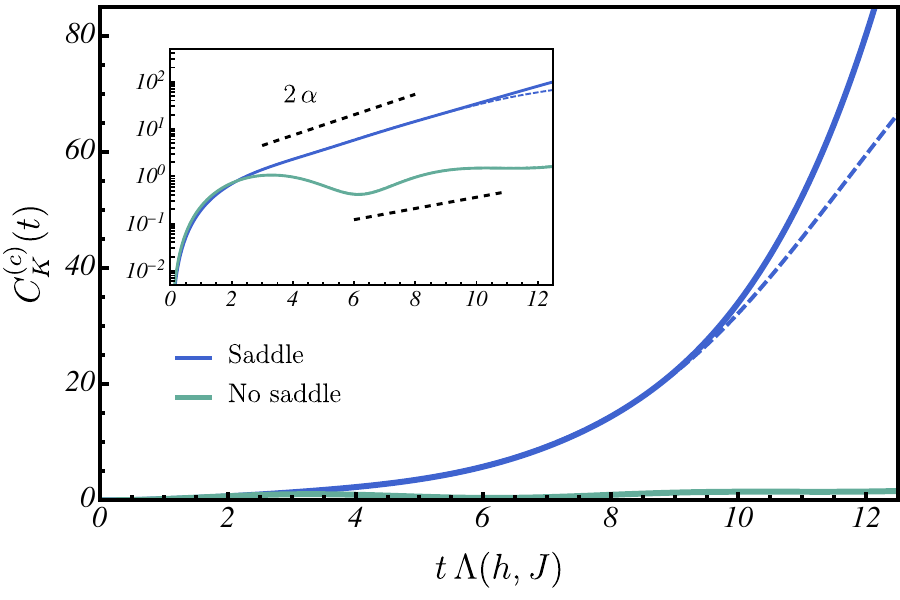}\qquad 
    \includegraphics[width=.45\textwidth]{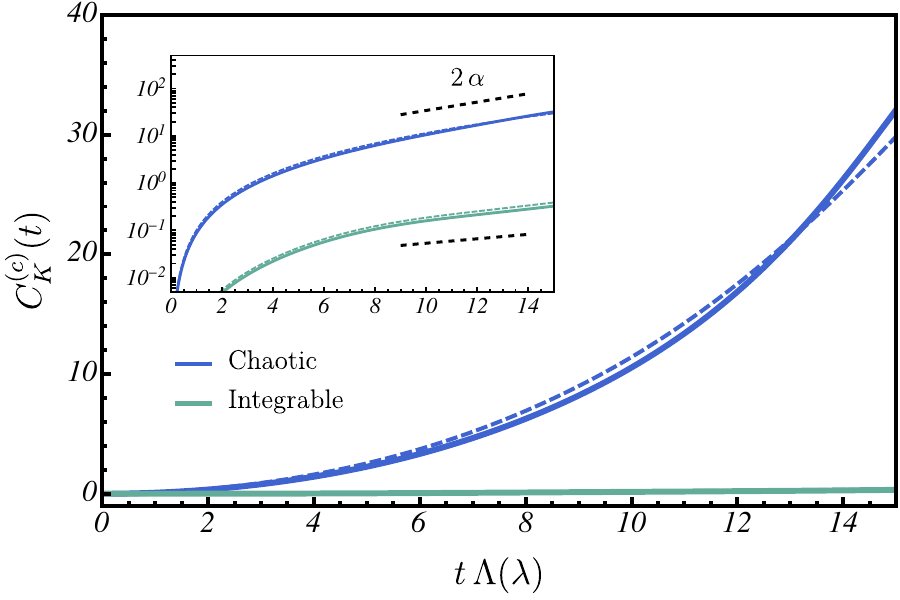}
    \caption{\textbf{(Left)} Classical K-complexity for the LMG model \eqref{LMG_classical} computed from Lanczos sequences up to $n=750$. Solid blue curves denote the saddle regime ($h=1/2$, $J=1$), while solid green curves denote the no-saddle regime ($h=2$, $J=1$) . Corresponding dashed lines show quantum K-complexity for the LMG model \eqref{LMG_rescaled_TSS} ($S=40$). Initial states are $\ketc{z}$ (classical) and $\hat{s}_z$ (quantum). Inset shows K-complexity on a logarithmic scale with the Krylov exponent $2 \alpha$ extracted from a linear fit of the corresponding Lanczos sequence. \textbf{(Right)} Classical K-complexity for the FP model \eqref{FP_classical} computed from Lanczos sequences up to $n=60$. Solid blue lines denote the chaotic regime ($\lambda=0$), while solid green lines denote the integrable regime ($\lambda=0.9$). Corresponding dashed lines show quantum K-complexity for the FP model \eqref{FP_rescaled} ($L=6$). Initial states are $\ketc{z_1 z_2}$ (classical) and $\hat{l}_z\otimes\hat{m}_z$ (quantum). Inset follows the same convention as the left panel. Both panels depict time ranges for which artificial truncation effects are not significant.}
    \label{fig:classical_Krylov_complexity}
\end{figure}

We have investigated the growth of Lanczos coefficients for various initial functions in both the saddle and no-saddle regimes of the classical LMG. Our results are depicted in Figure~\ref{fig:classical_Lanczos_LMG_multiple}, following the spectral bandwidth normalization discussed in Section~\ref{sec:Krylov_toolbox}, which is given by Eq.~\eqref{spectral_bandwidth_LMG}. We find that the normalized Lanczos coefficients grow faster when the classical phase space contains an instability. For coupling values such that there is no unstable saddle in phase space, we find a slower growth of the $b_n$ sequence, which in addition features enhanced fluctuations that hinder the complexity growth, as illustrated in Figure~\ref{fig:classical_Krylov_complexity}.

Given a classical Lanczos sequence truncated to a finite $n$, denoted $n_\star$, the corresponding K-complexity is computed using Eq.~\eqref{eq:Classical_KC}. Using a finite sequence requires caution, as there exists a time $t$ at which the Krylov wavefunctions begin to probe the boundary of the chain. Beyond this time, artificial truncation effects become significant, and the growth of K-complexity is altered. In practice, we require that $\left|\braketc{f_{n_\star}}{f(t)}\right|^2$ remains below a specific tolerance. With this in mind, we present the K-complexity data for the initial function $\ketc{z}$ in the left panel of Figure~\ref{fig:classical_Krylov_complexity}. The inset displays K-complexity on a logarithmic scale, revealing that in the saddle regime it features an exponential growth following the form predicted in Eq.~\eqref{eq:Ehrenfest_time_line1} after an initial transient, before truncation effects become relevant. This early-time exponential growth also persists in the no-saddle regime, though it is suppressed by the fluctuations accompanying the linear growth of the Lanczos coefficients, see Appendix~\ref{sec:linear_growth_flucuations}.

Is it possible, however, to go beyond this analysis and isolate the effect of the saddle point on the growth of the Lanczos coefficients and understand the role it plays in the fluctuations? This will be explored in Section~\ref{sec:classical_microcanonical_Lanczos}, where a microcanonical inner product is used to scan over energy shells in phase space.

\subsection{Classical Lanczos sequence for Feingold-Peres model}\label{sec:classical_FP}

The FP Hamiltonian in Eq.~\eqref{FP_rescaled} reduces to a classical Hamiltonian on the phase space $S^2_2$ ($N=2$) in the limit $L\to \infty$. Using the coordinates $(\theta_1,\theta_2,\varphi_1,\varphi_2)$ and the relations in Eqs.~\eqref{x_to_spherical_harmonics}, the classical Hamiltonian is given by
\begin{align}\label{FP_classical}
    H&=-\left(1+\lambda\right)\left(z_1+z_2\right)-4\left(1-\lambda\right) x_1 x_2\nonumber \\
    &=-\frac{\left(1+\lambda\right)}{\sqrt{3}}\left(Y_1^0+Z_1^0\right)-\frac{2\left(1-\lambda\right)}{3}\left(Y_1^1-Y_1^{-1} \right) \left(Z_1^1-Z_1^{-1} \right),
\end{align}
where the components of the first spin are denoted by $\bm{l}= (x_1,y_1,z_1)$ and those of the second by $\bm{m}=(x_2,y_2,z_2)$. From now on, we use the notation $Z_k^n(\theta_2,\varphi_2)\equiv (Y_2)_k^n(\theta_2,\varphi_2)$ to denote spherical harmonics in the phase space of the second classical spin. 

The classical integrability of the model is briefly discussed in Appendix~\ref{app:classical_FP}. For a fixed value of $\lambda$, integrability depends on the selected energy shell. The model appears predominantly chaotic for $0\lesssim \lambda \lesssim 0.25$, which is in agreement with the crossover from quantum chaos to integrability shown in Figure~\ref{fig:quantum_integrability_chaos_FP}.

Hamilton’s equations and their corresponding
linear stability analysis are detailed in Appendix~\ref{app:classical_FP}. A key feature of this system is that at least one saddle point is always present in phase space for $\lambda\in(-1,1)$. For $\lambda< 3/5$, it contains four saddles. At $\lambda=3/5$, a bifurcation occurs where two of these saddles are replaced by centers, leaving only two saddles for $\lambda> 3/5$ (see Table~\ref{tab:stability_fixed_points}).

\begin{table}[ht!]
\centering
\begin{subtable}{0.45\textwidth}
\centering
\begin{tabular}{|c|c|c|}\hline
 & $h <J$ & $h >J$\\\hline
$A_{1+}$ & saddle & center\\
$A_{1-}$ & center & center\\
$A_{2\pm}$ & centers & $\nexists$\\
\hline
\end{tabular}
\caption*{}
\end{subtable}
\begin{subtable}{0.45\textwidth}
\centering
\begin{tabular}{|c|c|c|}\hline
 & $-1<\lambda <3/5$ & $3/5<\lambda<1$\\\hline
$B_{1\pm}$ & saddles & centers\\
$B_{2\pm}$ & saddles & saddles\\
$B_{3\pm}$ & centers & $\nexists$\\
$B_{4\pm}$ & centers &$\nexists$\\
\hline
\end{tabular}
\caption*{}
\end{subtable}
\caption{\textbf{(Left)} Stability analysis of the fixed points of LMG Hamilton's equations \eqref{fixed_points_LMG}, see Appendix~\ref{app:classical_LMG} for their definition, and (\textbf{Right}) for FP Hamilton's equations \eqref{fixed_points_FP}, see Appendix~\ref{app:classical_FP}. A center is a fixed point at which eigenvalues of the Jacobian of Hamilton's vector field evaluated on it are complex. If at least one of the corresponding eigenvalues is positive the fixed point is a saddle.}
\label{tab:stability_fixed_points}
\end{table}

 As in the classical LMG model, the action of the classical Liouvillian operator on spherical harmonics can be obtained. For convenience, we introduce the following coefficients,
 \begin{subequations}\label{classical_FP_coefficients}
\begin{align}
    & \beta_\pm(l,m)\equiv \sqrt{(l\pm m+1)(l\mp m)}\, , \\
    & \eta_\pm(l,m)\equiv  \sqrt{\frac{(l\pm m+2)(l\pm m + 1)}{2l+3}}\,, \\
    & \theta_\pm(l,m) \equiv \sqrt{\frac{(l\pm m-1)(l\pm m)}{2l-1}}\, .
\end{align}
\end{subequations}
Using the formulas \eqref{spherical_harmonics_relation_1} and \eqref{spherical_harmonics_relation_2}, we show in Appendix~\ref{app:classical_Liouvillian_operator} that 
\begin{alignat}{2}\label{classical_Liouvillian_FP}
    & \mathcal{L}_c&&Y_l^m Z_k^n\nonumber \\
    &=&&-\left(1+\lambda\right) (m+n)Y_l^m Z_k^n+\left(1-\lambda\right)\Big[\frac{1}{\sqrt{2k+1}}\left(\beta_+ \eta_+Y_l^{m+1}Z_{k+1}^{n+1}-\beta_+ \theta_- Y_l^{m+1}Z_{k-1}^{n+1}\right.\nonumber \\
    & &&\left.-\beta_+ \eta_-Y_l^{m+1}Z_{k+1}^{n-1}+\beta_+ \theta_+ Y_l^{m+1}Z_{k-1}^{n-1}+\beta_- \eta_+Y_l^{m-1}Z_{k+1}^{n+1}-\beta_- \theta_-Y_l^{m-1}Z_{k-1}^{n+1}\right.\nonumber \\
    & &&\left.-\beta_- \eta_-Y_l^{m-1}Z_{k+1}^{n-1}+\beta_- \theta_+Y_l^{m-1}Z_{k-1}^{n-1}\right)+\frac{1}{\sqrt{2l+1}}\left(\eta_+\beta_+Y_{l+1}^{m+1}Z_{k}^{n+1} -\theta_-\beta_+ Y_{l-1}^{m+1}Z_{k}^{n+1}\right.\nonumber \\
    & &&- \eta_-\beta_+Y_{l+1}^{m-1}Z_{k}^{n+1}+ \theta_+\beta_+ Y_{l-1}^{m-1}Z_{k}^{n+1}+ \eta_+\beta_-Y_{l+1}^{m+1}Z_{k}^{n-1}- \theta_-\beta_-Y_{l-1}^{m+1}Z_{k}^{n-1}\nonumber\\ 
    & &&\left.-\eta_-\beta_-Y_{l+1}^{m-1}Z_{k}^{n-1}+ \theta_+\beta_-Y_{l-1}^{m-1}Z_{k}^{n-1}\right)\Big] ,
\end{alignat}
where, for conciseness, the arguments of the coefficients have been suppressed. For example, $\beta_+ \eta_+$ stands for $\beta_+(l,m)\eta_+(k,n)$. This expression is used in the right panel of Figure~\ref{fig:classical_quantum_Lanczos_LMG_FP} to compute the classical Lanczos coefficients associated with the initial function $\ketc{x_1 z_2}$. These are plotted alongside their quantum counterpart associated with the rescaled FP model in Eq.~\eqref{FP_rescaled} with initial operator $\hat{l}_x\otimes \hat{m}_z$ for increasing values of $L$. The quantum and classical coefficients align up to $n\simeq L$. We note that computing these coefficients requires more computational effort than for the LMG model for the same value of $n_\star$. This is primarily because the grid dimensionality is doubled, $(l,m)\to (l,m,k,n)$, thereby increasing the number of nonzero coefficients in the grid over which the classical Liouvillian acts.

\begin{figure}[t!]
    \centering
    \includegraphics[scale=0.295]{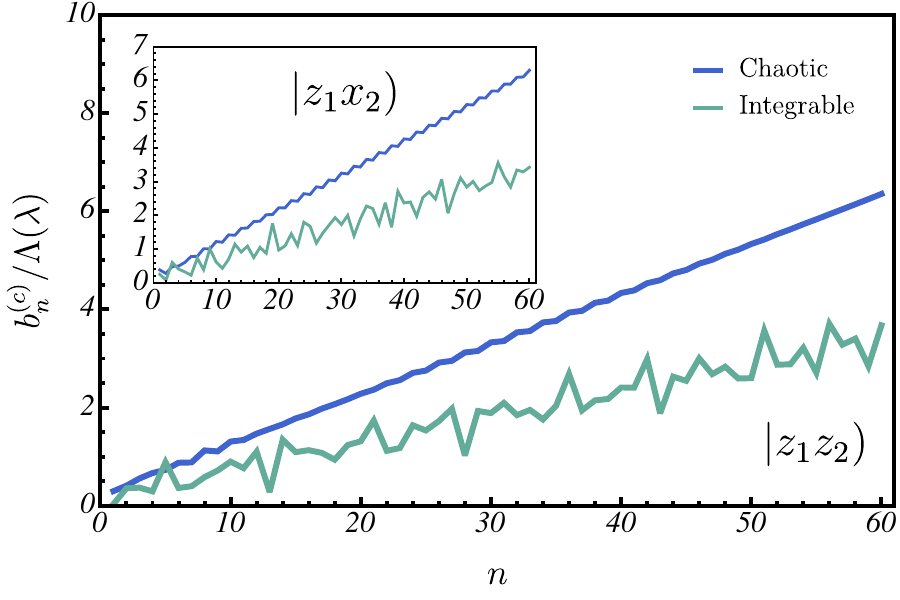}\qquad
    \includegraphics[scale=0.295]{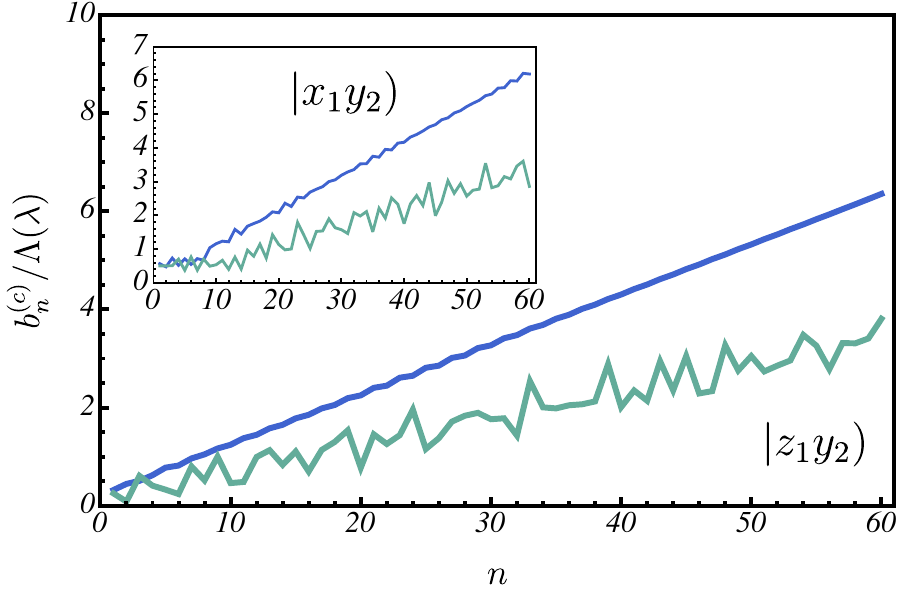}\qquad
    \includegraphics[scale=0.295]{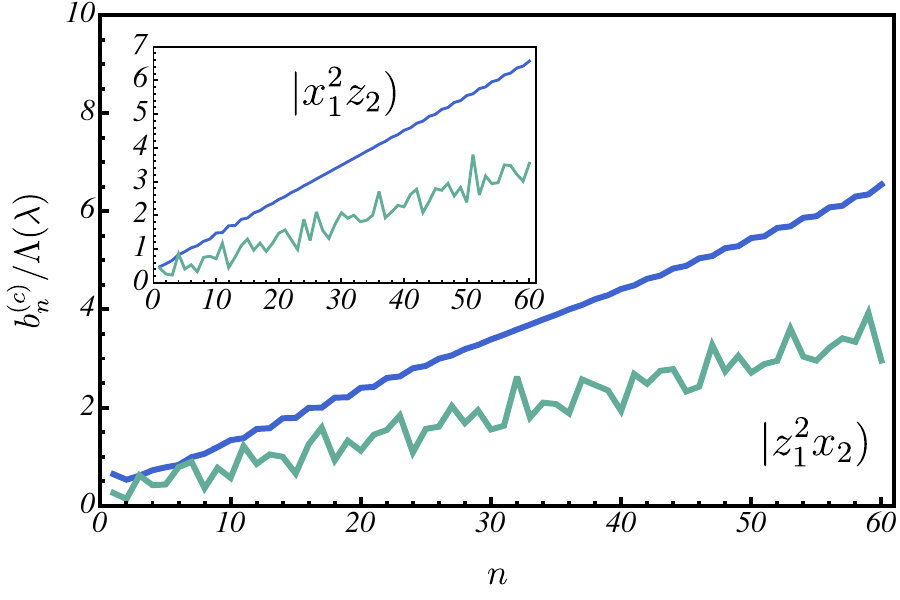}
    \caption{Classical Lanczos coefficients for the FP model \eqref{FP_classical} in both chaotic (blue lines with $\lambda=0$) and integrable (green lines with $\lambda=0.9$) regimes for multiple initial functions up to $n=60$. The Lanczos coefficients are normalized by the spectral bandwidth $\Lambda(\lambda)$, see Eq.~\eqref{spectral_bandwidth_FP}. For all cases illustrated, the Lanczos coefficients grow faster in the chaotic regime.}
    \label{fig:classical_Lanczos_FP_multiple}
\end{figure}

Similar to the LMG model, we have investigated the growth of Lanczos coefficients for various initial functions in both the chaotic and integrable regimes of the classical FP model. Following the spectral bandwidth normalization, which is given by Eq.~\eqref{spectral_bandwidth_FP}, the results are shown in Figure~\ref{fig:classical_Lanczos_FP_multiple}. It is clear that the Lanczos coefficients grow faster in the chaotic regime than in the integrable regime. In the chaotic case, this instability may be attributed not only to the onset of chaos as a genuine mechanism but also to the presence of four saddle points in the classical phase space, see Table~\ref{tab:stability_fixed_points}. Conversely, growth is slower in the integrable regime, where fluctuations around the mean linear trend become more pronounced. At this value of $\lambda$, the number of saddle points in the phase space is halved, resulting in fewer local instabilities to drive the linear growth of the Lanczos coefficients.

Results for the classical K-complexity of the initial function $\ketc{z_1 z_2}$ are depicted in the right panel of Figure~\ref{fig:classical_Krylov_complexity}. The inset demonstrates that K-complexity grows exponentially at early times in both regimes. This growth is slower in the integrable case due to the presence of fluctuations.

As a partial conclusion, our analysis of both the LMG and FP models indicates that results for the LMG in the saddle regime are very similar to those of the FP model in the chaotic regime: both feature almost perfectly linear Lanczos coefficients and K-complexity described by an exponential law. This suggests that, at the level of classical Krylov dynamics, the presence of classical saddle points can mimic the behavior of chaotic systems, even in the presence of integrable structure. By contrast, the no-saddle and integrable regimes exhibit a slower mean linear growth of the Lanczos sequence, which is moreover accompanied by pronounced fluctuations. As discussed in Appendix~\ref{sec:linear_growth_flucuations}, these fluctuations suppress the early-time growth of K-complexity, with the degree of suppression increasing with their magnitude. 

\subsection{Classical microcanonical Lanczos algorithm}\label{sec:classical_microcanonical_Lanczos}

The saddle points in the phase space of both classical LMG and FP models are localized at well-defined energies. We now investigate the growth of Lanczos coefficients within fixed energy shells. To this end, we return to the general form of the inner product given in Eq.~\eqref{eq:Classical_Inner_product_generic}. We choose the distribution $p$ to be a microcanonical indicator function, denoted $p_{E,\Delta E}$, which vanishes if a phase-space point lies outside the energy window $[E-\Delta E/2, E+\Delta E/2]$. Specifically, we define the microcanonical inner product as
\begin{equation}\label{microcanonical_inner_product_classical}
    \braketc{f}{g}_{E,\Delta E}=\frac{1}{K_{E,\Delta E}}\int_{S^2_N} f^* g \, p_{E,\Delta E} \, \omega_N\, ,
\end{equation}
where the distribution $p_{E,\Delta E}$ is given by
\begin{equation}\label{microcanonical_distribution}
p_{E,\Delta E}(\Omega)=\left\{
    \begin{array}{cl}
        1 & \text{if }H(\Omega)\in[E-\Delta E/2, E+\Delta E/2] \\
        0 & \text{otherwise}\, ,
    \end{array}
\right.
\end{equation}
and $K_{E,\Delta E}=\int_{S^2_N}\, p_{E,\Delta E} \, \omega_N$ is the volume of the energy shell. Here, $\Omega\in S^2_N$ denotes a point in phase space. Physically, $p_{E,\Delta E}$ represents a uniform probability distribution over a narrow energy shell of width $\Delta E\geq 0$ centered at energy $E$. 

However, a new difficulty arises: the spherical harmonics no longer form an orthonormal basis of $\mathcal{F}(S^2_N)$ with respect to this microcanonical inner product. While this does not prevent their use, it significantly slows the computation of the Lanczos coefficients. For a fixed energy shell, all inner products of the form $\braketc{Y_l^m}{Y_{l'}^{m'}}_{E,\Delta E}$ (for the LMG model) or $\braketc{Y_{l}^{m} Z_{k}^{n}}{Y_{l'}^{m'} Z_{k'}^{n'}}_{E,\Delta E}$ (for the FP model) must be explicitly evaluated before executing the algorithm. In our numerical computations, we have evaluated these integrals using Monte Carlo methods.

Nonetheless, by exploiting the symmetries of the classical LMG and FP Hamiltonians, one can derive selection rules that force certain microcanonical inner products to vanish. In Appendix~\ref{app:spherical_harmonics_selection_rules}, we show that for the classical LMG model,
\begin{subequations}\label{spherical_harmonics_LMG}
\begin{align}
    & \braketc{Y_l^m}{Y_{l'}^{m'}}_{E,\Delta E} \in \mathbb{R}\, ,\label{LMG_microcanonical_formula_0}\\
    & \braketc{Y_l^m}{Y_{l'}^{m'}}_{E,\Delta E}=0\quad \text{if }  l + l' + m + m' \text{ is odd}\, .\label{LMG_microcanonical_formula}
\end{align}
\end{subequations}
While \eqref{LMG_microcanonical_formula} is not as restrictive as a full orthogonality relation, it significantly reduces the computational burden by halving the number of nonzero inner products. Similar relations can be obtained for the classical FP model (not shown in this work).

With the microcanonical inner product in hand, the Lanczos algorithm remains unchanged and can be performed as described in Section~\ref{sec:classical_Lanczos_algo_practical}, using the classical Liouvillian operators from Eqs.~\eqref{classical_Liouvillian_LMG} and \eqref{classical_Liouvillian_FP}. We omit results for the classical FP model due to the expensive numerical cost of computing the microcanonical inner products $\braketc{Y_{l}^{m} Z_{k}^{n}}{Y_{l'}^{m'} Z_{k'}^{n'}}_{E,\Delta E}$. Even when accounting for the selection rules that force certain terms to vanish, we could only reach $n \simeq 10-15$, which is insufficient to draw meaningful physical conclusions. 

\begin{figure}[t!]
    \centering
    \includegraphics[width=.45\textwidth]{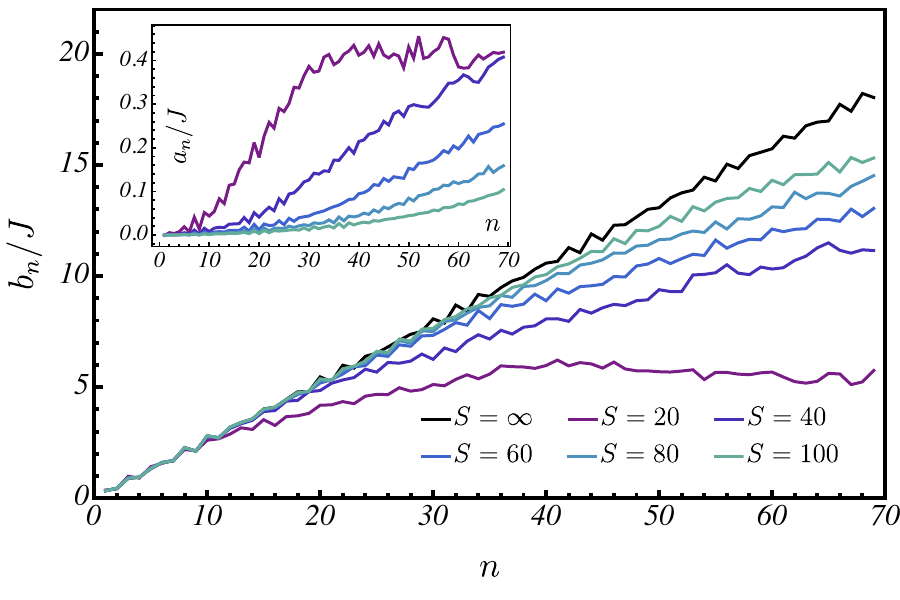}
    \caption{Classical microcanonical Lanczos coefficients for the classical LMG model \eqref{LMG_classical} (black curve) and its quantum counterparts in Eq.~\eqref{LMG_rescaled_TSS} (colored curves) for various values of collective spin size. The parameters are $h=1$ and $J=2$. In the classical case, the initial function is $\ketc{z}$, and in the quantum case, the initial operator is $\hat{s}_z$. The classical coefficients were computed using the inner product defined in Eq.~\eqref{microcanonical_inner_product_classical} at energy $E=-0.95$ with a shell width $\Delta E=0.25$ (same energy shell for the quantum case). Each inner product of the form $\braketc{Y_l^m}{Y_{l'}^{m'}}_{E,\Delta E}$ was computed via Monte Carlo integration using $5\times10^6$ random points on the sphere $S^2$. The inset shows the nonzero quantum $a_n$ coefficients, which vanish in the limit $S\to \infty$, see proof in Section~\ref{sec:quantum_Lanczos_microcanonical}.}
    \label{fig:microcanonical_classical_LMG}
\end{figure}

Figure~\ref{fig:microcanonical_classical_LMG} shows the classical microcanonical Lanczos coefficients for the LMG model computed with the inner product in Eq.~\eqref{microcanonical_inner_product_classical} and with initial function $\ketc{z}$, together with their quantum counterparts from the LMG model in Eq.~\eqref{LMG_rescaled_TSS} (see Section~\ref{sec:quantum_Lanczos_microcanonical} for the quantum microcanonical Lanczos algorithm) with initial operator $\hat{s}_z$. The inset shows the $a_n$ coefficients for increasing values of $S$, which vanish in the limit $S\to \infty$ as proved in Section~\ref{subsec:Classical_Krylov_Generic_setup}. Note that the coefficients here are normalized by $J$ rather than the spectral bandwidth used in the previous sections. While the latter normalization was intended to compare growth across different dynamical regimes associated with Hamiltonians of different spectral bandwidth, the current microcanonical analysis keeps the Hamiltonian parameters fixed, varying only the energy shell.

\begin{figure}[ht!]
    \centering
    \includegraphics[width=.45\textwidth]{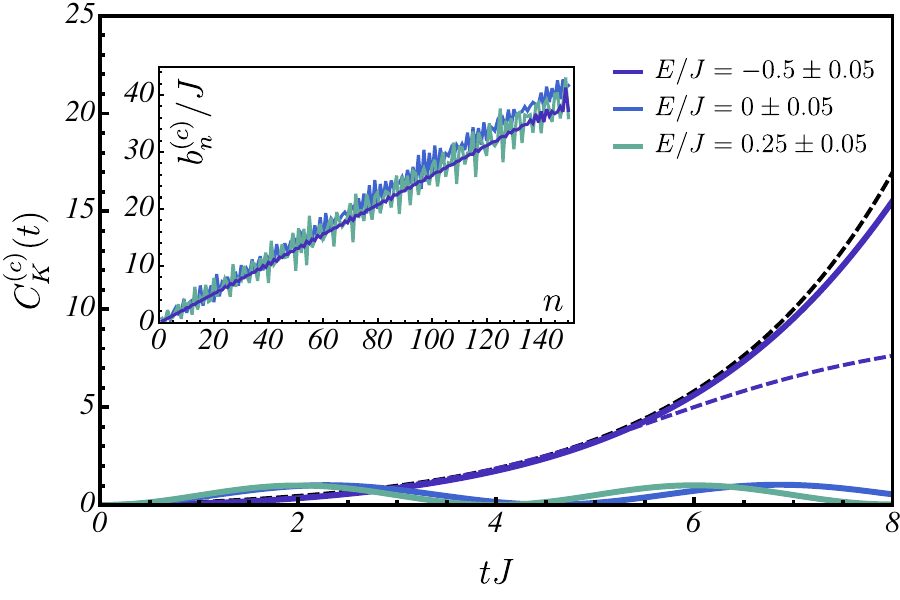}\qquad 
    \includegraphics[width=.45\textwidth]{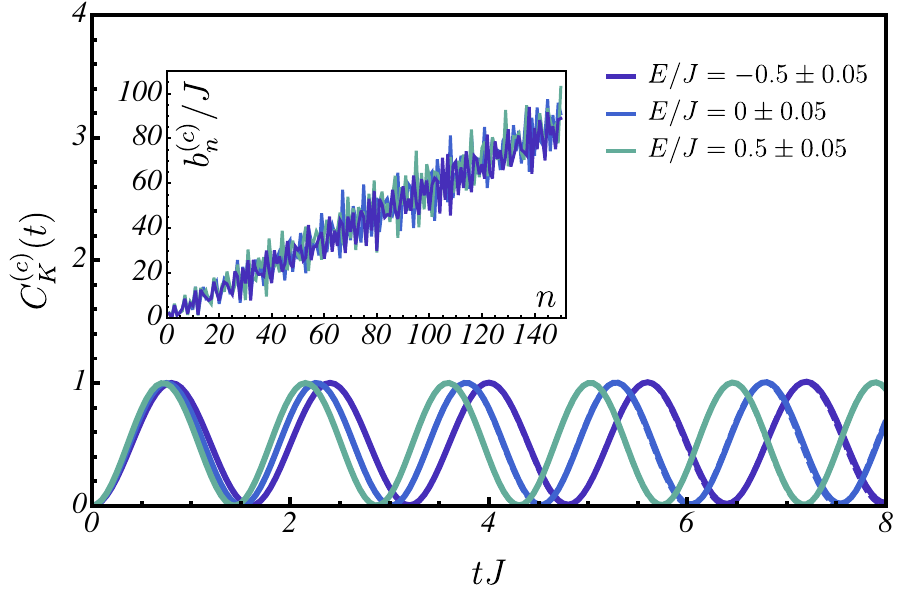}
    \caption{Classical microcanonical K-complexity for the LMG model \eqref{LMG_classical} for the initial function $\ketc{z}$ across various energy shells, computed from Lanczos sequences up to $n=150$ (shown in the insets). \textbf{(Left)} Saddle regime ($h=1/2$, $J=1$), including the energy shell containing the saddle point ($E/J=-0.5$). Colored dashed lines represent the quantum K-complexity for the LMG model \eqref{LMG_rescaled_TSS} ($S = 60$). In energy shells away from the saddle, the quantum K-complexity matches the classical results exactly and the two are indistinguishable (within the time range plotted here). The black dashed line denotes the exponential growth $e^{2\alpha t}$. \textbf{(Right)} No-saddle regime ($h=2$, $J=1$). Similarly, the quantum K-complexity matches the classical results exactly and the two cannot be distinguished. Both panels depict time ranges for which artificial truncation effects are not significant.}
    \label{fig:classical_microcanonical_Krylov_complexity}
\end{figure}

Leveraging the microcanonical inner product, we investigate the Lanczos coefficients within fixed energy shells. We recall that a key feature of the classical LMG model is the presence of a saddle point in phase space at the energy $E=-h$ for $h<J$ (see Appendix~\ref{app:classical_LMG}). Our results for the initial function $\ketc{z}$ are depicted in Figure~\ref{fig:classical_microcanonical_Krylov_complexity} for both saddle and no-saddle regimes. In each regime, the growth of the Lanczos coefficients is linear, and the corresponding slope is nearly identical. Crucially, however, the fluctuations around this linear trend exhibit distinct profiles that significantly affect the early-time behavior of the classical K-complexity. 

Within the energy shell containing the saddle, fluctuations are nearly absent -- even more so than in Figure~\ref{fig:classical_Lanczos_LMG_multiple} -- resulting in almost perfectly linear growth. Consequently, the corresponding K-complexity closely follows an exponential law at early times. In other energy shells (while still within the saddle regime), fluctuations increase abruptly, and cause an early-time suppression of K-complexity. In the no-saddle regime, the slopes of the linear trends remain again similar across energy shells, but fluctuations are enhanced and present throughout, leading to a suppression of K-complexity at early times.

These results confirm the observations made in Section~\ref{sec:classical_LMG}, where it was noticed that the saddle point reduces fluctuations in the Lanczos growth and drives the K-complexity to grow exponentially at early times. In its absence, fluctuations increase and the K-complexity is suppressed, preventing it from following an exponential growth. This naturally raises the following question: why does the saddle point act as a \textit{stabilizer}, effectively rendering the linear growth of the classical Lanczos coefficients nearly perfect? We leave this question for future investigation.

\section{Quantum Krylov Complexity}\label{sec:quantum_K_complexity}

In this section, we present a quantum analysis of K-complexity in the LMG and FP models. For these collective spin models, the effective Planck constant is inversely proportional to the total spin, so the quantum analysis is equivalent to a finite-size analysis. We focus on the late-time regime of K-complexity, complementing the early-time behavior examined in Section~\ref{sec:ClassicalKC} through the classical limit.

We begin by deriving the maximum Krylov dimension associated with a given operator, which is determined by its transformation properties under the discrete symmetries of the two models. Subsequently, we investigate the infinite-temperature K-complexity profile as a function of time and analyze its saturation value for several operators in both LMG and FP models. In particular, we examine the relation between the saturation value and the ESQPT in the LMG model, as well as the crossover from integrability to spectral chaos in the FP model. In the latter case, we observe a suppression of the K-complexity saturation value in the integrable regime, along the lines of Ref.~\cite{Adrian_3}. By contrast, in the LMG model, the ESQPT does not govern the late-time behavior of K-complexity, which is mostly controlled by the underlying $\mathfrak{su}(2)$ algebra.

Analogously to the classical analysis presented in Section~\ref{sec:ClassicalKC}, we then isolate the effect of the ESQPT in the LMG model by resolving individual spectral sectors through the implementation of the Lanczos algorithm with a microcanonical inner product. Our results show that (i) the microcanonical K-complexity saturation value closely tracks the local density of states, exhibiting a similar singularity at the ESQPT energy, and (ii) the time-dependent microcanonical complexity profile displays strong persistent complexity oscillations in energy windows away from the instability, where the LMG spectrum is known to be regular, as reviewed in Section \ref{sec:LMG_FP_models}. Together with the classical microcanonical analysis provided in Section~\ref{sec:classical_microcanonical_Lanczos}, these results complete the finite-energy analysis of K-complexity in the LMG model, demonstrating that its integrable nature can be recovered at both early and late times once individual energy shells away from the instability are resolved.

\subsection{Quantum Krylov dimension}\label{sec:K_dim}

Classical Lanczos coefficients, studied in Sections \ref{sec:classical_LMG} and \ref{sec:classical_FP}, grow indefinitely because the space of all smooth complex-valued functions on $S^2_N$ is infinite-dimensional. By contrast, the corresponding quantum operator space is finite-dimensional for the finite spin-$S$ systems considered here. Therefore, as anticipated in Section~\ref{sec:Krylov_toolbox}, the Lanczos algorithm necessarily terminates after a finite number of iterations, $n=K$, where $K$ is the Krylov dimension. Making use of the symmetries of the LMG and FP models, we derive exact upper bounds for the Krylov dimension. These bounds provide a useful benchmark to assess the validity of the numerical Lanczos algorithm. 

An arbitrary operator $\hat{\mathcal{O}}$ that satisfies 
\begin{equation}
    \hat{U} \hat{\mathcal{O}} \hat{U}^\dagger=\epsilon\,  \hat{\mathcal{O}} \, ,
\end{equation}
is said to be negatively charged under $\hat{U}$ if $\epsilon=-1$, and positively charged under $\hat{U}$ if $\epsilon=+1$. If this relation is not satisfied for any $\epsilon$, the operator $\hat{\mathcal{O}}$ is said to be uncharged. 

For the LMG model, we recall that $\hat{U}=e^{- i \pi \hat S_x}$. Consequently, eigenstates of the Hamiltonian can be written as $\ket{E, \alpha}$, with $\alpha=\pm$, such that $\hat H \ket{E,\alpha}=E\ket{E,\alpha}$ and $\hat U \ket{E,\alpha}=\alpha\ket{E,\alpha}$. We show in Appendix~\ref{app:LMG_symmetries} that the dimensions of the $\pm$ sectors, denoted $D_\pm$, depend on whether $S$ is even, odd or half-integer. Let $\hat{\mathcal{O}}$ be an operator with charge $\epsilon$ under $\hat U$. Then, its matrix elements $\mel{E,\alpha}{\hat{\mathcal{O}}}{E',\beta}$ vanish if $\alpha \beta = -\epsilon$. 

Similar considerations apply to the FP model. In that case, we recall that $\hat U_1=e^{- i  \pi( \hat L_z+\hat M_z)}$ and $\hat U_2$ exchanges the two collective spins. Eigenstates of the Hamiltonian can be written as $\ket{E, \alpha_1, \alpha_2}$, with $\alpha_i=\pm$ ($i=1,2$), such that $\hat H \ket{E,\alpha_1, \alpha_2}=E\ket{E,\alpha_1, \alpha_2}$, $\hat{U}_1 \ket{E, \alpha_1, \alpha_2}=\alpha_1\ket{E, \alpha_1, \alpha_2}$ and $\hat{U}_2 \ket{E, \alpha_1, \alpha_2}=\alpha_2\ket{E, \alpha_1, \alpha_2}$. Let $\hat{\mathcal{O}}$ be an operator with charges $\epsilon_1$ under $\hat U_1$ and $\epsilon_2$ under $\hat U_2$, which we denote by $(\epsilon_1,\epsilon_2)$. The matrix elements $\mel{E,\alpha_1,\alpha_2}{\hat{\mathcal{O}}}{E,\beta_1,\beta_2}$ vanish whenever either $\alpha_1 \beta_1 = -\epsilon_1$ or $\alpha_2 \beta_2 = -\epsilon_2$ is satisfied.

The presence of the operator $\hat V$, which anti-commutes with the FP Hamiltonian, is crucial here because it can generate degeneracies in the Liouvillian spectrum. Since the subspaces $\mathcal{H}_{--}$ and $\mathcal{H}_{-+}$ are exchanged by $\hat V$, if $\ket{E,-,-}\in\mathcal{H}_{--}$, then $\ket{-E,-,+}\in\mathcal{H}_{-+}$ and vice versa. Consequently, the two eigenoperators $\ket{E,-,-}\bra{E',-,+}$ and $\ket{-E',-,-}\\ \bra{-E,-,+}$ share the same phase $E-E'$, which is doubly degenerate, except for eigenoperators of the form $\ket{E,-,-}\bra{-E,-,+}$. The eigenspaces representatives associated with these nondegenerate phases contribute to the Krylov space as long as the anti-diagonal entries of the operator $\hat{\mathcal{O}}$ in the energy basis, denoted $\overline{\text{diag}}(\hat{\mathcal{O}})$, are nonzero. The corresponding Krylov dimension in the case where $\overline{\mathrm{diag}}(\hat{\mathcal{O}}) \neq 0$ is denoted by $\widetilde{K}$.

The resulting formulas for the Krylov dimension are summarized in Table~\ref{tab:Krylov_dimension_LMG} for the LMG model and in Table~\ref{tab:Krylov_dimension_FP} for the FP model, covering all possible charges of the initial operator. A detailed derivation of these formulas is provided in Appendix~\ref{app:LMG_Krylov_dim} for the LMG model and in Appendix~\ref{app:FP_Krylov_dim} for the FP model. Strictly speaking, these values correspond to upper bounds on the Krylov dimension, which will be saturated if all the matrix elements of the operator in the energy basis that are not constrained to vanish by selection rules are nonzero. We have verified numerically that this is the case for the collective spin operators that we considered in our study. To this end, we implemented an arbitrary-precision version of the Lanczos algorithm. The use of arbitrary-precision arithmetic is necessary because the Lanczos algorithm exhibits significant numerical instability when using standard floating-point representations. 

\begin{table}[t!]
    \renewcommand{\arraystretch}{1.4}
    \centering
    \begin{tabular}{|c|c|c|}
            \hline
            Charge & $K \, (S\in \mathbb{N})$ & $K \, (S\in \mathbb{N}/2)$ \\ \hline
            $-$ & $2S(S+1)$ & $2(S+1/2)^2$  \\ \hline
            $+$ & $2S^2+1$ & $2 S^2+1/2$ \\ \hline
            $0$ & $4 S^2+2S+1$ & $4 S^2+2S+1$ \\ \hline
    \end{tabular}
    \caption{Krylov dimensions of the LMG model for all possible charges of the initial operator $\hat{\mathcal{O}}$, for both integer and half-integer values of $S$. These Krylov dimensions are obtained assuming saturation of the corresponding upper bound, as explained in the text.}
    \label{tab:Krylov_dimension_LMG}
\end{table}

\begin{table}[t!]
    \renewcommand{\arraystretch}{1.4}
    \centering
    \begin{tabular}{|c|c|c|}
            \hline
            Charge & $K$ & $\widetilde{K}$ \\ \hline 
            $(-,-)$ & $2 L (L+1)(2L^2+2L+1)$ & $\nexists$ \\ \hline
            $(-,+)$ & $2 L (L+1)(2L^2+2L+1)$ & $\nexists$ \\ \hline
            $(+,-)$ &  $L (L+1) (2 L^2+2 L-1)$ & $ L (L+1)(2L^2+2L+1)$  \\ \hline
            $(+,+)$ &  $L (L+1) (2 L^2+2 L-1)+1$ & $L (L+1)(2L^2+2L+1)+1$ \\ \hline
            $(0,-)$ &  $L (L+1)(6L^2+6L+1)$ & $3 L (L+1)(2L^2+2L+1)$ \\ \hline
            $(0,+)$ &  $L (L+1)(6L^2+6L+1)+1$ & $3 L (L+1)(2L^2+2L+1)+1$ \\ \hline
            $(-,0)$ &  $2 L (L+1)  (2 L^2+2 L+1)$ & $\nexists$ \\ \hline
            $(+,0)$ &  $2 L (L+1)(2L^2+2L-1)+1$ & $2 L (L+1)(2L^2+2L+1)+1$ \\ \hline
            $(0,0)$ &  $8 L^2(L+1)^2+1$ & $4 L (L+1)(2L^2+2L+1)+1$  \\ \hline
    \end{tabular}
    \caption{Krylov dimensions of the FP model for all possible charges of the initial operator $\hat{\mathcal{O}}$, for $L$ in $\mathbb{N}$. The first column corresponds to the case where all the anti-diagonal elements of $\hat{\mathcal{O}}$ in the energy basis are zero, while the second column corresponds to the case where they are all nonzero. The symbol ``$\nexists$'' indicates that there cannot exist an operator with the given charge and anti-diagonal property. As also explained in the main text, the values of $K$ and $\widetilde{K}$ are obtained assuming saturation of the Krylov dimension upper bound.}
    \label{tab:Krylov_dimension_FP}
\end{table}

The general observation is that the more symmetry charges the initial operator carries, the smaller the Krylov dimension of the space generated under its time evolution. Importantly, in both LMG and FP models, the Krylov dimension depends only on the charge of the initial operator and not on the parameters of the model. For the LMG model, this indicates that the Krylov dimension is insensitive to the presence of the ESQPT, while for the FP model that it is insensitive to whether the system is in the integrable or chaotic regime.

A noticeable difference arises between the LMG and FP models when the initial operator is fully uncharged. In the former, the bound $D^2-D+1$ derived in Ref.~\cite{Adrian_1} is saturated, whereas in the latter it is not, in both cases where $\overline{\text{diag}}(\hat{\mathcal{O}})=0$ and $\overline{\text{diag}}(\hat{\mathcal{O}})\neq 0$. 

\subsection{Krylov complexity of Lipkin-Meshkov-Glick and Feingold-Peres models}\label{sec:late-time_K-complexity}

We now investigate the K-complexity of both LMG and FP models. This quantity, defined in Eq.~\eqref{eq:KC_from_wavefn}, can be interpreted as a measure of operator growth and complexification under time evolution. For systems with a finite-dimensional Hilbert space, K-complexity is strictly bounded by $K$. Once the probe operator has fully explored its associated Krylov space, K-complexity saturates to a positive value (see Refs.~\cite{Adrian_1,barbon2019} for a detailed discussion on the relevant time scales).

We provide evidence that the late-time saturation value of K-complexity $\overline{C_K}$ defined in Eq.~\eqref{eq:KC_LongTimeAvg}, computed with the infinite-temperature inner product in Eq.~\eqref{eq:Operator_Inner_product} with $\hat{\rho}=\mathbbm{1}$, fails to distinguish between saddle and no-saddle regimes in the LMG model, and between the chaotic and integrable regimes in the FP model. We show the critical role of quasi-degeneracies in the convergence of K-complexity toward $\overline{C_K}$. Subsequently, we demonstrate that for ``non-generic'' initial operators (defined below), this saturation value is fully determined by the corresponding autocorrelation functions, leaving only ``generic'' operators as potential probes of the saturation value of K-complexity.

Within certain parameter regimes and at finite values of the collective spin $S$ or $L$, both the LMG and FP Hamiltonians exhibit quasi-degeneracies in their spectra (see Section~\ref{sec:LMG_FP_models}) that result in a significantly slowed convergence of K-complexity toward $\overline{C_K}$. Indeed, the Lanczos coefficients $b_n$ are related to the spectrum of the Liouvillian operator: in systems with a finite Krylov dimension, coefficients associated with larger $n$ resolve increasingly fine energy differences \cite{Rabinovici:2025otw}. Consequently, quasi-degeneracies -- manifesting as very small energy differences in the Liouvillian spectrum -- yield very small Lanczos coefficients at large $n$. This bottleneck prevents the Krylov wavepacket from reaching the far end of the Krylov chain, thereby extending the time scale required to attain the saturation value. 

Quasi-degeneracies in both the LMG and FP models are associated with eigenstates from different symmetry sectors of the $\mathbb{Z}_2$ collective spin-inversion symmetry. As discussed in Section~\ref{sec:K_dim}, symmetries of the Hamiltonian impose selection rules on the matrix elements of the initial operator in the energy basis. Consequently, only operators belonging to specific symmetry sectors -- those that connect quasi-degenerate states -- are sensitive to the slowed convergence of K-complexity. This occurs for negatively charged and uncharged operators (with respect to $\hat{U}$ and $\hat{U}_1$ in the LMG and FP models, respectively).

\begin{figure}[t!]
    \centering
    \includegraphics[width=.45\textwidth]{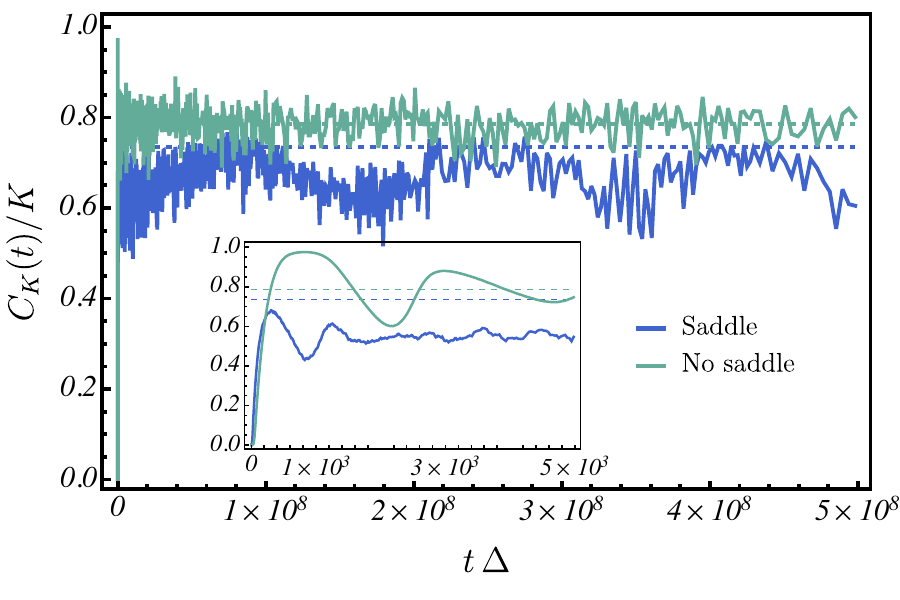}
    \includegraphics[width=.45\textwidth]{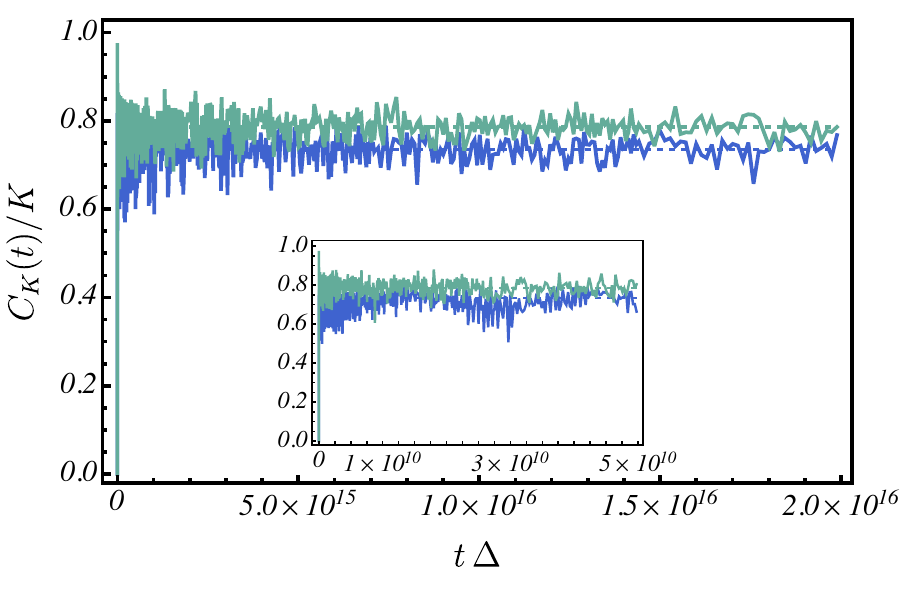}
    \caption{Late-time K-complexity for the LMG model \eqref{LMG_rescaled_TSS} ($S=40$) associated with the initial operator $\hat{s}_z$ across various time scales in both saddle (blue curves with $h=1/2$, $J=1$) and no-saddle (green curves with $h=2$, $J=1$) regimes. Time is rescaled by the mean-level spacing $\Delta\simeq \Lambda(h,J)/S$. Horizontal dashed lines indicate the late-time saturation values $\overline{C_K}$ computed via Eq.~\eqref{eq:KC_LongTimeAvg}.}
    \label{fig:C_K_very_large_time_LMG}
\end{figure}

For brevity, we provide numerical data only for the LMG model, presented in Figure~\ref{fig:C_K_very_large_time_LMG}. This figure displays the time evolution of K-complexity for the initial operator $\hat{s}_z$ in both saddle and no-saddle regimes over various time scales. While both cases exhibit quantitatively similar saturation values when normalized by the corresponding Krylov space dimension, the primary observation is that the K-complexity requires significantly longer times to reach saturation in the saddle regime. In this regime, where $h<J$, eigenstates below the saddle energy $E=-h$ belonging to opposite symmetry sectors are quasi-degenerate at finite $S$. Consequently, the longest characteristic time scale is proportional to the inverse of the smallest energy difference in the Liouvillian spectrum. The numerical results also reveal persistent complexity oscillations across multiple long time scales that correspond to the inverse spacings of these quasi-degenerate levels. 

Figure~\ref{fig:Krylov_wavefunctions_LMG} in Appendix~\ref{app:more_CK} provides a complementary view of this effect through the convergence of the Krylov wavefunctions $|\varphi_n(t)|^2$ toward their analytical late-time average value $Q_{0n}$ defined in Eq.~\eqref{Q_0n}. In the saddle regime (left panel), the Krylov wavepacket requires significantly longer times to develop a substantial projection on the far edge of the Krylov chain and to converge toward $Q_{0n}$.

\begin{figure}[ht!]
    \centering
    \includegraphics[width=.45\textwidth]{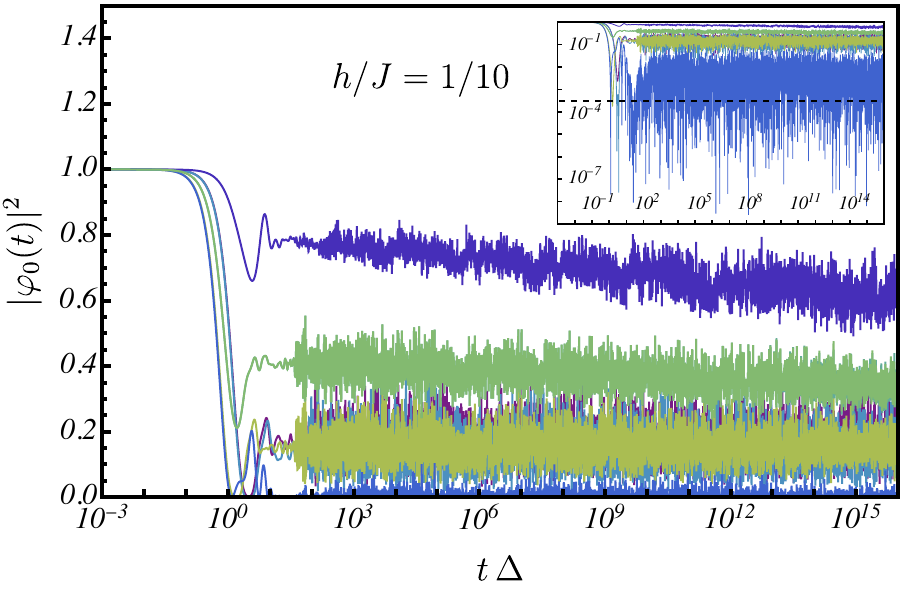}\qquad
    \includegraphics[width=.45\textwidth]{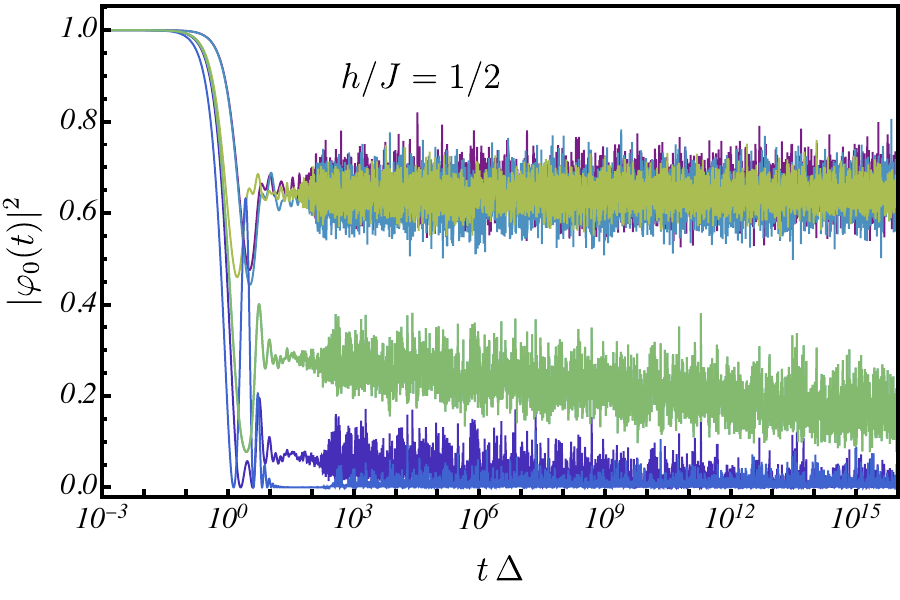}\\
    \includegraphics[width=.45\textwidth]{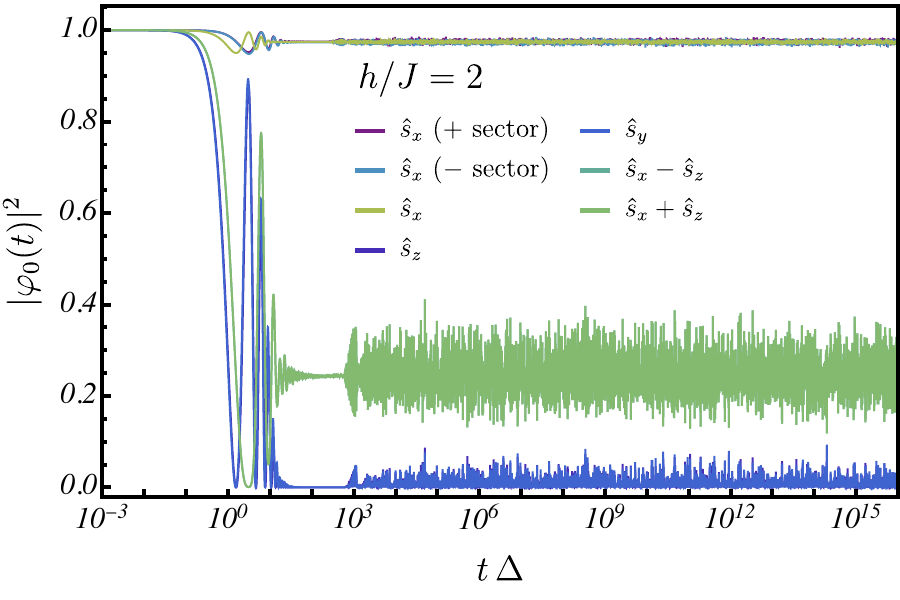}\qquad
    \includegraphics[width=.45\textwidth]{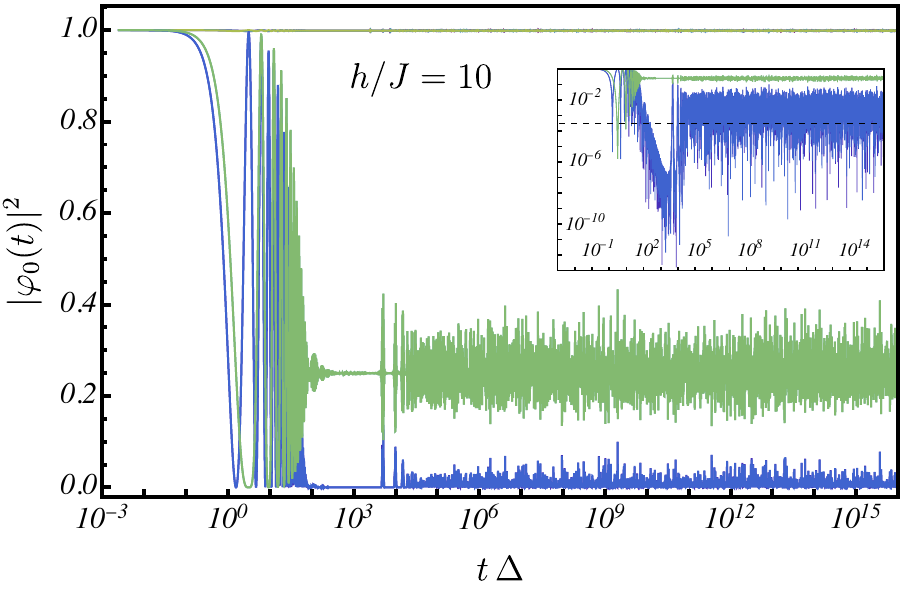}
    \caption{\textbf{(Main)} Squared modulus of the autocorrelation function for the LMG model \eqref{LMG_rescaled_TSS} ($S=40$) for various operators as a function of time (rescaled by the mean level spacing). Results are shown for various values of the ratio $h/J$. The late-time average of each curve corresponds to the time-averaged Krylov wavefunction $Q_{00}$. \textbf{(Insets)} Saturation values of generic operators are largely insensitive to $h/J$ and generally deviate from chaotic estimates. The horizontal black dashed lines denotes the value $1/K_-$, where $K_-=2S(S+1)$ is the upper bound on the Krylov dimension of $\hat{s}_y$ (see Table~\ref{tab:Krylov_dimension_LMG}), the simplest generic operator of the LMG model. Its autocorrelation function deviates from the typical plateau of $1/K$ characteristic of chaotic systems.}
\label{fig:two_points_functions_LMG}
\end{figure}

\begin{figure}[t!]
    \centering
    \includegraphics[width=.45\textwidth]{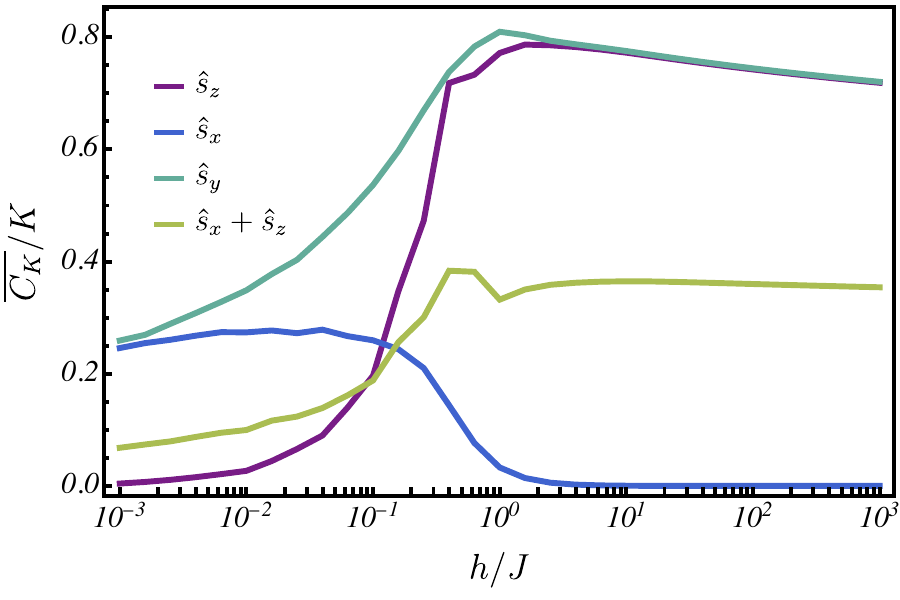}\qquad
    \includegraphics[width=.45\textwidth]{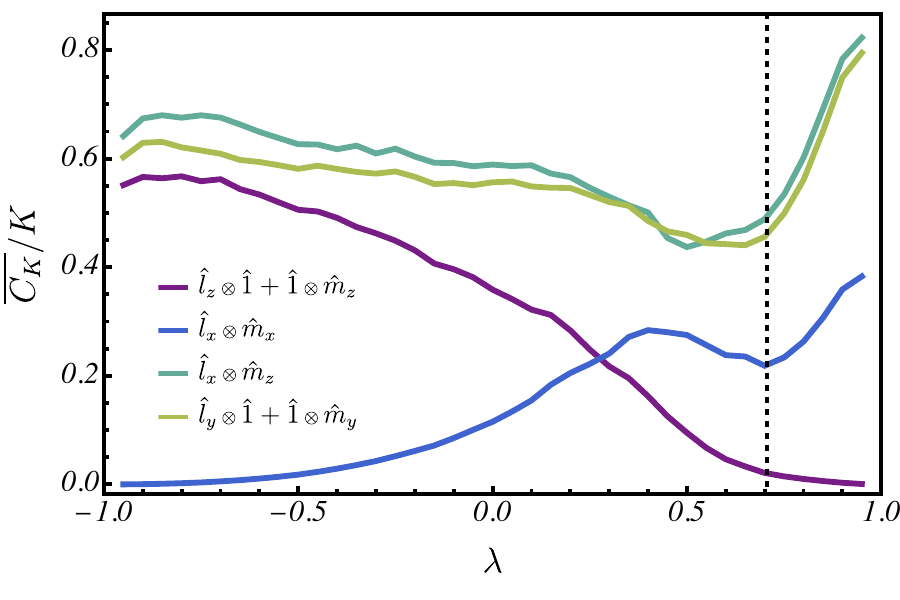}
    \caption{Late-time saturation value of K-complexity normalized by the Krylov dimension $K$, $\overline{C_K}/K$, as a function of the model parameters for generic and non-generic operators. \textbf{(Left)} LMG model \eqref{LMG_rescaled_TSS} ($S=40$) and $J=1$. \textbf{(Right)} FP model \eqref{FP_rescaled} ($L=6$). While $\overline{C_K}/K$ for non-generic operators is dictated by their respective autocorrelation functions, generic operators serve as a potential probe of the different dynamical regimes. However, the numerical results show that they remain insensitive to the saddle/no-saddle regimes in the LMG model -- where different operators exhibit distinct saturation values -- and to the chaotic/integrable crossover in the FP model -- where $\overline{C_K}/K$ starts to increase in the supposedly integrable regime (see after the vertical dashed line).}
    \label{fig:large_time_CK_multiple}
\end{figure}

We now demonstrate that for certain operators, $\overline{C_K}$ is controlled by the autocorrelation function $\varphi_0(t)=\braketc{\mathcal{O}_0}{\mathcal{O}(t)}$\footnote{We anticipate that specific relations can be derived for autocorrelation functions of operators of the form $\hat{s}_\mu\pm \hat{s}_\nu$ where $\mu,\nu=x,y,z$ and $\mu\neq \nu$.

\textit{Case 1}: if either index is $x$, the cross-correlation vanishes, $\text{Tr}(\hat{s}_x \hat{s}_\mu(t))=0$. This follows from the fact that $\hat{H}$ commutes with the symmetry operator $\hat{U}=e^{- i  \pi \hat s_x}$, while $\hat{U}^\dagger \hat{s}_\mu \hat{U}=-\hat{s}_\mu$ if $\mu=y,z$.

\textit{Case 2}: if $\mu=y$, $\nu=z$ (and vice versa), it can be shown that $\text{Tr}(\hat{s}_\mu \hat{s}_\nu(t))+\text{Tr}(\hat{s}_\nu \hat{s}_\mu(t))=0$. This result stems from the property that $\hat{s}_z$ remains real-symmetric and $\hat{s}_y$ remains purely imaginary and antisymmetric in the energy basis of $\hat{H}$.

Consequently, the autocorrelation function of the composite operators $\hat{s}_\mu\pm \hat{s}_\nu$ is always equal to the sum of the individual autocorrelation functions, see example for $\hat{s}_x\pm\hat{s}_z$ in Figure~\ref{fig:two_points_functions_LMG}. Similar relations can be derived for the FP model, which we omit here for brevity.}. Consider the LMG model in Eq.~\eqref{LMG_rescaled_TSS} rescaled by the energy unit $J$ (i.e. $\hat{H}/J$). In the limit $h/J\to 0$, it commutes with $\hat{s}_z$, causing the autocorrelation function of $\hat{s}_z$ to approach $1$. In this regime, $\overline{C_K}$ of $\hat{s}_z$ is expected to be very small, as the Krylov wavepacket remains localized at $n=0$ and thus fails to reach higher $n$ sites of the chain. Conversely, in the limit $h/J\to \infty$, the Hamiltonian commutes with $\hat{s}_x$, and a similar phenomenology applies. This behavior is illustrated in Figure~\ref{fig:two_points_functions_LMG}, which shows the late-time dynamics of autocorrelation functions of $\hat{s}_z$ and $\hat{s}_x$, and other operators, as well as in the left panel of Figure~\ref{fig:large_time_CK_multiple}, where the normalized value $\overline{C_K}/K$ is computed for various ratios of $h/J$, thereby confirming our analysis.

Analogous reasoning applies to the FP model in Eq.~\eqref{FP_rescaled}. In the limit $\lambda\to -1$, the Hamiltonian commutes with the three operators $\hat{l}_x\otimes \mathbbm{1}$, $\mathbbm{1}\otimes \hat{m}_x$, and $\hat{l}_x\otimes \hat{m}_x$. In the limit $\lambda\to 1$, it commutes with $\hat{l}_z\otimes \mathbbm{1}$, $\mathbbm{1}\otimes \hat{m}_z$, and $\hat{l}_z\otimes \hat{m}_z$. As shown in Figure~\ref{fig:two_points_functions_FP}, when the autocorrelation functions of $\hat{l}_z\otimes \mathbbm{1}+\mathbbm{1}\otimes \hat{m}_z$ and $\hat{l}_x\otimes \hat{m}_x$ remain close to $1$, the corresponding normalized value $\overline{C_K}/K$ is suppressed, as seen in the right panel of Figure~\ref{fig:large_time_CK_multiple}.

This raises the question of the late-time saturation value for ``generic'' operators -- those that do not commute with the Hamiltonian in any parameter limit. Figures~\ref{fig:two_points_functions_LMG} and \ref{fig:two_points_functions_FP} illustrate the late-time dynamics of the autocorrelation functions for several such generic operators in the LMG and FP models, respectively. In such cases, we observe that the autocorrelation function plateau does not strongly depend on the coupling values, but it deviates from the chaotic prediction of an exponentially suppressed plateau with parametrically small fluctuations \cite{Altland:2020ccq,Altland:2021rqn}. In Figure~\ref{fig:large_time_CK_multiple}, the normalized saturation value $\overline{C_K}/K$ is computed for these same operators. For the LMG model, these numerical results highlight an important feature: the normalized value $\overline{C_K}/K$ for generic operators remains essentially insensitive to the underlying dynamical regime. Specifically, it is largely unaffected by the presence of an ESQPT. The time evolution of $C_K(t)$ and of the corresponding $Q_{0n}$ for selected values of the ratio $h/J$ in both the saddle and no-saddle regimes are shown for the same operators in Figure~\ref{fig:Krylov_complexity_LMG_appendix}.

For the FP model, the normalized saturation value of generic operators can be understood in light of the energy spectrum's nature at a given value of $\lambda$. The mean level-spacing ratio $\langle r \rangle$ was plotted in Figure~\ref{fig:quantum_integrability_chaos_FP} for $L=60$ -- a system size that is too large for a numerical implementation of the Lanczos algorithm. For the $L=6$ case used in Figure~\ref{fig:large_time_CK_multiple}, both the mean ratio $\langle r \rangle$ and the level-spacing distributions at values of $\lambda$ corresponding to Poisson or GOE statistics are shown in Figure~\ref{fig:integrability_FP_L6}, revealing a less pronounced transition between quantum chaos and integrability. For this value of $L$, a phenomenological description of the observed trend for $\overline{C_K}/K$ can be provided. This description is articulated around the fact that disorder in the Lanczos sequence localizes the Krylov wavefunction, and thereby suppresses K-complexity (through a similar mechanism as discussed in Appendix~\ref{sec:linear_growth_flucuations}). Crucially, this localization effect is further enhanced by the Poissonian statistics of the Hamiltonian spectrum \cite{Adrian_2}.

Despite the relatively small number of eigenvalues available at this system size, the spectrum exhibits a tendency toward Poisson statistics around $\lambda\in[0.5, 0.7]$, where $\overline{C_K}/K$ is as expected minimal (reflecting the localization of the $Q_{0n}$ shown in Figure~\ref{fig:Krylov_complexity_FP_appendix}). As $\lambda$ increases toward $1$, the energy spectrum becomes increasingly regular --  becoming exactly equally spaced at $\lambda=1$ -- and the level-spacing distribution develops a peak at a finite value of $s$. Consequently, departing from the Poisson regime for $\lambda\gtrsim 0.7$ (a threshold that is $L$-dependent and shifts toward $1$ as $L$ increases, see the vertical dashed line in Figure~\ref{fig:large_time_CK_multiple}) leads to an increase in $\overline{C_K}/K$. On the other hand, when reducing $\lambda$ below $0.5$, two distinct effects come into play. First, the distribution combines eigenvalues from uncorrelated symmetry sectors -- in two of which the statistics are GOE -- breaking level-repulsion. Second, quasi-degeneracies introduce additional level spacings close to zero (analogous to the behavior observed for the LMG model in Figure~\ref{fig:quantum_integrability_chaos_1}). Consequently, this deviation from Poissonian statistics once again drives an increase in $\overline{C_K}/K$.

\subsection{Microcanonical Lanczos algorithm}\label{sec:quantum_Lanczos_microcanonical}

We now employ a microcanonical inner product to refine the previous analysis. In this setting, the Krylov dynamics unfolds within a restricted energy shell, allowing us to isolate clearly the effects of saddle points and spectral features that are otherwise obscured by the infinite-temperature inner product. We begin by discussing the two possible choices of microcanonical inner product and their structural consequences for the Lanczos algorithm and the microcanonical Krylov dimension. We then present numerical results for the K-complexity of the LMG model using these two inner products.

\subsubsection{Choice of an inner product and microcanonical Krylov dimension}

In Section~\ref{sec:Krylov_toolbox}, we commented on the possible choices of microcanonical inner products, which generally correspond to inner products of the form given in Eq.~\eqref{eq:Operator_Inner_product}, where the density matrix $\hat{\rho}$ is a projector onto an energy shell. More precisely, let $\hat H$ be a Hamiltonian and let $\{\ket{E_i} \}_{i=1,\cdots,N(E,\Delta E)}$ denote the set of $N(E,\Delta E)$ eigenstates whose energies lie within the energy shell $[E - \Delta E/2, E + \Delta E/2]$. The corresponding (non-normalized) microcanonical density matrix $\hat{\rho}_{E,\Delta E}$, which depends on both the central energy $E$ and the shell width $\Delta E$, is defined as
\begin{equation}\label{microcanonical_density_matrix}
    \hat \rho_{E, \Delta E} =  \sum_{i=1}^{N(E,\Delta E)} \ket{E_i}\bra{E_i}.
\end{equation}

Once the microcanonical density matrix has been defined, two possible choices of inner product can be considered: Eq.~\eqref{eq:Operator_Inner_product} or its regularized version, Eq.~\eqref{eq:Operator_Inner_product_sym}, in which the density matrix is split among the operator insertions. We recall that the Lanczos algorithm generates two sequences of Lanczos coefficients, $\{a_n\}_{n=0}^{K-1}$ and $\{b_n\}_{n=1}^{K-1}$. For the inner product \eqref{eq:Operator_Inner_product} with $\hat{\rho}_{E,\Delta E}$, the coefficients $a_n$ are generically nonzero, and play an essential role in the computation of K-complexity and related quantities, such as $\overline{C_K}$ and $Q_{0n}$. By contrast, for the regularized inner product \eqref{eq:Operator_Inner_product_sym}, the coefficients $a_n$ vanish identically. Importantly, the two inner products coincide in the classical limit (see Eq.~\eqref{eq:Op_products_equal_classically}). Correspondingly, the nonzero $a_n$ coefficients obtained from the inner product \eqref{eq:Operator_Inner_product} vanish in this limit (see Eq.~\eqref{eq:an_coeffs_vanish_classically}). This behavior was illustrated for the LMG model in the inset of Figure~\ref{fig:microcanonical_classical_LMG}. As a consequence, the remaining nonzero Lanczos coefficients $b_n$ computed from the two inner products must converge toward one another in the classical limit. The same conclusion then applies to all quantities derived from the $b_n$ coefficients.

Structurally, we note that the inner product \eqref{eq:Operator_Inner_product} with $\hat{\rho}_{E,\Delta E}$ does not strictly restrict the operator to the energy shell. Indeed, while the trace projects onto states within the shell, one of the operator indices is unrestricted over the full Hilbert space, i.e. $\braketc{\mathcal{A}}{\mathcal{B}}_{E,\Delta E}\propto \sum_i \mel{E_i}{\hat{\mathcal{A}}^\dagger \hat{\mathcal{B}}}{E_i}$. Consequently, the operator algebra itself is not confined to the shell. On the other hand, the inner product \eqref{eq:Operator_Inner_product_sym} fully restricts both operator indices to the energy shell, i.e. $\braketc{\mathcal{A}}{\mathcal{B}}_{E,\Delta E}^{(w)}\propto \sum_{i,j} \mel{E_i}{\hat{\mathcal{A}}^\dagger}{E_j} \mel{E_j}{\hat{\mathcal{B}}}{E_i}$. The effect of having one operator index unrestricted to the shell will be investigated in the next section.

Once a choice of inner product has been made, an upper bound on the corresponding Krylov dimension can be derived. The latter now depends on the energy shell parameters $E$ and $\Delta E$, and we denote it by $K_{E, \Delta E}$ and $K^{(w)}_{E, \Delta E}$ for the inner products in Eqs.~\eqref{eq:Operator_Inner_product} and \eqref{eq:Operator_Inner_product_sym}, respectively. We find that
\begin{subequations}
\begin{align}
    & K_{E, \Delta E} \leq N D-N+1\, ,\\ 
    & K^{(w)}_{E, \Delta E} \leq N^2-N+1\, ,
\end{align}
\end{subequations}
where $N$ is shorthand for $N(E,\Delta E)$. In the limiting case where the energy shell covers the entire spectrum of the Hamiltonian, $N=D$, the density matrix $\hat \rho_{E, \Delta E}$ reduces to the identity operator, and the familiar bound $K\leq D^2-D+1$ is recovered in both cases, where $D$ is the Hilbert space dimension. These expressions can be justified as follows: the microcanonical Krylov dimension can be interpreted as the number of distinct phases $\omega_{ij}$ of the Liouvillian operator (see Section~\ref{sec:Krylov_toolbox}) for which at least one index corresponds to an eigenstate $\ket{E_i}$ within the shell. For the regularized inner product, both indices must lie within the shell. In either case, all degenerate zero phases contribute only a single Krylov dimension, leading to the subtraction terms appearing in the bounds above. The key observation is that, apart from the Hilbert space dimension, both bounds depend only on the number of states $N$ within the shell, and therefore on the corresponding DOS $\sigma(E, \Delta E)$. Specifically, both bounds are monotonically increasing functions of $N$, and hence attain their largest values in regions where the DOS is maximal. As a direct consequence, microcanonical Krylov dimensions are highly sensitive to the ESQPT which manifest themselves as singular features in the DOS. This is the case for the LMG model, as discussed in Section~\ref{sec:LMG}.

We have numerically verified that both bounds are satisfied when the Lanczos algorithm is applied to the LMG model using the corresponding microcanonical inner products. In particular, for uncharged operators, the bounds are saturated\footnote{Under the usual assumption that the initial operator is dense in the energy basis.}, just as in the infinite-temperature case, see Section~\ref{sec:K_dim}.

\subsubsection{Microcanonical Krylov complexity}

We now present numerical results for the microcanonical K-complexity of the LMG model. Note that we do not address the FP model in this section because the accessible system sizes $L$ for which the microcanonical Lanczos algorithm can be computed are too small to extract meaningful physical features. For fixed model parameters, the spectral bandwidth $\Lambda(h,J)$ can be estimated using Eq.~\eqref{spectral_bandwidth_LMG}. Strictly speaking, this expression corresponds to the spectral bandwidth of the classical Hamiltonian and becomes an exact expression for the quantum spectral bandwidth in the limit $S\to \infty$, where the lowest and highest energies converge to the minimum and maximum values of the corresponding classical Hamiltonian.

\begin{figure}[t!]
    \centering
    \includegraphics[width=.45\textwidth]{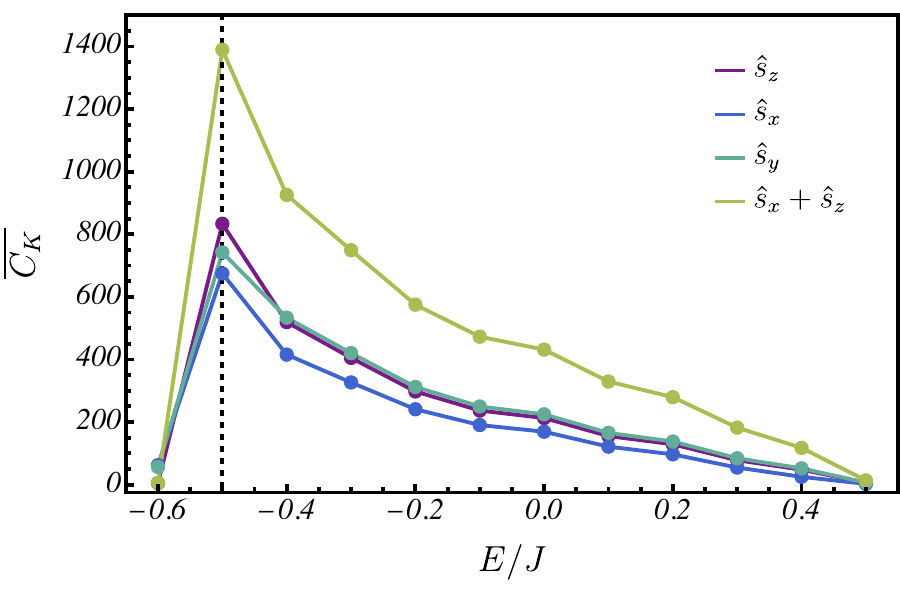} \qquad
    \includegraphics[width=.45\textwidth]{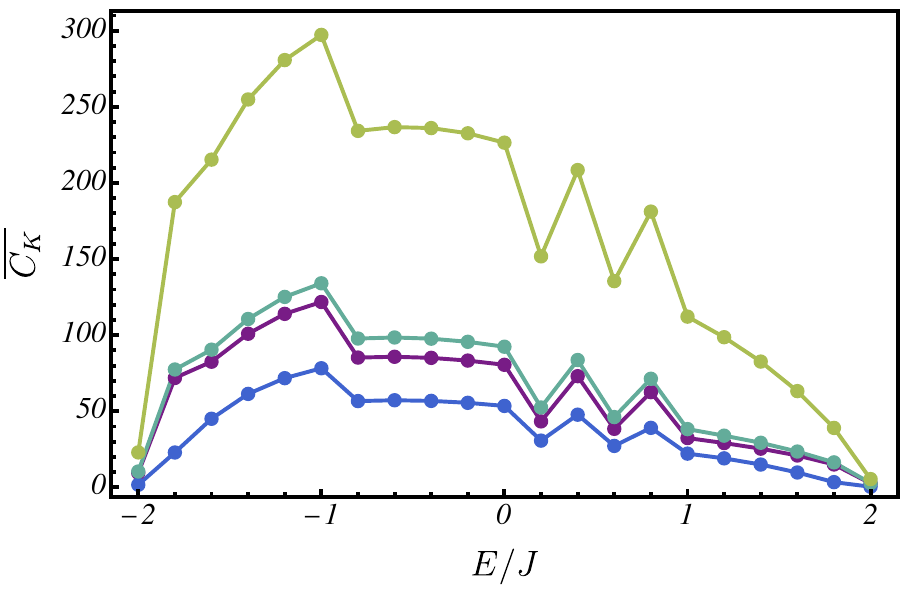}
    \caption{$\overline{C_K}$ as a function of the microcanonical energy shell $E$ for the LMG model \eqref{LMG_rescaled_TSS} ($S=60$). $\overline{C_K}$ is not normalized by $K$ since $K$ is different in every energy shell. Results are shown for various initial operators indicated in the legend. \textbf{(Left)} Saddle point at $E/J=-h$ (vertical black dotted line) in the saddle regime with parameters $h=1/2$, $J=1$ and $\Delta E=0.1$. \textbf{(Right)} No-saddle regime with parameters $h=2$, $J=1$ and $\Delta E=0.2$.}
    \label{fig:microcanonical_C_K_LMG}
\end{figure}

We first perform the Lanczos algorithm using the microcanonical inner product \eqref{eq:Operator_Inner_product} with $\hat{\rho}_{E,\Delta E}$. From the resulting Lanczos coefficients $\{a_n, b_n\}$, we compute the late-time saturation value of K-complexity. The results are shown in Figure~\ref{fig:microcanonical_C_K_LMG} for a range of energy shells and several initial operators\footnote{As opposed to the infinite-temperature case in Section~\ref{sec:late-time_K-complexity} where the one-point function of initial operators was zero for the studied operators, this is no longer true for both microcanonical inner products introduced in the previous section. In every numerical simulation of this section, the one-point function is removed from the initial operator, making it traceless. This procedure is needed as a nonzero operator one-point functions can influence the late-time behavior of
K-complexity if it is not subtracted, see \cite{Adrian_2} for more details.}, as well as in both saddle and no-saddle regimes (we recall that the DOS is singular in the former). 

A qualitative similarity can be identified between the microcanonical $\overline{C_K}$ and the DOS of the LMG model shown in Figure~\ref{fig:DOS_LMG_FP}. The connection is as follows: the more eigenstates contained within a given energy shell, the larger the corresponding microcanonical Krylov dimension. This, in turn, increases the length of the Krylov chain and consequently the value of the microcanonical K-complexity. In short, the microcanonical $\overline{C_K}$ is a direct manifestation of the underlying DOS of the considered model.

However, the manner in which K-complexity approaches its saturation value also depends sensitively on the choice of the energy shell. Figures~\ref{fig:LMG_microcanonical_1} and \ref{fig:LMG_microcanonical_2} show the time evolution of K-complexity in both saddle and no-saddle regimes, where we have considered an energy shell centered at the critical energy $E=-h$. The numerical results clearly show that, within this shell, K-complexity saturates to a well-defined value at late times. By contrast, in the remaining energy shells, as well as in all shells in the no-saddle regime, the late-time value corresponds instead to the average of persistent coherent oscillations. The origin of these oscillations can be understood from the fact that energies of the LMG model are approximately commensurate throughout most of the spectrum, but become strongly distorted in the vicinity of the critical point, where the density of states diverges, see Figure~\ref{fig:LMG_FP_sectors}. As a result, coherent revivals persist away from the critical energy.

This distinction mirrors the expected late-time behavior of  K-complexity in integrable/quadratic systems, on the one hand, and chaotic systems, on the other \cite{Adrian_2, Adrian_3}. This observation suggests that the dynamics within the energy shell containing the saddle point acquires features typically associated with chaotic behavior.

The corresponding $Q_{0n}$ distributions are illustrated in Figures~\ref{fig:LMG_microcanonical_Q0n_1} and \ref{fig:LMG_microcanonical_Q0n_2}, and exhibit an interesting feature. For all energy shells except the one containing the saddle point, the distributions are effectively supported over a smaller portion of the available Krylov space. Beyond this region, $Q_{0n}$ decay exponentially with $n$. Within its support, $Q_{0n}$ remains approximately uniform, implying that the late-time saturation value of K-complexity is approximately proportional to one-half of this effective Krylov dimension. In the case of the shell containing the saddle point, the microcanonical Krylov dimension becomes arbitrarily large in the $S\to \infty$ limit causing the support of $Q_{0n}$ to extend throughout the available Krylov space.

We next perform the Lanczos algorithm using the regularized microcanonical inner product \eqref{eq:Operator_Inner_product_sym} with $\hat{\rho}_{E,\Delta E}$. Since the Krylov dimension $K^{(w)}_{E,\Delta E}$ scales approximately as the square of the number of eigenstates contained within the energy shell, larger values of the collective spin $S$ are required than in the case of the non-regularized microcanonical inner product. As a result, the exact diagonalization of the Hamiltonian becomes a new numerical challenge.

The time evolution of K-complexity in both the saddle and no-saddle regimes is shown in Figure~\ref{fig:LMG_microcanonical_regularized} for a selection of energy shells. The same qualitative picture emerges as before: within the shell containing the saddle point, K-complexity saturates in a manner reminiscent of chaotic systems, whereas in the remaining shells -- and throughout the no-saddle regime -- it exhibits persistent coherent oscillations.

The corresponding $Q_{0n}$ distributions are shown in Figure~\ref{fig:LMG_microcanonical_regularized_Q0n}. In all energy shells displaying integrable-like behavior, the $Q_{0n}$ grow and oscillate, whereas these oscillations are suppressed in the shell containing the critical point. The apparent exponential decay observed in Figures~\ref{fig:LMG_microcanonical_Q0n_1} and \ref{fig:LMG_microcanonical_Q0n_2} is absent. This suggests that the decay reflects the manner in which the Krylov dimension $K_{E,\Delta E}$ (and so the $Q_{0n}$) converges to $K^{(w)}_{E,\Delta E}$ in the $S\to \infty$ limit.

As a partial conclusion, the microcanonical K-complexity provides a more refined picture than its infinite-temperature counterpart and offers a useful framework for resolving the apparent tension between chaotic Krylov dynamics and the integrability of the LMG model. By isolating the influence of the saddle point to its energy shell, this approach reveals its role as a potential enhancer of chaotic-like behavior in Krylov dynamics. At the same time, it promotes the microcanonical K-complexity to an energy-resolved diagnostic tool capable of distinguishing the dynamics near the saddle point from the integrable behavior that dominates elsewhere in the spectrum.

\section{Discussion}\label{sec:discussion}

In this paper, we have developed the correspondence between classical and quantum K-complexity. The former is defined on phase space endowed with its canonical Poisson structure and a phase-space inner product, while the latter lives in operator Hilbert space, endowed with a choice of inner product between two operators. A natural choice is the Frobenius norm, but in this paper, we have demonstrated the usefulness of other inner products in distinguishing chaotic from integrable behavior, in particular using a microcanonical inner product, defined in Eq.~\eqref{eq:Operator_Inner_product}, or its regularized version in Eq. \eqref{eq:Operator_Inner_product_sym}, with the microcanonical density matrix given in Eq.~\eqref{microcanonical_density_matrix}.
The correspondence hinges on Dirac's quantization principle, which associates ($i\hbar$-times) the Poisson bracket between two phase-space functions to the commutator of two quantum operators. In particular, the commutator of an operator with the Hamiltonian corresponds to the application of the classical Liouvillian to the phase-space function associated with the operator. 

We have shown that the classical Krylov space, its Lanczos sequence as well as the associated K-complexity are recovered in an appropriate $\hbar\to 0$ limit of quantum K-complexity. In order to establish this correspondence, we employed Wigner's phase-space formalism, or more precisely the Stratonovich-Weyl formalism, whereby every operator is associated with a symbol (a phase-space distribution), and quantum commutators become Moyal commutators. With the aid of this formalism, we showed that the quantum Lanczos algorithm, and thus the Krylov subspace and K-complexity, pass smoothly into their classical analogues, $b_n^{(q)}\to b_n^{(c)}$, $a_n^{(q)}\to a_n^{(c)}$, $C_K(t)\to C_K^{(c)}(t)$, which can therefore rightfully be considered as the classical version of their quantum cousins. A crucial fact we established is that each choice of quantum inner product reduces, in this limit, to its natural classical counterpart, i.e. the Frobenius inner product to the uniform Liouville measure on $\mathcal{M}$, and the microcanonical inner product to its classical microcanonical analogue. As a consequence, the classical diagonal Lanczos coefficients $a_n^{(c)}$ all vanish, in agreement with the general phase-space argument given in Section~\ref{subsec:Classical_Krylov_Generic_setup}. A particularly useful result we derived is that classical K-complexity reproduces its quantum counterpart at early times, with the matching breaking down at a system-dependent Krylov-Ehrenfest time $t_{*}$. In chaotic systems, this scale takes the familiar logarithmic form $t_{*}\sim \log\left(1/\hbar\right)/\lambda_K$, in direct analogy with the standard scrambling/Ehrenfest time prominent in discussions of OTOCs, with the Lyapunov exponent of the standard formula replaced by the Krylov exponent, $\lambda_K$, which bounds it from above and saturates for maximally chaotic dynamics. For integrable systems, the matching instead breaks down on polynomial timescales $t_{*}\sim \hbar^{-\alpha}$, reflecting the absence of exponential operator growth.

As a key tool of the analysis, both at the classical and quantum level, we have made extensive use of a microcanonical, energy-resolved notion of K-complexity \cite{microcanonical_Krylov}, which turns the latter from a single number to a function of the energy shell. Varying the energy shell is the microcanonical analogue of choosing the temperature of the canonical framework, at least morally. However, the ability to precisely locate K-complexity to a given energy shell gives added power to the formalism, allowing one to isolate energy-specific saddle points that may be present in the classical phase space of the system.

As a physical application to the quantum-classical Krylov correspondence, we considered K-complexity in collective spin models, which have a well-defined semiclassical limit in terms of the effective Planck constant $h_{\mathrm{eff}} = \hbar / S$ for large spin-$S$. For these models, we developed the classical Krylov space using coherent-state type constructions, which map the classical Lanczos formalism to a hopping model on a latticized phase space. The integrable LMG model was previously shown to exhibit linear Lanczos growth and early-time exponential growth of K-complexity originating not in genuine chaos but in an unstable classical saddle, a phenomenon termed saddle-dominated scrambling \cite{saddle_Krylov}. Our results are consistent with that interpretation, and we show further that classical K-complexity, and in particular its microcanonical formulation, is able to isolate saddle-dominated Krylov growth in integrable theories to energy windows that correspond to classical saddles, while unveiling the underlying integrability in the other energy shells. Regarding the FP model, infinite-temperature K-complexity, computed both at early times using the phase space formalism and at late times using the standard quantum Lanczos algorithm, is sensitive to the chaotic-to-integrable transition of the model as a function of its coupling strength. Nevertheless, it would be important to extend our numerics to higher system sizes to perform a similar reliable microcanonical analysis for the FP model.

An important advantage of the formalism employed in this paper lies in its ability to investigate Krylov dynamics in an energy-resolved fashion, which allows one to distinguish true chaotic physics from saddle-dominated ``fake chaos'', that is, chaotic behavior that is an artifact of the phase-space structure of an integrable model. We showed that this behavior is confined to the energy shells of the classical saddles, whereas genuine chaos is expected to leave chaotic signatures across the entire spectrum. The energy profile of the microcanonical K-complexity therefore provides a natural diagnostic to tell the two apart. By contrast, the more commonly employed ``grand-canonical'' K-complexity, which makes use of an inner product with inverse temperature $\beta$, often fails to distinguish these two notions,  especially in the oft-employed infinite temperature limit. In this context, microcanonical K-complexity is a considerably finer-grained diagnostic of true chaos, although an explicit demonstration against a genuinely chaotic model, such as the FP model, is left to future work. We further noted that the microcanonical inner product we employed in Section \ref{sec:quantum_Lanczos_microcanonical} at the quantum level is not the unique choice associated with a given energy, and we showed that the differently regularized versions have the same classical limit, which is therefore unambiguous. It would be interesting to investigate systematically the impact of the choice of inner product on the phenomenology of quantum microcanonical K-complexity. Nevertheless, our results for the LMG model suggest that the ``normal'' and regularized inner products give mutually consistent results -- see Section~\ref{sec:quantum_Lanczos_microcanonical} and Appendix~\ref{app:more_CK}.

A further point that deserves future study is the impact of the seed operator on the Krylov dynamics. In this work we were able to study this issue in detail in both the LMG and FP models, owing to their high degree of solvability, tracing the late-time saturation values back to properties of the seed operator, such as their symmetry properties, including their commutation properties with the Hamiltonian, governed by the algebra $\mathfrak{su}(2)$. More generally, it would be desirable to identify canonical simple, symmetry-neutral operators, or potentially an ensemble thereof in order to investigate late-time issues more universally \cite{Craps:2024suj}.

The rich imprints left on Krylov dynamics by phase-space features connected to classical saddles uncovered in this paper suggest further investigations of the manifestations of other phase-space structures, in particular in mixed phase spaces including KAM islands. More ambitiously, our results here may just be a first glimpse into using Krylov methods and techniques to characterize and organize semiclassical dynamics, extending semiclassical notions of operator growth to a much broader set of physical manifestations of chaos and integrability. A natural starting point for investigation is the extension of our approaches to establish a semiclassical theory of spread complexity, or K-complexity in the Schrödinger picture \cite{Balasubramanian:2022tpr,Caputa:2024vrn}. Important steps have been taken for the density matrix itself, via the usual Wigner transform \cite{Basu:2024tgg,Basu:2025mmm}, but a more general analysis similar to the one we have explored for operator complexity in this work remains an important future target for investigation.

\paragraph{Note added.} The recent paper \cite{Camargo:2026szl} proposes an alternative approach to the phenomenon of saddle-dominated scrambling in K-complexity and the LMG model, by introducing the so-called $\text{logK}$-complexity. We are especially thankful to its authors for insightful discussions and for sharing their results with us before publication.

\paragraph{Data availability statement.} The data and code supporting the findings of this study are openly available on GitHub (\url{https://github.com/nicolasdero/KrylovQuantumClassical}) and are archived on Zenodo \cite{KrylovQuantumClassical}.

\acknowledgments
We thank Alexander Altland, Vijay Balasubramanian, Ben Craps, Felix Haehl, Marcos Mariño, Eliezer Rabinovici, Klaus Richter, Ruth Shir and Juan-Diego Urbina for insightful discussions. JS and ASG thank the Institut d'Études Scientifiques de Cargèse for hospitality in the final phases of preparation of this paper. This research is supported in part by the Fonds National Suisse de la Recherche Scientifique (Schweizerischer Nationalfonds zur Förderung der wissenschaftlichen Forschung) through the Project Grant 200021 215300 and the NCCR51NF40-141869 The Mathematics of Physics (SwissMAP).
ASG is supported by UK Research and Innovation (UKRI) under the UK government’s Horizon Europe Funding Guarantee EP/X030334/1. Some of the numerical computations for this work were performed using the Baobab HPC service at the University of Geneva.

\appendix

\section{Details on the Lanczos algorithm and Krylov complexity}\label{appx:Lanczos}

In the first part of this appendix, we describe the numerical implementation of the Lanczos algorithm and briefly discuss its numerical limitations. We then examine the effect of fluctuations around the linear growth of the Lanczos coefficients on the early-time growth of the K-complexity.

\subsection{Lanczos algorithm}\label{appx:Lanczos_Lanczos}

The Lanczos algorithm for operators evolving in the Heisenberg picture is given by the following algorithmic procedure:

\begin{algorithm}[ht!]\label{alg:lanczos}
\begin{algorithmic}
\Require $\ketc{\mathcal{O}}$, $\mathcal{L}$
\State $\ketc{\mathcal{O}_0}=\ketc{\mathcal{O}}/\sqrt{\braketc{\mathcal{O}}{\mathcal{O}}}$  \Comment{Initial operator is normalized}
\State $a_0=\melc{\mathcal{O}_0}{\mathcal{L}}{\mathcal{O}_0}$
\State $\ketc{A_1}=(\mathcal{L}-a_0) \ketc{\mathcal{O}_0}$ \Comment{First Lanczos step is manually performed}
\State $b_1=\sqrt{\braketc{A_1}{A_1}}$
\If{$b_1=0$}
    \State \textbf{stop}
\Else
\State $\ketc{\mathcal{O}_1}=\frac{1}{b_1}\ketc{A_1}$
\State $a_1=\melc{\mathcal{O}_1}{\mathcal{L}}{\mathcal{O}_1}$
\EndIf
\For{$n \geq 2$}
    \State $\ketc{A_n}=(\mathcal{L}-a_{n-1})\ketc{\mathcal{O}_{n-1}}-b_{n-1}\ketc{\mathcal{O}_{n-2}}$
    \State $b_n=\sqrt{\braketc{A_n}{A_n}}$
    \If{$b_n=0$} \Comment{Process stops when $n=K$}
        \State \textbf{stop}
    \Else \Comment{If the current $b_n\neq 0$, algorithm continues}
        \State $\ketc{\mathcal{O}_n}=\frac{1}{b_n}\ketc{A_n}$
        \State $a_n=\melc{\mathcal{O}_n}{\mathcal{L}}{\mathcal{O}_n}$
    \EndIf
\EndFor
\end{algorithmic}
\end{algorithm}

As mentioned in Section~\ref{sec:Krylov_toolbox}, the algorithm constructs an orthonormal Krylov basis $\{|\mathcal{O}_n)\}_{n=0}^{K-1}$ and the Lanczos coefficients $\{a_n\}_{n=0}^{K-1}$ and $\{b_n\}_{n=1}^{K-1}$. From these, one can compute the K-complexity $C_K(t)$ defined in Eq.~\eqref{eq:KC_from_wavefn}, its late-time saturation value $\overline{C_K}$ in Eq.~\eqref{eq:KC_LongTimeAvg} or the late-time averages of the Krylov wavefunctions $Q_{0n}$ in Eq.~\eqref{Q_0n}. In practice, one uses the fact that the Liouvillian $\mathcal{L}$ is a tridiagonal matrix in the Krylov basis, see Eq.~\eqref{eq:Krylov_elements_recursion}.

The Lanczos algorithm above is known to suffer from numerical instabilities that gradually destroy the orthogonality of the computed Krylov basis \cite{full_ortho_algorith}. This loss of orthogonality can be eliminated using full orthogonalization \cite{full_ortho_algorith} or partial re-orthogonalization \cite{Lanczos_PRO} algorithms, both of which have been applied to many-body models in \cite{Adrian_1}. 

Nevertheless, these methods remain fundamentally limited by machine precision. In particular, when the Krylov dimension is significantly smaller than the upper bound $D^2-D+1$ (for example, due to symmetries) an additional numerical error can arise: the computed Krylov basis elements may acquire spurious projections outside of Krylov space, causing the iteration to effectively escape it and generate an oversized, nearly orthonormal basis \cite{Escape_Krylov_space}. A more detailed discussion of these numerical issues is given in \cite{Rabinovici:2025otw}. 

In this work, we instead use the basic Lanczos algorithm presented above together with arbitrarily high precision arithmetic \cite{KrylovQuantumClassical}. We verify convergence by confirming that the length of the resulting Lanczos sequence agrees with the expected Krylov dimension (see Section~\ref{sec:K_dim}).

\subsection{Linear growth of the Lanczos coefficients and fluctuations}\label{sec:linear_growth_flucuations}

\begin{figure}[hb!]
    \centering
    \includegraphics[width=0.45\linewidth]{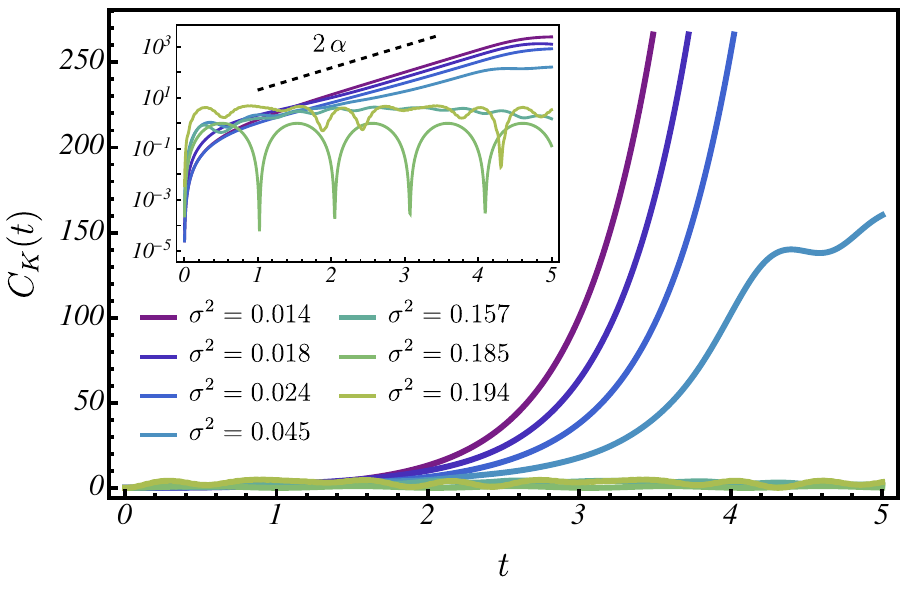}
    \caption{Early-time K-complexity computed from Lanczos sequences of the form in Eq.~\eqref{linear_Lanczos} for different fluctuation magnitudes. We set $\alpha=1$ and use sequences of length $n=5000$. As the fluctuations increase, the growth rate of the K-complexity decreases, eventually leading to saturation on the depicted time scale. The inset shows the growth on a logarithmic scale, highlighting the deviation from exponential growth.}
    \label{fig:Krylov_complexity_disorder}
\end{figure}

In Sections~\ref{sec:classical_LMG}, \ref{sec:classical_FP} and \ref{sec:classical_microcanonical_Lanczos}, we study the linear growth of the classical Lanczos coefficients and the corresponding early-time growth of K-complexity. As discussed in those sections, this linear trend is typically accompanied by fluctuations, whose magnitude depends on the dynamical regime. Here, we briefly examine how these fluctuations affect the growth of K-complexity.

We model the Lanczos coefficients as growing linearly with superimposed fluctuations of tunable magnitude. Specifically, we consider a sequence of the form
\begin{equation}\label{linear_Lanczos}
    b_n = \alpha n + w_n\, ,
\end{equation}
where $\{w_n\}$ is a set of independent and identically distributed random variables. We assume $w_n\sim \mathcal{N}(0,s^2)$, corresponding to Gaussian fluctuations with zero mean and variance $s^2$. As shown in Ref.~\cite{Adrian_2}, stronger disorder in the Lanczos coefficients enhances wavefunction localization on the Krylov chain. Following Ref.~\cite{sigma_localization}, the strength of the fluctuations is quantified by
\begin{equation}
\sigma^2=\mathrm{Var}\left(x_i\right), \qquad x_i=\log \left|\frac{b_i}{b_{i+1}} \right|.
\end{equation}
Since the coefficients $b_n$ fluctuate, the random variables $\{x_i\}$ also fluctuate around a zero mean. Consequently, increasing the variance of $w_n$ increases $\sigma^2$, indicating stronger localization.

The resulting early-time growth of the K-complexity for Lanczos sequences of the form \eqref{linear_Lanczos} is shown in Figure~\ref{fig:Krylov_complexity_disorder}. It is clear that as the fluctuations become stronger, the K-complexity increasingly deviates from the exponential growth of Eq.~\eqref{eq:Ehrenfest_time_line1}. This is particularly evident in the inset, where the slope of the curve falls below the Krylov exponent $2\alpha$. In the limit of sufficiently large fluctuations, the K-complexity is effectively suppressed and ceases to grow on the scale shown. Physically, this occurs because a Lanczos coefficient that becomes sufficiently close to zero acts as a bottleneck, preventing further ballistic spreading of the wavepacket along the Krylov chain.

\section{Details on the LMG model}\label{appx:LMG}

In this appendix, we derive the dimensions of the parity sectors $D_\pm$ of the LMG model, and show that they are given by Eqs.~\eqref{number_U_sector} for even, odd, and half-integer values of $S$. We then use these identities to derive an upper bound on the Krylov dimension of the LMG model as a function of the symmetry charge of the initial operator. Finally, we analyze the phase space of the classical LMG model, identify the saddle point, and derive the spectral bandwidth of the classical Hamiltonian, which is used extensively to normalize the classical Lanczos coefficients in Section~\ref{sec:classical_LMG}.

\subsection{Parity sectors}\label{app:LMG_symmetries}

Let $\left\{\ket{m}\hspace{.1cm}|\hspace{.1cm}-S\leq m\leq S~\right\}$ denote the usual basis of the spin-$S$ Hilbert space consisting of the eigenstates of $\hat{S}_z$ ($S$ is fixed in the following discussion). We first prove that the action of $\hat U=e^{- i \pi \hat S_x}$ on these basis states is given by 
\begin{equation}\label{app:U_action}
    \hat U \ket{m}=e^{-i \pi S} \ket{-m} .
\end{equation}
The action of $\hat U$ on the collective spin components is given by
\begin{equation}
    \hat U \hat{S}_x \hat{U}^\dagger = \hat{S}_x\, , \qquad \hat U \hat{S}_y \hat{U}^\dagger = -\hat{S}_y\, ,\qquad \hat U \hat{S}_z\hat{U}^\dagger = -\hat{S}_z\, .\label{app:transporm_spin_component}
\end{equation}
Using these identities one has
\begin{equation}\label{app:R4}
    \hat{S}_z \hat U \ket{m}=-m \,\hat U\ket{m}, \qquad \hat{\bm{S}}^2 \hat U\ket{m}=S(S+1) \,\hat U\ket{m},
\end{equation}
which suggests that $\hat U\ket{m}\propto \ket{-m}$ up to a phase, i.e.
\begin{equation}\label{app:global_phase}
    \hat U\ket{m}=e^{ i \alpha(m)}\ket{-m} ,\qquad\alpha(m)\in [0,2\pi)\, .
\end{equation}
The first two identities in \eqref{app:transporm_spin_component} can be combined to obtain a similar identity for the ladder operators $\hat{S}_\pm\equiv \hat{S}_x\pm  i \hat{S}_y$,
\begin{equation}\label{app:R_+-}
    \hat U  \hat{S}_\pm \hat{U}^\dagger = \hat{S}_\mp\, ,
\end{equation}
from which it follows that
\begin{equation}\label{app:transit_proof_1}
    \hat{S}_\mp \hat U \ket{m}=a_\pm(m)\, \hat U \ket{m\pm 1}=a_\pm(m)\, e^{ i \alpha(m\pm 1)}\ket{-m\mp 1} ,
\end{equation}
where we have used that $\hat{S}_\pm \ket{m}=a_\pm(m)\ket{m\pm 1} $ for some coefficients $a_\pm (m)$ which satisfy the relation $a_\pm (m)=a_\mp(-m)$. From Eq.~\eqref{app:global_phase}, we also get that 
\begin{equation}\label{app:transit_proof_2}
    \hat{S}_\mp \hat U\ket{m}=e^{ i \alpha(m)} \hat{S}_\mp \ket{-m} =a_\mp(-m) e^{ i \alpha(m)} \ket{-m\mp 1} .
\end{equation}
Eqs.~\eqref{app:transit_proof_1} and \eqref{app:transit_proof_2} together imply that $e^{ i \alpha(m\pm 1)}=e^{ i \alpha(m)}$. That is, $ \alpha(m)$ does not depend on $m$. We therefore denote it by $\alpha(S)$, since it can depend only on the spin quantum number $S$. To determine $\alpha(S)$, we recall that by definition, $e^{- 2i\pi \hat{S}_x}=(-1)^{2S} \mathbbm{1}$. Hence, performing a double rotation yields
\begin{equation}
    \hat{U}^2 \ket{m}=e^{2 i \alpha(S)}\ket{m}=(-1)^{2S}\ket{m}.
\end{equation}
Therefore, $e^{2 i \alpha(S)}=(-1)^{2S}=e^{2 i \pi S }$, that is $\alpha(S)=\pm\pi S+k\pi$. If $S\in\mathbb{N}$, the sign is irrelevant. However, if $S\in\mathbb{N}/2$, the properties of the Wigner-$D$ matrices \cite{spherical_harmonics_relations} fix the remaining sign ambiguity and $\alpha(S)=-\pi S+2k\pi$ (we don't show this explicitly here). Therefore, taking $k=0$, we conclude that $\alpha(S)=-\pi S$, which completes the proof of Eq.~\eqref{app:U_action}.

We now use Eq.~\eqref{app:U_action} to determine the dimension of the $\pm$ sectors of the LMG Hamiltonian in Eq.~\eqref{LMG_rescaled_TSS}. Normalized eigenstates of $\hat U$ have the form,
\begin{equation}
    \ket{\chi_\pm(m)}\equiv \frac{1}{\sqrt{2}} \left(\ket{m}\pm\ket{-m}\right) .
\end{equation}
Note, however, that $\ket{\chi_+(0)}$ is not normalized. We denote by $D_\pm$ the dimension of the $\pm$ sector, i.e. the number of $\pm$ eigenvalues of $\hat U$ if $S\in\mathbb{N}$ or of $\pm i$ eigenvalues if $S\in \mathbb{N}/2$. We first assume that $S\in \mathbb{N}$. In that a case, Eq.~\eqref{app:U_action} takes the form
\begin{equation}
    \hat U \ket{\chi_\pm(m)}=\pm(-1)^S\ket{\chi_\pm(m)} .
\end{equation}
For a fixed value of $S$, there are $S$ pairs $(m, -m)$ and so $S$ eigenstates $\ket{\chi_+(m)}$ that belong to the $+$ sector and $S$ eigenstates $\ket{\chi_-(m)}$ that belong to the $-$ sector. If $S$ is even, we find that $\hat U \ket{\chi_+(0)}=\ket{\chi_+(0)}$ and $\ket{\chi_+(0)}$ belongs to the $+$ sector. On the other hand, if $S$ is odd, we find that $\hat U \ket{\chi_+(0)}=-\ket{\chi_+(0)}$ and $\ket{\chi_+(0)}$ belongs to the $-$ sector.

We next consider $S\in \mathbb{N}/2$, for which Eq.~\eqref{app:U_action} takes the form 
\begin{equation}
    \hat U \ket{\chi_\pm(m)}= \left\{
    \begin{array}{ll}
         \pm i \ket{\chi_\pm(m)}& \mbox{if  }\, 2S\equiv 1\, (\text{mod } 4) \\
        \mp i \ket{\chi_\pm(m)}& \mbox{if  }\, 2S\equiv 3\, (\text{mod } 4)\, .
    \end{array}
\right.
\end{equation}
In this case, there are $S+1/2$ pairs $(m, -m)$ and $m\neq 0$. Each pair contributes one eigenstate with eigenvalue $i$ and one with eigenvalue $-i$.

Summarizing, we obtain\begin{equation}\label{number_U_sector}
    S\text{ even: }\left\{
    \begin{array}{l}
        D_+ = S + 1 \\
        D_- = S\, ,
    \end{array}
\right. \quad S\text{ odd: }\left\{
    \begin{array}{l}
        D_+ = S \\
        D_- = S + 1 \, ,
    \end{array}
\right. \quad S\in \mathbb{N}/2:\hspace{.1cm}\left\{
    \begin{array}{l}
        D_+ = S + 1/2\\
        D_- = D_+\, .
    \end{array}
\right.
\end{equation}

\subsection{Krylov dimension for operators charged under \texorpdfstring{$\hat U$}{U}}\label{app:LMG_Krylov_dim}

Here, we provide details about the derivation of the formulas given in Table~\ref{tab:Krylov_dimension_LMG} for the upper bound on the Krylov dimension associated with initial operators of various charges. In the following, we assume that all the matrix elements of the operator in the energy basis that are not constrained to vanish by selection rules are nonzero.

Let us start with the case where $\hat{\mathcal{O}}$ is negatively charged under $\hat{U}$. The only nonzero matrix elements in the energy basis are $\mel{E,+}{\hat{\mathcal{O}}}{E,-}$ and $\mel{E,-}{\hat{\mathcal{O}}}{E,+}$. Since the spectrum of the LMG Hamiltonian is nondegenerate within symmetry sectors\footnote{\label{footnote:app_quasi_degeneracies}Strictly speaking, some of them correspond to quasi-degeneracies, which for finite $S$ correspond to close but different energies.}, the phases $\omega_{ij}=E_i-E_j$, with $(i,j)\in (+,-)$ and $(-,+)$, are also nondegenerate. Consequently, the Krylov dimension is given by the number of these matrix elements, 
\begin{equation}\label{K_dim_LMG_-}
    K_-=2 D_+ D_-=\left\{
    \begin{array}{ll}
        2S(S+1)\, , & S\in \mathbb{N} \\
        2(S+1/2)^2\, , & S\in \mathbb{N}/2\, .
    \end{array}
\right.
\end{equation}

We now consider the case where $\hat{\mathcal{O}}$ is positively charged under $\hat U$. The only nonzero matrix elements in the energy basis are $\mel{E,+}{\hat{\mathcal{O}}}{E,+}$ and $\mel{E,-}{\hat{\mathcal{O}}}{E,-}$. However, the phases $\omega_{ij}=E_i-E_j$, with $(i,j)\in (+,+)$ and $(-,-)$, are degenerate since there are $D_+$ and $D_-$ vanishing phases in the $+$ and $-$ sectors, respectively. Bearing in mind that all zero phases contribute one dimension to the Krylov space \cite{Adrian_1}, the Krylov dimension is given by
\begin{equation}\label{K_dim_LMG_+}
    K_+=D_+^2+D_-^2-D+1=\left\{
    \begin{array}{ll}
        2S^2+1\, , & S\in \mathbb{N} \\
        2S^2+1/2\, , & S\in \mathbb{N}/2\, ,
    \end{array}
\right.
\end{equation}
where $D=D_++D_-=2S+1$ is the dimension of the Hilbert space of the LMG model. We note that this expression corresponds to applying the upper bound $D^2 - D + 1$ separately to each symmetry sector.

Finally, in the case where $\hat{\mathcal{O}}$  is uncharged, all matrix elements contribute to the Krylov dimension. In this case, the Krylov dimension is given by
\begin{equation}\label{K_dim_LMG_0}
    K_0=D^2-D+1=4S^2+2S+1\, .
\end{equation}

\subsection{Linear stability analysis of the classical LMG model}\label{app:classical_LMG}

We recall that the classical LMG model expressed in terms of the classical spin components $(x,y,z)$ is given by $H=-\frac{J}{2}z^2-h x$. Using Eq.~\eqref{eq:classical_Liouvillian_operator} for the time evolution of observables, the corresponding Hamilton's equations are 
\begin{equation}\label{Hamilton_LMG}
    \left\{
    \begin{array}{ll}
        \dot{x}&=J y z\\
        \dot{y}&=-J x z+h z\\
        \dot{z}&=-h y\, .
    \end{array}
\right.
\end{equation} 
The quantity $\bm{s}^2=x^2+y^2+z^2$ is conserved (and equal to $1$ by construction), and one of the variables can thus be eliminated. Using the canonical variables $(q, p)$ (defined in Section~\ref{sec:general_prescription}), Hamilton's equations reduce to a $2$-dimensional system which is trivially Liouville integrable \cite{Arnold}.

The system possesses the four fixed points
\begin{subequations}\label{fixed_points_LMG}
\begin{align}
    &A_{1\pm}=\left(\pm 1,\, 0,\, 0 \right),\\
    &A_{2\pm}=\left(\frac{h}{J},\, 0,\, \pm \sqrt{1-\left( \frac{h}{J}\right)^2} \right).
\end{align}
\end{subequations}
We note that the points $A_{2\pm}$ require $h\leq J$ to exist. The linear stability of these fixed points is summarized in Table~\ref{tab:stability_fixed_points}. The main aspect is that in the regime $h<J$, $A_{1+}$ is a saddle point with energy $E(A_{1+})=-h$. For $h>J$, the phase space does not contain any saddle point. We refer the reader to Ref.~\cite{Silvia_LMG} for a visual representation of the phase space of LMG. At $h=J$, the system undergoes a bifurcation where the two centers $A_{2\pm}$ merge with the saddle point $A_{1+}$ to form a single center at $h>J$.

The extrema of the classical Hamiltonian and their corresponding energies are
\begin{equation}\label{max_energies_LMG}
h\leq J:\hspace{.25cm} \left\{
    \begin{array}{lll}
        \text{max:}\,&A_{1-},\, &E(A_{1-})=h \\
        \text{min:}\,&A_{2\pm},\, &E(A_{2\pm})=-\frac{h^2+J^2}{2J} \, ,
    \end{array}
\right. \qquad
    h\geq J:\hspace{.25cm} \left\{
    \begin{array}{lll}
            \text{max:}\,&A_{1-},\, &E(A_{1-})=h\\
        \text{min:}\,&A_{1+},\, &E(A_{1+})=-h\, .
    \end{array}
\right.
\end{equation}
We note that for $h<J$, $A_{2\pm}$ are degenerate global minima, which can be understood in the light of the double-well shaped potential associated with the classical LMG Hamiltonian, see discussion in Section~\ref{sec:symmetries_spectrum_LMG}. Similarly, the quantum version of the model in Eq.~\eqref{LMG_rescaled_TSS} admits a doubly degenerate ground state in the ordered phase. From these extremal energies, the classical spectral bandwidth is given by
\begin{equation}\label{spectral_bandwidth_LMG}
\Lambda(h,J)=\hspace{.25cm} \left\{
    \begin{array}{lll}
        \frac{(h+J)^2}{2J}\, , \quad h\leq J \\
        2h\, ,\quad h\geq J \, .
    \end{array}
\right.
\end{equation}

\section{Details on the FP model} \label{appx:FP}

In this appendix, we define the symmetry sectors of the FP model and their respective dimensions for integer values of $L\in \mathbb{N}$. We then use these dimensions to derive an upper bound on the Krylov dimension of the FP model as a function of the symmetry charge of the initial operator. Finally, we analyze the phase space of the classical FP model, briefly comment on classical chaos, identify saddle points, and derive the spectral bandwidth of the classical Hamiltonian, which is used extensively to normalize the classical Lanczos coefficients in Section~\ref{sec:classical_FP}.

\subsection{Symmetry sectors}\label{app:FP_symmetries}

As argued in Section~\ref{sec:symmetries_spectrum_FP}, in the $(L,L)$ (with $L\in \mathbb{N}$) sector of the Hilbert space, the FP Hamiltonian in Eq.~\eqref{FP_rescaled} is invariant under the two commuting symmetries $\hat{U}_1$ and $\hat{U}_2$. Consequently, this sector decomposes into the four orthogonal invariant subspaces
\begin{equation}
    \mathcal{H}_{(L,L)}=\mathcal{H}_{++}\oplus\mathcal{H}_{+-}\oplus\mathcal{H}_{-+}\oplus\mathcal{H}_{--}\, .
\end{equation}
These subspaces are spanned by the following simultaneous eigenstates of $\hat{U}_1$ and $\hat{U}_2$ \cite{FP_modern}, where $m_2> m_1$
\begin{subequations}
\begin{align}
    &\mathcal{H}_{++} = \left\{m_1+m_2 \text{ even} \hspace{.15cm}\textit{\&}\hspace{.15cm} \frac{1}{\sqrt{2}}\left(\ket{m_1,m_2}+\ket{m_2,m_1}\right)\right\}, \\
    &\mathcal{H}_{-+} = \left\{m_1+m_2 \text{ odd} \hspace{.15cm}\textit{\&}\hspace{.15cm}\frac{1}{\sqrt{2}}\left(\ket{m_1,m_2}+\ket{m_2,m_1}\right)\right\}, \\
    &\mathcal{H}_{+-} = \left\{m_1+m_2 \text{ even}\hspace{.15cm}\textit{\&}\hspace{.15cm} \frac{1}{\sqrt{2}}\left(\ket{m_1,m_2}-\ket{m_2,m_1}\right)\right\}, \\
    &\mathcal{H}_{--} = \left\{m_1+m_2 \text{ odd}\hspace{.15cm}\textit{\&}\hspace{.15cm} \frac{1}{\sqrt{2}}\left(\ket{m_1,m_2}-\ket{m_2,m_1}\right)\right\}.
\end{align}
\end{subequations}
The states $\ket{m_1,m_2}$ with $m_1=m_2$ are simultaneous eigenstates of $\hat{U}_1$ and $\hat{U}_2$ with eigenvalue $+1$, and thus belong to the $(+, +)$ sector. The dimensions of the four symmetry sectors are given by
\begin{subequations}
    \begin{align}
        & D_{++}\equiv \dim \mathcal{H}_{++}=\left(L+1\right)^2,\\
        & D_{-+}\equiv\dim \mathcal{H}_{-+}=L\left(L+1\right),\\
        & D_{+-}\equiv\dim \mathcal{H}_{+-}=L^2,\\
        & D_{--}\equiv\dim \mathcal{H}_{--}=L\left(L+1\right).
    \end{align}
\end{subequations}

\subsection{Krylov dimension for operators charged under \texorpdfstring{$\hat{U}_1$}{U1} and \texorpdfstring{$\hat{U}_2$}{U2}}\label{app:FP_Krylov_dim}

Here, we provide details about the derivation of the formulas given in Table~\ref{tab:Krylov_dimension_FP} for the upper bound on the Krylov dimension associated with initial operators of various charges. In the following, we assume that all the matrix elements of the operator in the energy basis that are not constrained to vanish by selection rules are nonzero.

Before proceeding, we first note that $\hat{U}_1$ imposes additional structure on $\hat{\mathcal{O}}$, as expressed by the following property: if an operator $\hat{\mathcal{O}}$ is negatively charged under $\hat{U}_1$, then all its anti-diagonal elements must vanish. Indeed, one has
\begin{equation}
    \mel{m_1,m_2}{\hat{\mathcal{O}}}{m_1',m_2'}=-e^{- i \pi(m_1+m_2-m_1'-m_2')}  \mel{m_1,m_2}{\hat{\mathcal{O}}}{m_1',m_2'} ,
\end{equation}
where we have used the explicit action of $\hat U_1$ on the states $\ket{m_1,m_2}$. Thus, if $m_1+m_2-m_1'-m_2'$ is even, the matrix elements $\mel{m_1,m_2}{\hat{\mathcal{O}}}{m_1',m_2'}$ vanish. Setting $m_1'=-m_1$ and $m_2'=-m_2$ shows that all anti-diagonal elements of $\hat{\mathcal{O}}$ are zero in this case. This justifies the use of the symbol $\nexists$ in the first, second, and seventh rows of Table~\ref{tab:Krylov_dimension_FP}. More precisely, if all anti-diagonal elements of $\hat{\mathcal{O}}$ are zero (respectively nonzero) in the computational basis, then all anti-diagonal elements in the energy basis (not constrained to be zero by the selection rules) are also zero (respectively nonzero)\footnote{This was numerically verified.}.

We start by considering the Krylov dimension in the case where $\hat{\mathcal{O}}$ has charge $(-,-)$. According to the selection rules, the nonzero matrix elements are
\begin{alignat*}{2}
    & \mel{E,+,+}{\hat{\mathcal{O}}}{E',-,-},\qquad && \mel{E,-,-}{\hat{\mathcal{O}}}{E',+,+},\\
    & \mel{E,+,-}{\hat{\mathcal{O}}}{E',-,+}, && \mel{E,-,+}{\hat{\mathcal{O}}}{E',+,-}.
\end{alignat*}
The corresponding eigenspace representatives involve eigenoperators of the form $\ket{E,+,+}\\\bra{E',-,-}$ and $\ket{E,+,-}\bra{E',-,+}$ (and their Hermitian conjugate) all associated with nondegenerate phases\footnote{See footnote \ref{footnote:app_quasi_degeneracies}.}. The Krylov dimension is therefore simply equal to the number of nonzero matrix elements, namely $K_{--}=2D_{++}D_{--}+2D_{+-}D_{-+}$. The case $(-,+)$ is strictly analogous and leads to the same Krylov dimension.

The cases where $\hat{\mathcal{O}}$ has charge $(+,-)$ or $(+,+)$ require a more careful analysis. In the former, the nonzero matrix elements are
\begin{alignat*}{2}
    & \mel{E,+,+}{\hat{\mathcal{O}}}{E',+,-},\qquad && \mel{E,-,-}{\hat{\mathcal{O}}}{E',-,+},\\
    & \mel{E,+,-}{\hat{\mathcal{O}}}{E',+,+}, && \mel{E,-,+}{\hat{\mathcal{O}}}{E',-,-}.
\end{alignat*}
The eigenspace representatives associated with matrix elements in the first column above involve eigenoperators of the form $\ket{E,+,\alpha}\bra{E',+,-\alpha}$ and $\ket{-E',+,-\alpha}\bra{-E,+,\alpha}$ (with $\alpha=\pm$), which share the same phase which is therefore doubly degenerate. For each matrix element in the second column above, the eigenspace representatives involve eigenoperators of the form $\ket{E,-,\alpha}\bra{E',-,-\alpha}$ and $\ket{-E',-,\alpha}\bra{-E,-,-\alpha}$, which share the same phase, except for those of the form $\ket{E,-,\alpha}\bra{-E,-,-\alpha}$ for which the phase is unique. Their number is $D_{--}+D_{-+}=2L(L+1)$ unless $\overline{\text{diag}}(\hat{\mathcal{O}})= 0$. The Krylov dimension is therefore given by
\begin{subequations}
\begin{align}
        &K_{+-}=D_{++}D_{+-}+D_{--}(D_{-+}-1)\,,
        \\
        &\widetilde{K}_{+-}=\frac{1}{2}(2D_{++}D_{+-}+2D_{--}D_{-+}-2L(L+1))+2L(L+1)\,.
\end{align}
\end{subequations}

The nonzero matrix elements in the $(+,+)$ case are 
\begin{alignat*}{2}
    & \mel{E,+,+}{\hat{\mathcal{O}}}{E',+,+},\qquad && \mel{E,-,-}{\hat{\mathcal{O}}}{E',-,-},\\
    & \mel{E,+,-}{\hat{\mathcal{O}}}{E',+,-}, && \mel{E,-,+}{\hat{\mathcal{O}}}{E',-,+}.
\end{alignat*}
For each matrix element in the first column, the eigenspace representatives involve eigenoperators of the form $\ket{E,+,\alpha}\bra{E',+,\alpha}$ and $\ket{-E',+,\alpha}\bra{-E,+,\alpha}$, which share the same phase, except for those of the form  $\ket{E,+,\alpha}\bra{-E,+,\alpha}$ for which the phase is unique\footnote{A subtlety arises for the eigenoperator $\ket{0, +, +}\bra{0,+,+}$ (if $L$ is even) or $\ket{0, +, -}\bra{0,+,-}$ (if $L$ is odd) which also has a zero phase.}. Their number is $D_{++}+D_{+-}-1=2L(L+1)$. The eigenspace representatives associated with the matrix elements in the second column involve eigenoperators of the form $\ket{E,-,\alpha}\bra{E',-,\alpha}$ and $\ket{-E',-,-\alpha}\bra{-E,-,-\alpha}$, which share the same phase. In addition, every eigenoperator of the form $\ket{E,\alpha_1,\alpha_2}\bra{E,\alpha_1,\alpha_2}$ contributes a zero phase. Their total number is $D_{++}+D_{+-}+D_{--}+D_{-+}=(2L+1)^2$. Hence, the Krylov dimension is given by
\begin{subequations}
\begin{align}
    &K_{++}=\frac{1}{2}(D_{++}(D_{++}-1)+D_{+-}(D_{+-}-1)+D_{--}^2+D_{-+}^2-(2L+1)^2+1)+1,\\
        &\widetilde{K}_{++}=\frac{1}{2}(D_{++}^2+D_{+-}^2+D_{--}^2+D_{-+}^2- (2L+1)^2 - 2L(L+1))+2L(L+1)+1.
\end{align}
\end{subequations}

We now briefly comment on the Krylov dimension for operators with charges $(0,-)$, $(0,+)$, $(-,0)$, $(+,0)$ and $(0,0)$. The total number of matrix elements in each case can be determined by adding those of the corresponding charged cases. For instance, the number of nonzero matrix elements for an operator with charge $(0,-)$ is given by the sum of the nonzero matrix elements from the $(+,-)$ and $(-,-)$ cases. This summation may affect the Krylov dimension, and it is necessary to verify whether eigenoperators from different eigenspace representatives share the same phase.

We begin by examining the cases where the operator has charge $(0,-)$ or $(0,+)$. We find that the union of eigenspace representatives of $(+,+)$ and $(-,+)$, and of $(-,-)$ and $(+,-)$ does not produce any additional degenerate phases. In this case, the respective Krylov dimensions are simply additive,
\begin{subequations}
\begin{alignat}{2}
&K_{0-}=K_{+-}+K_{--}\, ,\qquad && K_{0+}=K_{++}+K_{-+} \,, \\
& \widetilde{K}_{0-}=\widetilde{K}_{+-}+K_{--}\, , \qquad &&\widetilde{K}_{0+}=\widetilde{K}_{++}+K_{-+}\, .
\end{alignat}
\end{subequations}
In the expressions for $\widetilde{K}_{0+}$ and $\widetilde{K}_{0-}$, we added $K_{--}$ and $K_{-+}$, and not $\widetilde{K}_{--}$ and $\widetilde{K}_{-+}$, because operators with $\overline{\text{diag}}(\hat{\mathcal{O}})\neq 0$ cannot have a negative $\hat{U}_1$ charge.  

Degeneracies, however, arise when considering the union of the eigenspace representatives of $(0,-)$ and $(0,+)$. In particular, each eigenoperator from $(0,-)$ can be paired with a corresponding eigenoperator from $(0,+)$ which share the same phase. The total number of these pairings is $D_{++}D_{-+}+D_{+-}D_{--}=L (L+1)(2L^2+2L+1)$. Hence, the Krylov dimension of the fully uncharged case is no longer additive and is given by
\begin{subequations}
    \begin{align}
        &K_{00}=K_{0+}+K_{0-}-2L (L+1)(2L^2+2L+1)\, , \label{K_dim_FP_00}\\
        & \widetilde{K}_{00}=\widetilde{K}_{0+}+\widetilde{K}_{0-}-2L (L+1)(2L^2+2L+1)\, .
    \end{align}
\end{subequations}

Finally, we discuss the cases where the operator has charge $(-,0)$ or $(+,0)$. In the former, the Krylov space is spanned by the union of the eigenspace representatives from the $(-,-)$ and $(-,+)$ cases. Each eigenoperator from $(-,-)$ can be paired with a corresponding eigenoperator from $(-,+)$ which share the same phase. This pairing does not occur in the $(+,0)$ case. Consequently, the Krylov dimensions are given by
\begin{subequations}
    \begin{alignat}{2}
        & K_{-0}=\frac{1}{2}(K_{-+}+K_{--})\, ,&&  \\
        & K_{+0}=K_{++}+K_{+-} \, , \qquad &&\widetilde{K}_{+0}=\widetilde{K}_{++}+\widetilde{K}_{+-}\, .
    \end{alignat}
\end{subequations}
The Krylov dimension of the fully uncharged case can be recovered from $K_{00}=K_{+0}+K_{-0}$, in agreement with the formula in Eq.~\eqref{K_dim_FP_00}. We note, however, that when $\overline{\text{diag}}(\hat{\mathcal{O}})\neq 0$, $\widetilde{K}_{00}$ cannot be expressed in terms of $\widetilde{K}_{-0}$ and $\widetilde{K}_{+0}$, since $\widetilde{K}_{-0}$ is not defined.

\subsection{Linear stability analysis of the classical FP model}\label{app:classical_FP}

We recall that the classical FP model expressed in terms of the classical spin components $(x_1,y_1,z_1,x_2,y_2,z_2)$ is given by $H=-\left(1+\lambda\right)\left(z_1+z_2\right)-4\left(1-\lambda\right) x_1 x_2$. Using Eq.~\eqref{eq:classical_Liouvillian_operator} for the time evolution of observables, the corresponding Hamilton's equations are 
\begin{equation}\label{Hamilton_FP}
    \left\{
    \begin{array}{ll}
        \dot{x}_1&=\left(1+\lambda\right) y_1\\
        \dot{y}_1&=-\left(1+\lambda\right)x_1+4\left(1-\lambda\right) z_1x_2\\
        \dot{z}_1&=-4\left(1-\lambda\right)y_1 x_2\\
        \dot{x}_2&=\left(1+\lambda\right) y_2\\
        \dot{y}_2&=-\left(1+\lambda\right)x_2+4\left(1-\lambda\right)x_1 z_2\\
        \dot{z}_2&=-4\left(1-\lambda\right)x_1 y_2\, .
    \end{array}
\right.
\end{equation}
Both classical spin magnitudes are conserved, $\bm{l}^2=1$ and $\bm{m}^2=1$. Using the canonical variables $(q_1,p_1,q_2,p_2)$ (defined in Section~\ref{sec:general_prescription}), Hamilton's equations reduce to a $4$-dimensional system. The system is also invariant under exchange of the two classical spins $\bm{l}\leftrightarrow \bm{m}$. 

We briefly comment on the integrability of the classical model. As is often the case for classical Hamiltonian systems, the phase space consists of a stochastic sea of chaotic motion within which islands of regular motion are embedded. Chaotic trajectories cannot penetrate these islands, while regular trajectories confined to an island cannot escape from it. First, we note that for $\lambda=-1$ and $1$, the system is integrable with conserved quantities $x_1, x_2$, and $z_1, z_2$, respectively. It was shown in the original study \cite{FP_1} that classical chaos is mostly developed in the energy shell $E=0$ in a region that includes $\lambda=0$. This was later refined in \cite{FP_modern} where the authors showed that in this same energy shell, the dynamics is chaotic in the range $0\lesssim \lambda \lesssim 0.25$. For larger values, the phase space becomes increasingly mixed and finally becomes regular near $\lambda=1$, in agreement with Figure~\ref{fig:quantum_integrability_chaos_FP} (for the quantum case). It remains, however, unclear if there exists a range in $\lambda$ at which the system is fully chaotic, i.e. every possible initial condition (irrespective of the energy shell) yielding a positive Lyapunov exponent.

For $\lambda\in (-1, 1)$, the system possesses eight fixed points given by 
\begin{subequations}\label{fixed_points_FP}
\begin{align}
    &B_{1\pm}=\left(0,\,0,\,\pm1,\,0,\,0,\,\pm 1 \right),\\
    &B_{2\pm}=\left(0,\,0,\,\pm1,\,0,\,0,\, \mp1 \right), \\
    &B_{3\pm}=\left(\sqrt{1-\gamma^2},\,0,\,\pm\gamma,\,\pm\sqrt{1-\gamma^2},\,0,\,\pm\gamma\right),\label{fixed_point_FP_1}\\
    &B_{4\pm}=\left(-\sqrt{1-\gamma^2},\,0,\,\pm\gamma,\,\mp\sqrt{1-\gamma^2},\,0,\,\pm\gamma\right),\label{fixed_point_FP_2}
\end{align}
\end{subequations}
where $\gamma\equiv \left(1+\lambda\right)/(4\left(1-\lambda\right))$. The four points $B_{3\pm}$ and $B_{4\pm}$ exist only for $\lambda\leq 3/5$, where they coincide with the point $B_{1\pm}$ at $\lambda=3/5$. Each pair of fixed points $\pm$ has the same linear stability, which is summarized in Table~\ref{tab:stability_fixed_points}. In contrast to the LMG model, the phase space of the FP model always contains at least one saddle point. More precisely, the fixed points $B_{2\pm}$ are saddles throughout this interval in $\lambda$, while $B_{1\pm}$ are saddles only for $\lambda\leq3/5$. At $\lambda = 3/5$, the system undergoes a bifurcation in which the four centers $B_{3\pm}$ and $B_{4\pm}$ merge with the saddles $B_{1\pm}$, to form two centers at $\lambda>3/5$. The corresponding saddle-point energies are $E(B_{1\pm})=\mp 2\left(1+\lambda\right)$ and $E(B_{2\pm})=0$.

For completeness, we also discuss the two marginal cases $\lambda=-1,1$. For $\lambda=-1$ ($\gamma=0$), the system has the four fixed points corresponding to $B_{3\pm}$ and $B_{4\pm}$ as well as the $16$ fixed points (with all possible sign combinations)
\begin{equation}
    B_{5}=\left(0,\,y_1,\,\sqrt{1-y_1^2},\,0,\,y_2,\,\sqrt{1-y_2^2} \right),\qquad y_1, y_2\in[-1,1]\, .
\end{equation}
Linearization does not determine stability as the eigenvalues of the Jacobian evaluated on these points are all zero. For $\lambda=1$, the system only admits the four fixed points $B_{1\pm}$ and $B_{2\pm}$ which are centers.

The extrema of the classical Hamiltonian and their corresponding energies are
\begin{subequations}
\begin{align}\label{max_energies_FP}
-1\leq \lambda\leq  3/5:\hspace{.25cm} &\left\{
    \begin{array}{lll}
        \text{max:}\,&B_{3-}, B_{4-},\, &E(B_{3-})=E(B_{4-})=-\frac{17 \lambda }{4}+\frac{1}{1-\lambda }+\frac{13}{4} \\
        \text{min:}\,&B_{3+}, B_{4+},\, &E(B_{3+})=E(B_{4+})=\frac{17 \lambda }{4}-\frac{1}{1-\lambda }-\frac{13}{4}\, ,
    \end{array}
\right. \\
3/5\leq \lambda \leq 1:\hspace{.25cm} &\left\{
    \begin{array}{lll}
            \text{max:}\,&B_{1-},\, &E(B_{1-})=2(1+\lambda)\\
        \text{min:}\,&B_{1+},\, &E(B_{1+})=-2(1+\lambda)\, .
    \end{array}
\right.
\end{align}
\end{subequations}
We note that for $-1\leq \lambda\leq3/5$, $(B_{3+}, B_{4+})$ and $(B_{3-}, B_{4-})$ are degenerate extrema. From these extremal energies, the corresponding spectral width is given by
\begin{equation}\label{spectral_bandwidth_FP}
\Lambda(\lambda)=\hspace{.25cm} \left\{
    \begin{array}{lll}
        -\frac{17\lambda}{2}+\frac{2}{1-\lambda}+\frac{13}{2}\, , \quad -1\leq \lambda\leq  3/5 \\
        4(1+\lambda)\, ,\quad 3/5\leq \lambda \leq 1 \, .
    \end{array}
\right.
\end{equation}

\section{Identities and relations for spherical harmonics}

In this appendix, we derive a collection of identities involving spherical harmonics, which, we recall, form a useful orthonormal basis of $\mathcal{F}(S^2)$. These identities are extensively used throughout Sections~\ref{sec:general_prescription}, \ref{sec:classical_LMG}, \ref{sec:classical_FP} and \ref{sec:classical_microcanonical_Lanczos}.

\subsection{Poisson bracket of \texorpdfstring{$Y_1^0$}{Y10} and \texorpdfstring{$Y_1^{\pm 1}$}{Y1(+-1)}}\label{app:spherical_harmonics_poisson_brackets}

We prove the Poisson bracket identities in Eqs.~\eqref{spherical_harmonics_relation_1} and \eqref{spherical_harmonics_relation_2} involving the spherical harmonics $Y_1^0$ and $Y_1^{\pm 1}$. These identities are later used to derive the classical Liouvillian operators for both the LMG and FP models.

For that purpose, we first enumerate several properties of spherical harmonics (see for example \cite{spherical_harmonics_relations}):
\begin{subequations}
\begin{align}
    & \pdv{}{\theta}Y_l^m(\theta,\varphi)=m \cot \theta\, Y_l^m(\theta,\varphi )+\sqrt{(l+m+1)(l-m)}e^{- i \varphi } Y_l^{m+1}(\theta ,\varphi)\, ,\label{app:formula_2}\\
    & \pdv{}{\varphi}Y_l^m(\theta,\varphi)= i m\,  Y_l^m(\theta,\varphi)\, ,\label{app:formula_1}\\
    &Y_l^{m-1}=-\left(\frac{2m \cot \theta \, e^{- i \varphi }}{\sqrt{l(l+1)-m(m-1)}}\, Y_l^m+\sqrt{\frac{(l+m+1)(l-m)}{l(l+1)-m(m-1)}}e^{-2 i \varphi }\,Y_l^{m+1}\right) .\label{app:formula_3}
\end{align}
\end{subequations}

Eq.~\eqref{spherical_harmonics_relation_1} is straightforward to prove: using the expression $Y_1^0(\theta,\varphi)=\sqrt{3}\cos \theta$, together with the definition of the Poisson bracket in Eq.~\eqref{Poisson_bracket}, and Eq.~\eqref{app:formula_1} above, the identity follows immediately.

The expression \eqref{spherical_harmonics_relation_2} requires some calculations. Given that $Y_1^{\pm 1}=\mp \sqrt{3/2}\sin \theta \\e^{\pm  i \varphi }$, it can easily be shown that
\begin{align}
     \left\{Y_1^{\pm 1},\, Y_l^m\right\}&=  i  m \sqrt{\frac{3}{2}}\cot \theta \, e^{\pm  i \varphi }\,Y_l^m\mp i \frac{\sqrt{(l+m+1)(l-m)}}{\sin \theta}e^{- i \varphi }\,Y_1^{\pm 1}Y_l^{m+1}\nonumber \\
     &\mp  i m\sqrt{\frac{3}{2}}\cot \theta \, e^{\pm  i \varphi }\,Y_l^m\, ,
\end{align}
where Eqs.~\eqref{Poisson_bracket} and \eqref{app:formula_2} were used. From this expression, we get that
\begin{equation}
    \left\{Y_1^1,\, Y_l^m\right\}=- i \frac{\sqrt{(l+m+1)(l-m)}}{\sin \theta}e^{- i \varphi }\,Y_1^{1}Y_l^{m+1}= i \sqrt{\frac{3}{2}}\sqrt{(l+m+1)(l-m)}\,Y_l^{m+1}\, .
\end{equation}
We also get that 
\begin{align}
     \left\{Y_1^{-1},\, Y_l^m\right\}&= i\sqrt{6} m \cot \theta \, e^{-  i \varphi }\,Y_l^m+ i \frac{\sqrt{(l+m+1)(l-m)}}{\sin \theta}e^{- i \varphi }\, Y_1^{- 1}Y_l^{m+1} \nonumber \\
     &= i  \sqrt{\frac{3}{2}}\left( 2 m \cot \theta \, e^{-  i \varphi }\,Y_l^m+ \sqrt{(l+m+1)(l-m)}e^{-2 i \varphi }\,Y_l^{m+1}\right)\nonumber\\
     &=- i \sqrt{\frac{3}{2}}\sqrt{(l-m+1)(l+m)}\,Y_l^{m-1}\, ,
\end{align}
which concludes the proof.

\subsection{Classical Liouvillian operators for the LMG and FP models}\label{app:classical_Liouvillian_operator}

We now show that the classical Liouvillian of the LMG model is given by Eq.~\eqref{classical_Liouvillian_LMG}. Given the definition of the classical Liouvillian operator \eqref{eq:classical_Liouvillian_operator} and that of the classical LMG Hamiltonian \eqref{LMG_classical}, we have
\begin{equation}\label{app:classical_Liouvillian_LMG}
    \mathcal{L}_cY_l^m=-\frac{J}{\sqrt{3}}m\,Y_1^0 Y_l^m-\frac{h}{2}\left(\sqrt{(l+m+1)(l-m)}\,Y_l^{m+1}+\sqrt{(l-m+1)(l+m)}\,Y_l^{m-1} \right) .
\end{equation}
Using Clebsch-Gordan coefficients, the product $Y_1^0 Y_l^m$ can be decomposed in terms of a sum involving single spherical harmonic terms \cite{spherical_harmonics_relations}. The exact calculation reads,
\begin{equation}\label{Product_spherical_harmonics_1}
    Y_1^0 Y_l^m = \sqrt{\frac{3}{2l+1}}\left(\sqrt{\frac{(l+m+1)(l-m+1)}{2l+3}}Y_{l+1}^m+\sqrt{\frac{(l-m)(l+m)}{2l-1}}Y_{l-1}^m \right).
\end{equation}
Inserting this expansion in \eqref{app:classical_Liouvillian_LMG} yields the expected expression of the classical Liouvillian operator in Eq.~\eqref{classical_Liouvillian_LMG}.

The derivation is similar for the classical Liouvillian operator of the FP model in Eq.~\eqref{classical_Liouvillian_FP}. Using the classical FP Hamiltonian \eqref{FP_classical}, we have
\begin{align}\label{app:classical_Liouvillian_FP}
    &\mathcal{L}_c Y_l^m Z_k^n\nonumber \\
    &=\left(1+\lambda\right) (m+n)Y_l^m Z_k^n-4\sqrt{\frac{\pi}{6}}\left(1-\lambda\right)\Big[\left(\beta_+(l,m)\, Y_l^{m+1}+\beta_-(l,m)\, Y_l^{m-1} \right)\left(Z_1^1-Z_1^{-1} \right)Z_k^n\nonumber \\
    &\hspace{.5cm}+\left(Y_1^1-Y_1^{-1} \right)Y_l^m\left(\beta_+(k,n)\, Z_k^{n+1}+\beta_-(k,n)\, Z_k^{n-1} \right)\Big] .
\end{align}
Coefficients $\beta_\pm$, $\eta_\pm$ and $\theta_\pm$ are defined in Eqs.~\eqref{classical_FP_coefficients}. We observe that terms of the form $Y_1^{\pm 1} Y_l^m$ and $Z_1^{\pm 1} Z_k^n$ arise and need to be expanded in terms of Clebsch-Gordan coefficients as before. One has 
\begin{equation}\label{Product_spherical_harmonics_2}
    Y_1^{\pm 1} Y_l^m = \sqrt{\frac{3}{2(2l+1)}}\left( \eta_\pm(l,m)\, Y_{l+1}^{m\pm 1}- \theta_\mp(l,m)\,Y_{l-1}^{m\pm 1}\right),
\end{equation}
and similarly for $Z_1^{\pm 1} Z_k^n$ replacing $l\to k$ and $m \to n$. Inserting this expansion in \eqref{app:classical_Liouvillian_FP} yields the expected expression of the classical Liouvillian operator in Eq.~\eqref{classical_Liouvillian_FP}.

\subsection{Microcanonical inner product and symmetries}\label{app:spherical_harmonics_selection_rules}

We finally derive the selection rules for the spherical harmonics in Eqs.~\eqref{spherical_harmonics_LMG}. Both are based on symmetries exhibited by the classical LMG Hamiltonian \eqref{LMG_classical}. First, its invariance under the transformation $\varphi\to 2\pi-\varphi$ (equivalent to $(x,y,z)\to(x,-y,z)$) implies \eqref{LMG_microcanonical_formula_0}, while its invariance under the transformation $\theta\to \pi-\theta$ (equivalent to $(x,y,z)\to(x,y,-z)$) implies \eqref{LMG_microcanonical_formula}.

The proof of the first selection rule is as follows. We have that
\begin{equation}
    \braketc{Y_l^m}{Y_{l'}^{m'}}_{E,\Delta E}
    \propto \int_0^\pi \dd \theta \sin \theta\, P_l^m(\cos \theta) P_{l'}^{m'}(\cos \theta) \int_0^{2\pi} \dd \varphi\, e^{ i (m' - m) \varphi} \,p_{E,\Delta E}(\theta, \varphi)\, ,
\end{equation}
where $P_l^m$ denote the associated Legendre polynomials. The imaginary part of the integral over $\varphi$ vanishes. Indeed, let $I=\int_0^{2\pi} \dd \varphi\, \sin\left[(m' - m) \varphi\right] p_{E,\Delta E}(\theta, \varphi)$. By the change of variable $\varphi\to 2\pi - \varphi$, one finds $I=0$. Hence, the inner products $\braketc{Y_l^m}{Y_{l'}^{m'}}_{E,\Delta E}$ are real. 

The second selection rule can be derived as follows. Under the transformation $\theta\to \pi-\theta$, spherical harmonics transform as $Y_l^m(\pi - \theta,\varphi)=(-1)^{l+m} Y_l^m(\theta,\varphi)$. Consequently, $\braketc{Y_l^m}{Y_{l'}^{m'}}_{E,\Delta E}=(-1)^{l-m+l'-m'}\braketc{Y_l^m}{Y_{l'}^{m'}}_{E,\Delta E}$ which can be directly obtained by a change of variable in the corresponding integral. This implies that $\braketc{Y_l^m}{Y_{l'}^{m'}}_{E,\Delta E}=0$ whenever $l+m - l' - m'$ is odd (equivalently, when $l+ l' +m+ m'$ is odd).

\section{More on the Krylov complexity of Lipkin-Meshkov-Glick and Feingold-Peres models}\label{app:more_CK}

\begin{figure}[ht!]
    \centering
    \includegraphics[width=.375\textwidth]{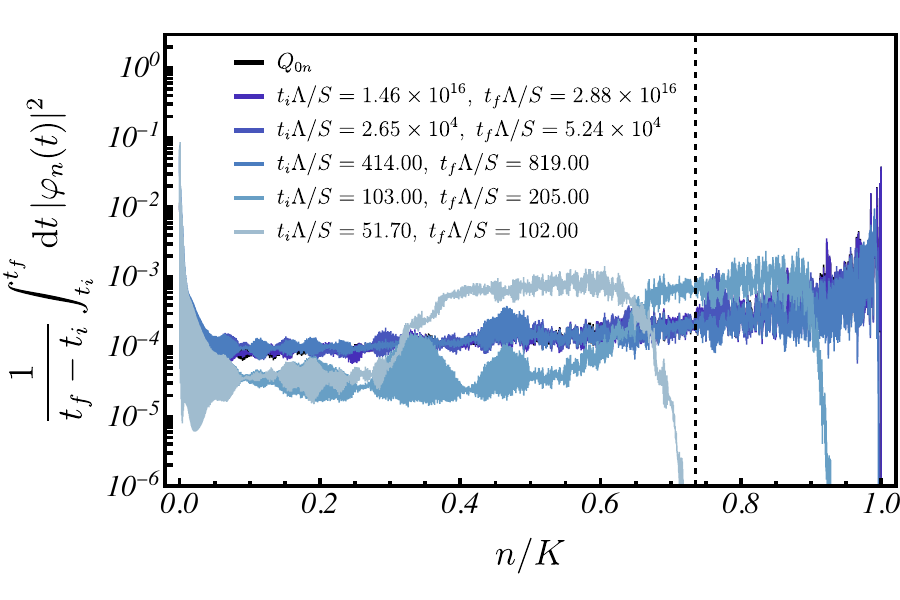}\qquad
    \includegraphics[width=.375\textwidth]{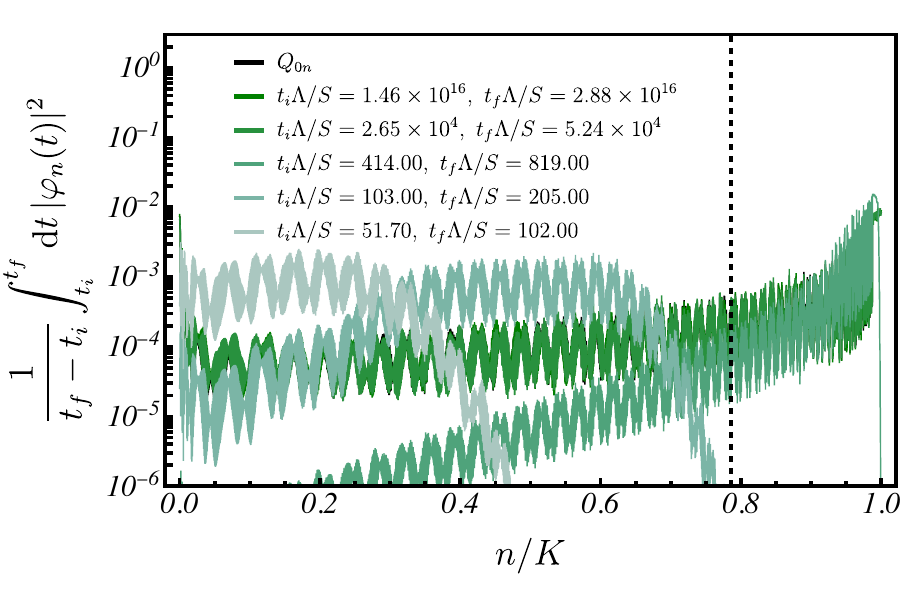}
    \caption{Convergence of the time-averaged Krylov wavefunctions $|\varphi_n(t)|^2$ to their late-time average value $Q_{0n}$ for increasingly large time windows. Results are presented for the LMG model \eqref{LMG_rescaled_TSS} ($S=40$) and the initial operator $\hat{s}_z$. The $Q_{0n}$ (black curves) are nearly identical to the time averages over the largest time windows. \textbf{(Left)} Saddle regime with $h=1/2$ and $J=1$. \textbf{(Right)} No-saddle regime with $h=2$ and $J=1$. In both panels, the black dashed vertical lines indicate the saturation value $\overline{C_K}$ derived from the $Q_{0n}$s. Here, $\Lambda=\Lambda(h,J)$ denotes the spectral bandwidth.}
    \label{fig:Krylov_wavefunctions_LMG}
\end{figure}

\begin{figure}[ht!]
    \centering
    \includegraphics[width=.375\textwidth]{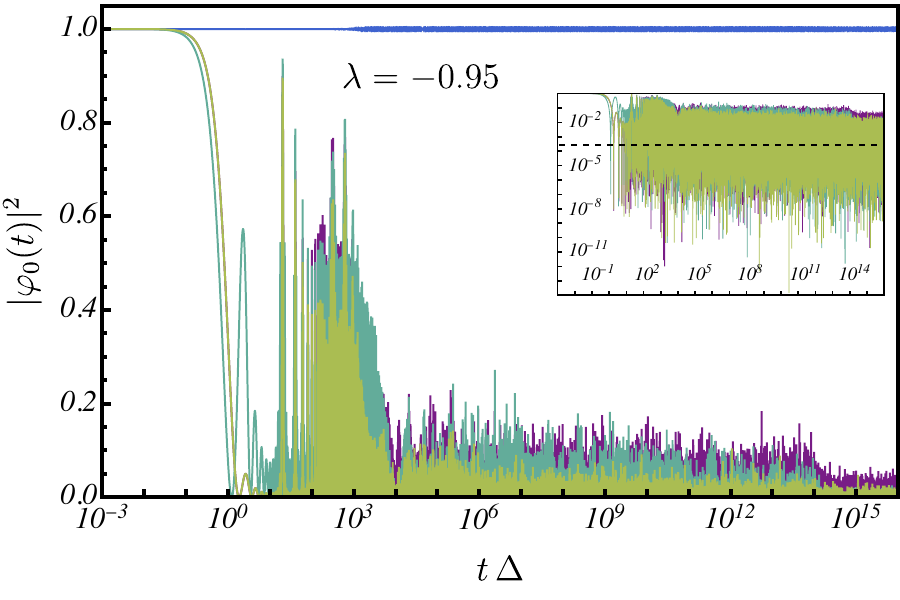}\qquad
    \includegraphics[width=.375\textwidth]{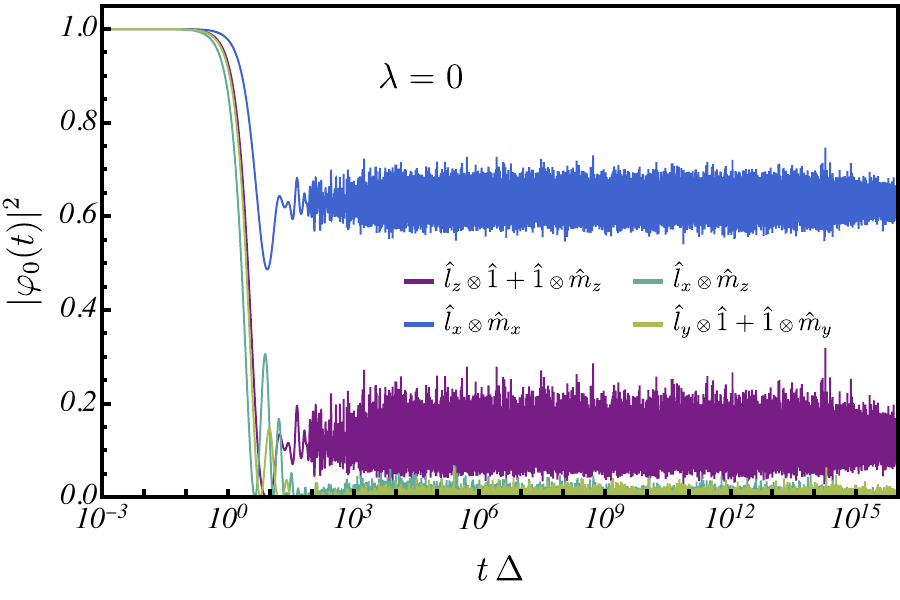}\\
    \includegraphics[width=.375\textwidth]{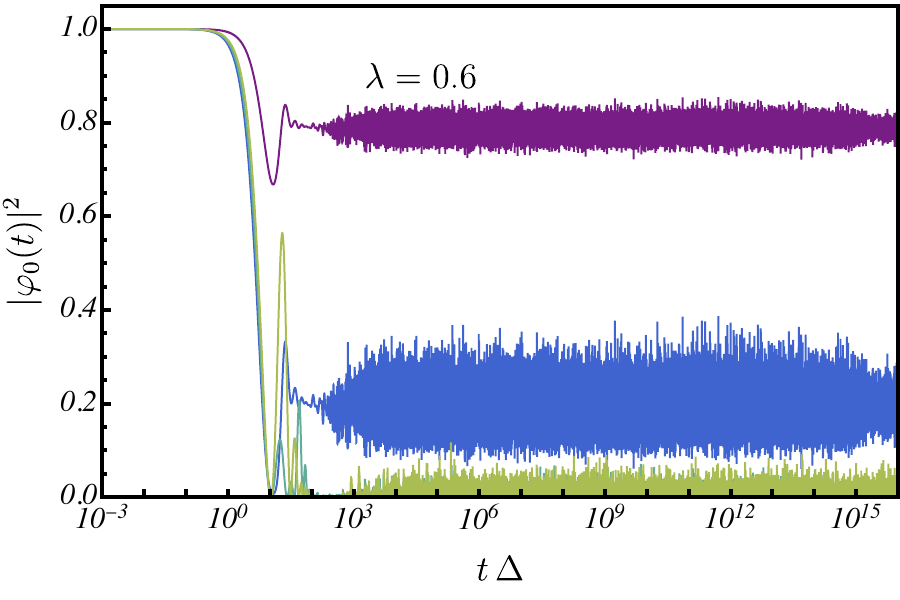}\qquad
    \includegraphics[width=.375\textwidth]{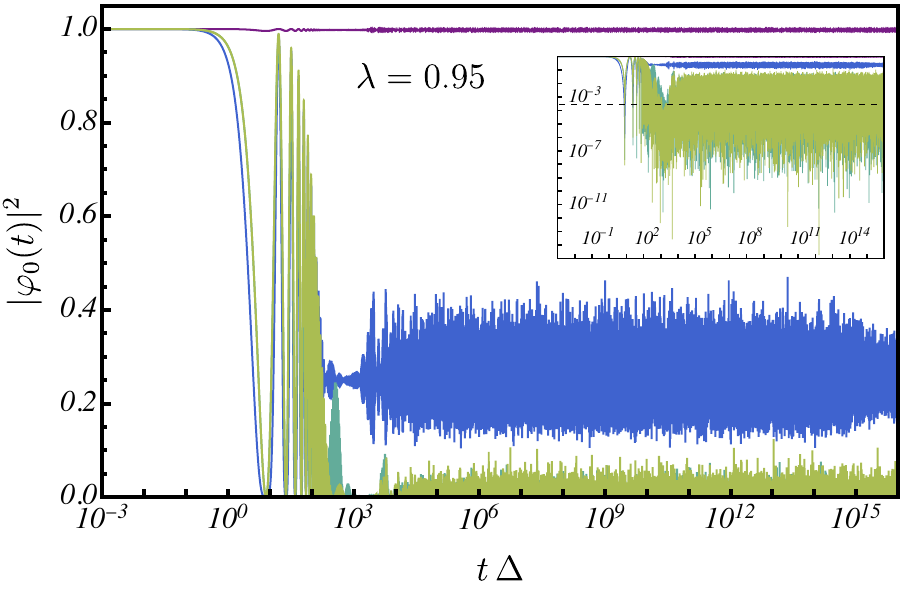}
    \caption{\textbf{(Main)} Squared modulus of the autocorrelation function for the FP model \eqref{FP_rescaled} ($L=5$) for various operators as a function of time (rescaled by the mean level spacing). Results are shown for various values of the parameter $\lambda$. \textbf{(Insets)} Similar comments to those in the insets of Figure~\ref{fig:two_points_functions_LMG} apply regarding the behavior of generic operators. In particular, the horizontal black dashed lines denote the value $1/K_{-0}=1/K_{-+}$ where $K_{-0}=K_{-+}$ are the upper bounds on the Krylov dimension of $\hat{l}_x\otimes \hat{m}_z$ and $\hat{l}_y\otimes \mathbbm{1}+\mathbbm{1}\otimes \hat{m}_y$ (see Table~\ref{tab:Krylov_dimension_FP}), two simple generic operators of the FP model. Their autocorrelation functions deviate from the typical plateau of $1/K$ characteristic of chaotic systems.}
    \label{fig:two_points_functions_FP}
\end{figure}

\begin{figure}[ht!]
    \centering
    \includegraphics[width=.375\textwidth]{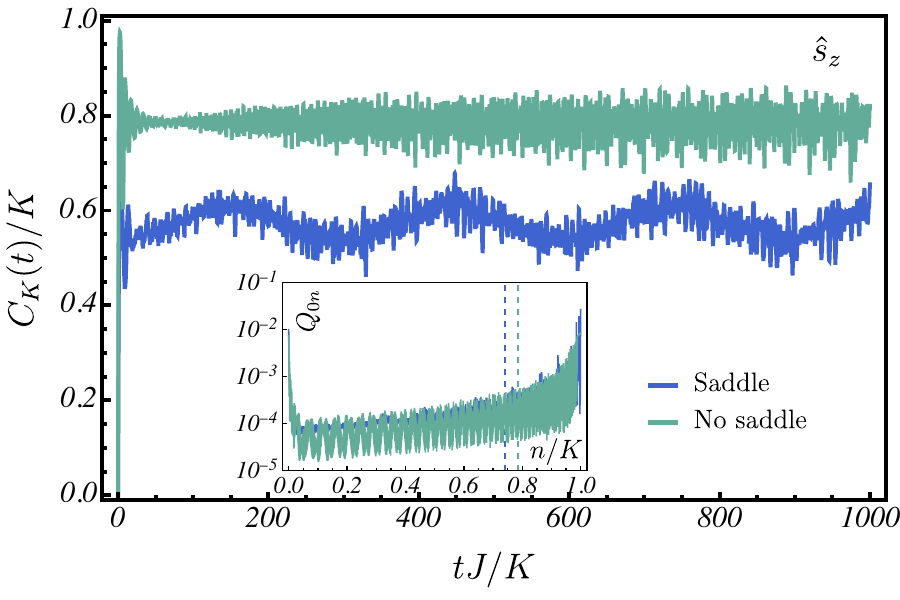}\qquad 
    \includegraphics[width=.375\textwidth]{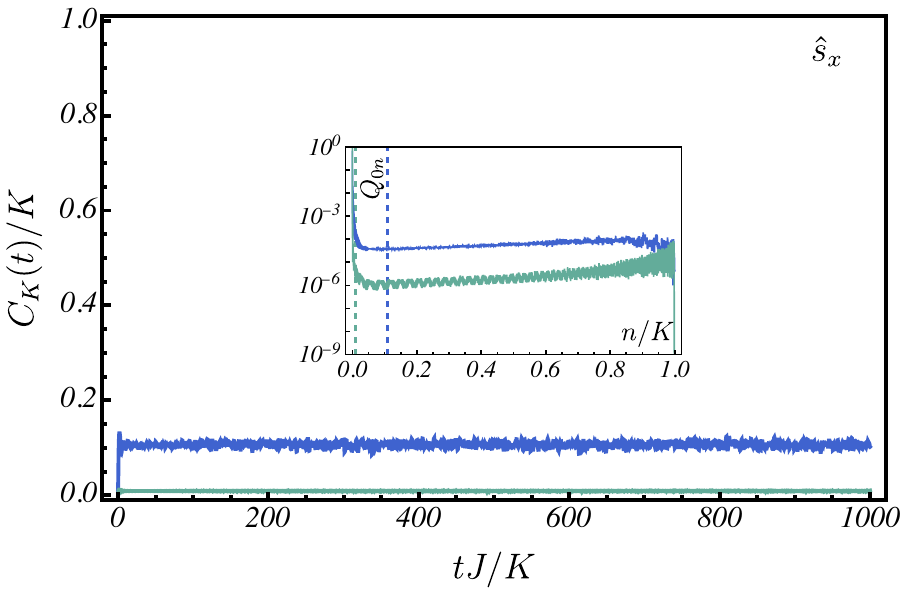}\\
    \includegraphics[width=.375\textwidth]{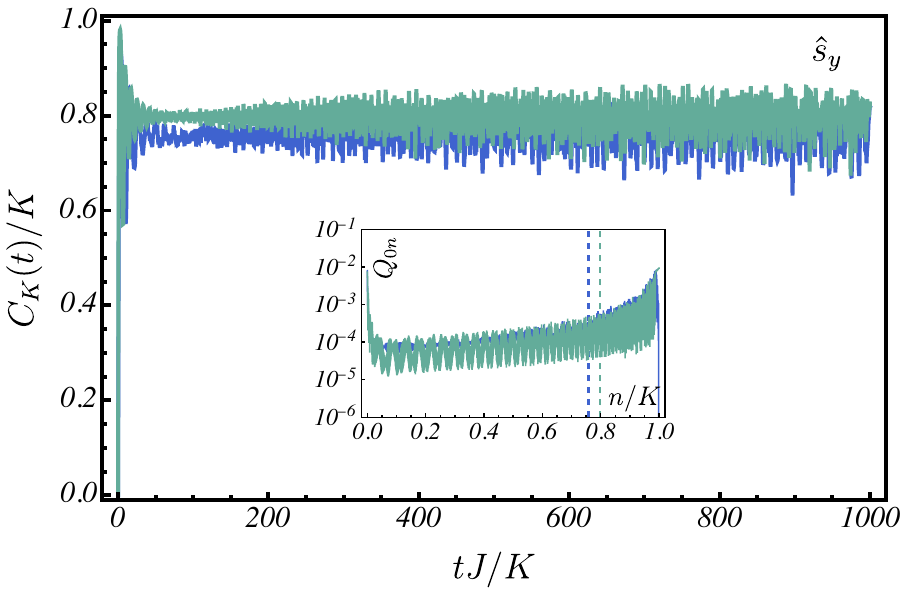}\qquad 
    \includegraphics[width=.375\textwidth]{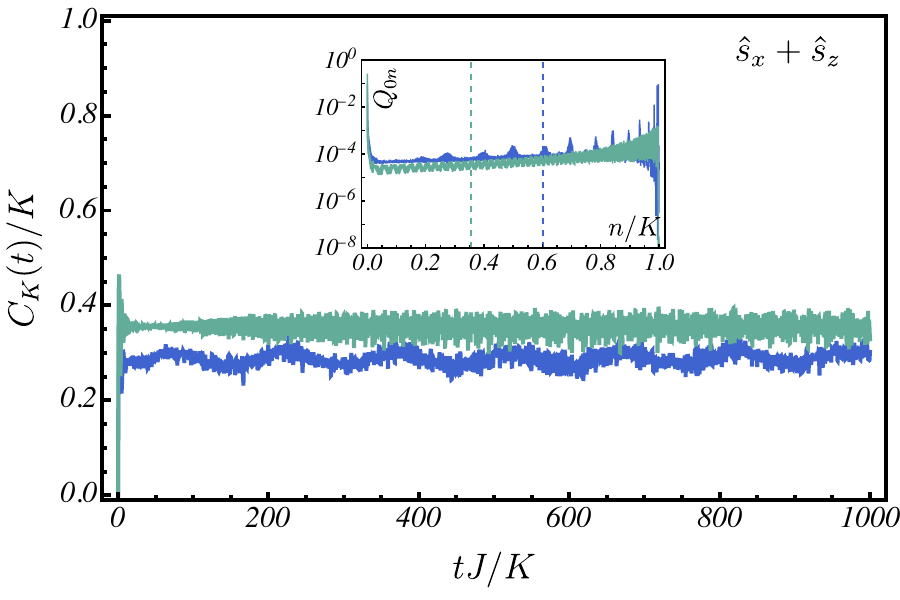}
    \caption{\textbf{(Main)} Time evolution of K-complexity, normalized by the Krylov dimension $K$, for the LMG model \eqref{LMG_rescaled_TSS} ($S=40$), in both saddle (blue curves with $h=1/2$, $J=1$) and no-saddle regimes (green curves with $h=2$, $J=1$). Time is rescaled by the Krylov dimension. \textbf{(Insets)} Corresponding $Q_{0n}$ as a function of the normalized index $n/K$. The vertical dashed lines denote the normalized late-time saturation value $\overline{C_K}/K$.}
    \label{fig:Krylov_complexity_LMG_appendix}
\end{figure}

\begin{figure}[ht!]
    \centering
    \includegraphics[width=.375\textwidth]{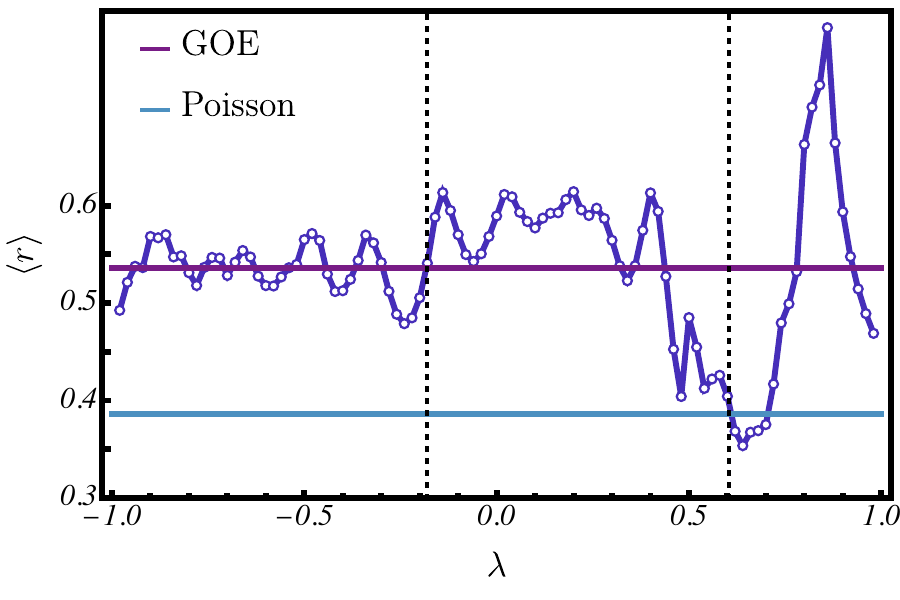}\qquad 
    \includegraphics[width=.375\textwidth]{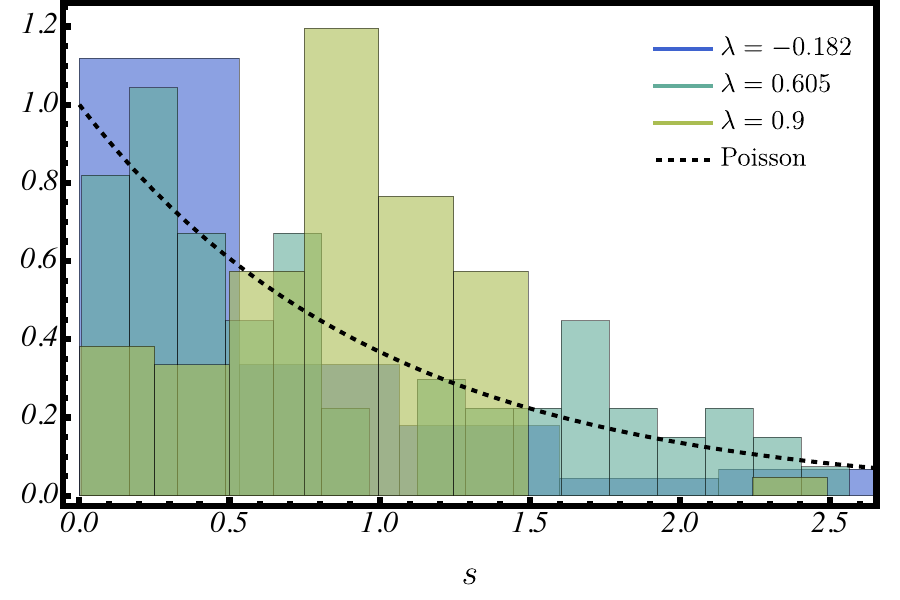}
    \caption{\textbf{(Left)} Mean ratio $\langle r\rangle$ of the FP Hamiltonian \eqref{FP_rescaled} ($L=6$) in the $\mathcal{H}_{-+}$ subspace, shown as a function of the parameter $\lambda$. The solid horizontal lines correspond to the Poisson and GOE predictions. The vertical dashed lines denote $\lambda=-0.182$ and $\lambda=0.605$, where $\langle r\rangle$ is exactly equal to the GOE and Poisson predictions, respectively. \textbf{(Right)} Level-spacing distribution of the full spectrum of the FP Hamiltonian \eqref{FP_rescaled} ($L=6$) at selected values of $\lambda$. Between $\lambda\simeq 0.5$ and $0.7$, the distribution is very close to a Poisson distribution.}
    \label{fig:integrability_FP_L6}
\end{figure}

\begin{figure}[ht!]
    \centering
    \includegraphics[width=.375\textwidth]{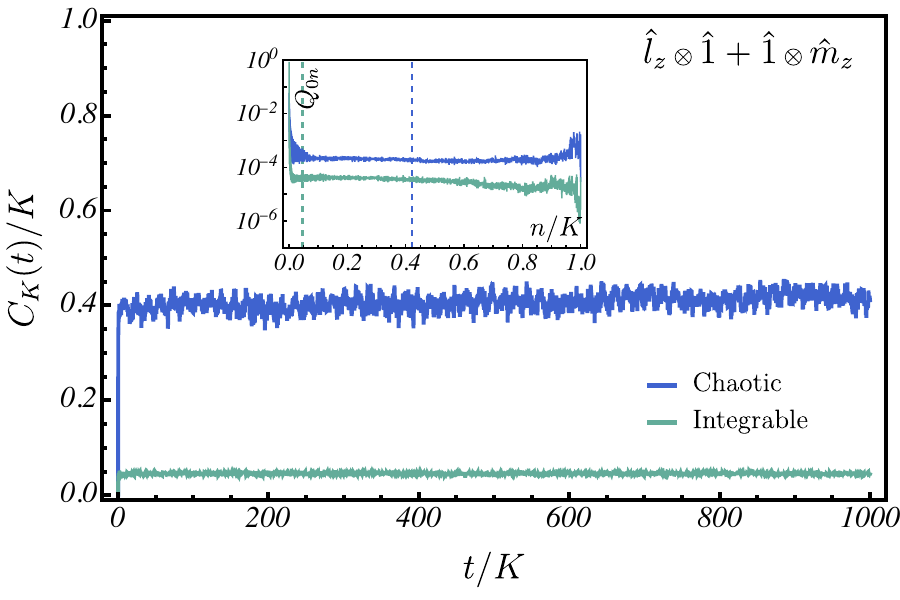}\qquad 
    \includegraphics[width=.375\textwidth]{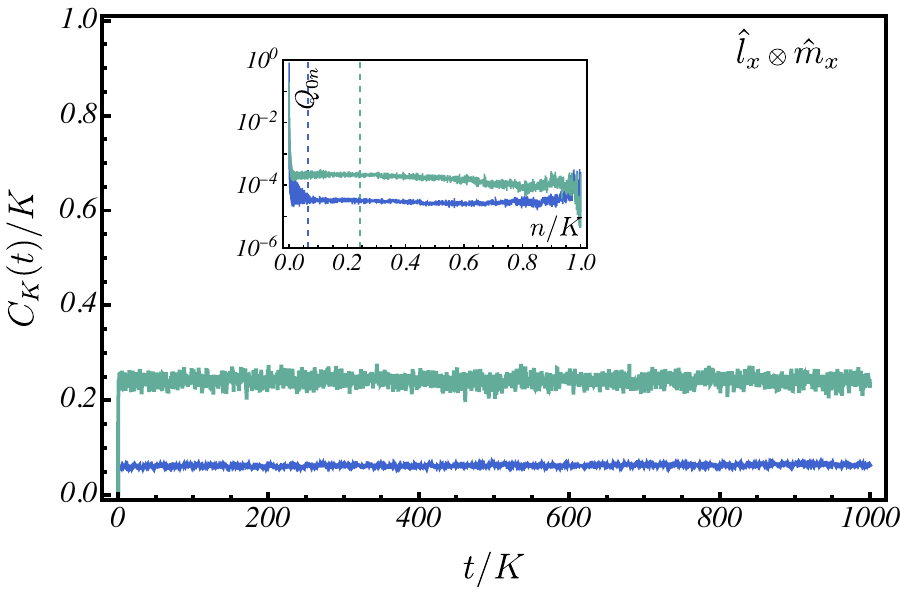}\\
    \includegraphics[width=.375\textwidth]{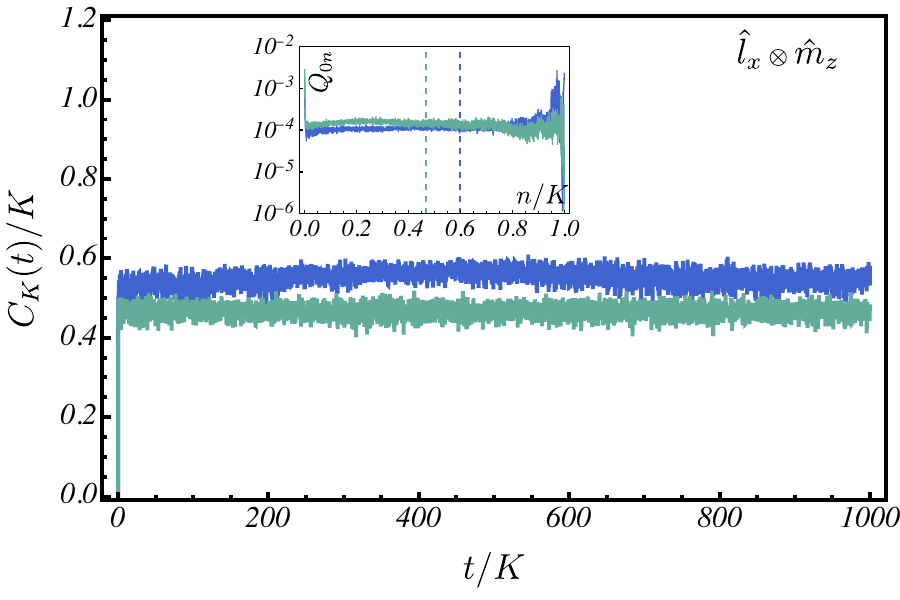}\qquad 
    \includegraphics[width=.375\textwidth]{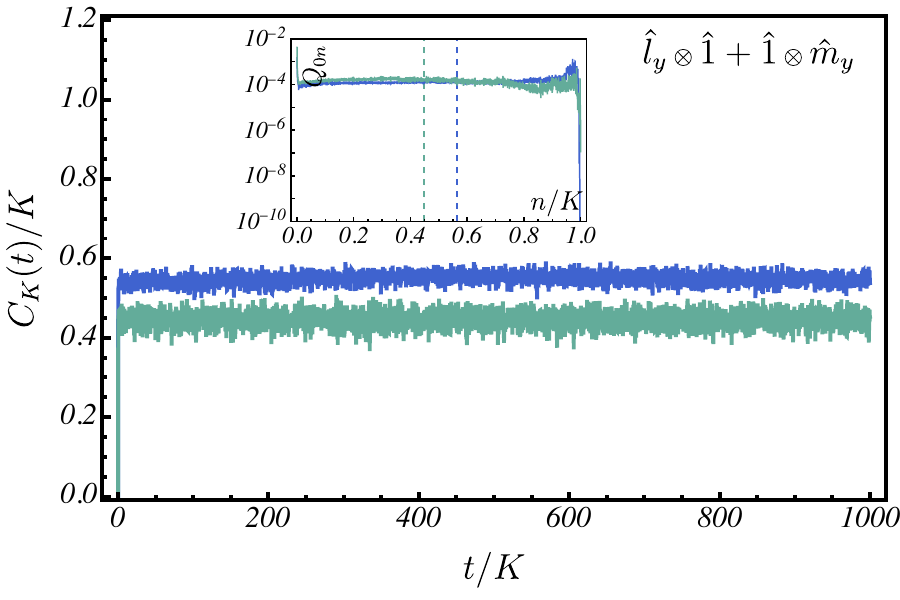}
    \caption{\textbf{(Main)} Time evolution of K-complexity, normalized by the Krylov dimension $K$, for the FP model \eqref{FP_rescaled} ($L=6$). Blue curves correspond to the chaotic regime ($\lambda=-0.182$) and green ones to the integrable regime ($\lambda=0.605$). Time is rescaled by the Krylov dimension. \textbf{(Insets)} Corresponding $Q_{0n}$ as a function of the normalized index $n/K$. The vertical dashed lines denote the normalized late-time saturation value $\overline{C_K}/K$.
    }
    \label{fig:Krylov_complexity_FP_appendix}
\end{figure}

\begin{figure}[ht!]
    \centering
    \includegraphics[width=.375\textwidth]{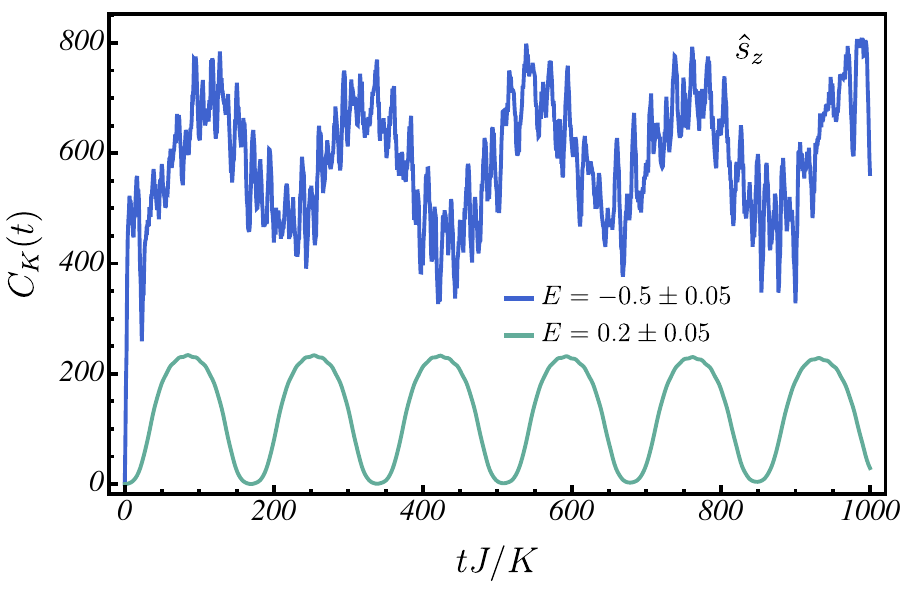}\qquad 
    \includegraphics[width=.375\textwidth]{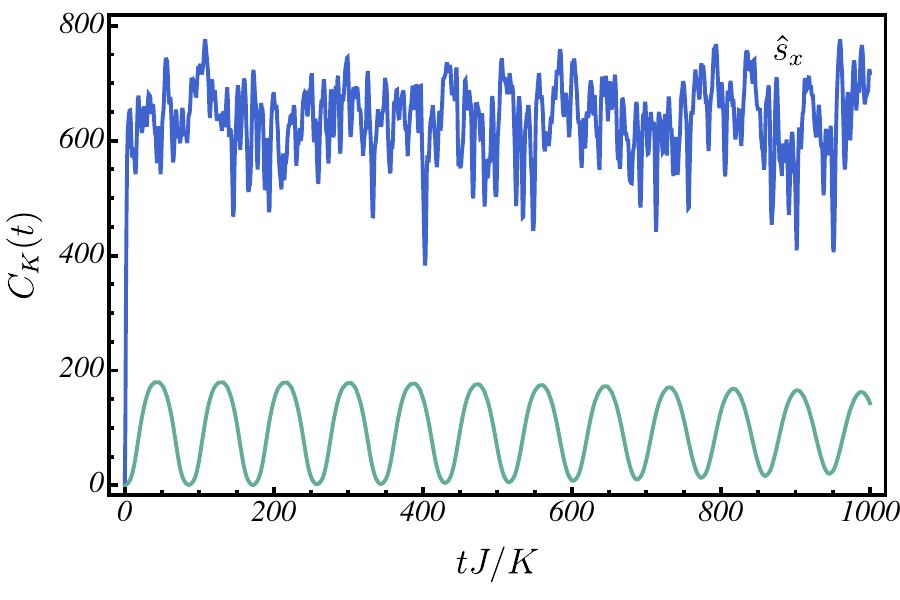}\\
    \includegraphics[width=.375\textwidth]{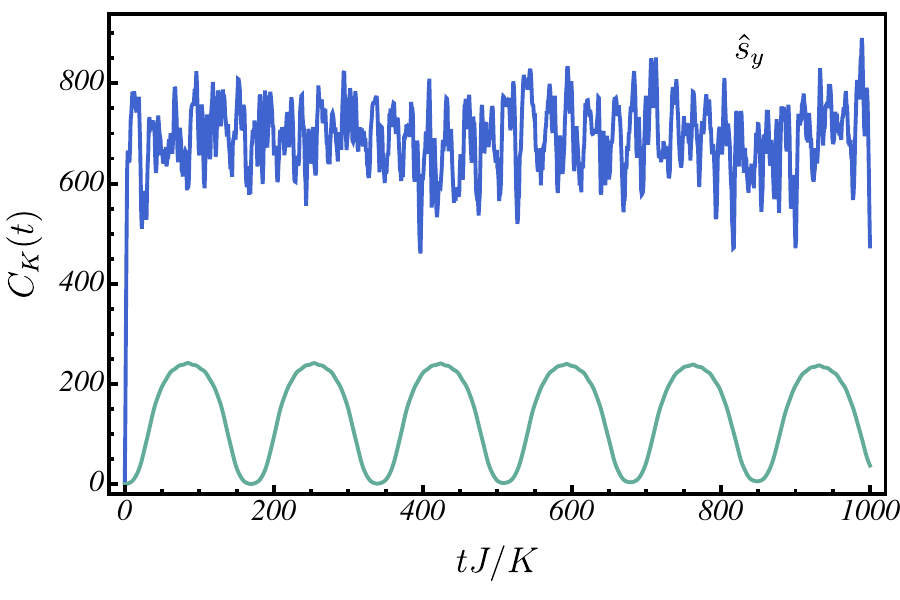}\qquad 
    \includegraphics[width=.375\textwidth]{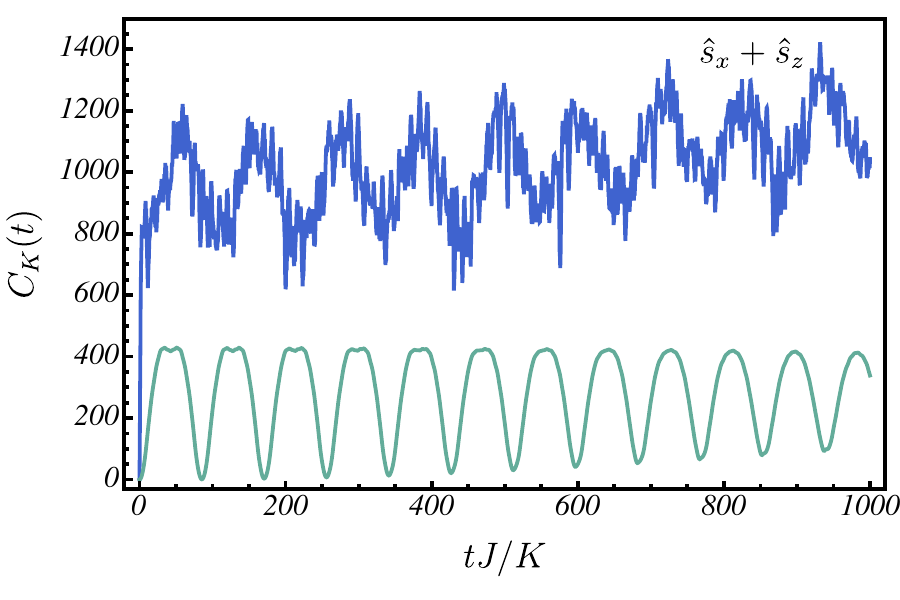}
    \caption{Time evolution of the microcanonical K-complexity (computed using the inner product \eqref{eq:Operator_Inner_product} with the density matrix \eqref{microcanonical_density_matrix}) for the LMG model \eqref{LMG_rescaled_TSS} ($S=60$) in the saddle regime ($h=1/2$, $J=1$). The blue curves correspond to an energy shell centered at the saddle point and the green ones to a shell centered away. Here, $K=1270$ in the shell containing the saddle point and $K=484$ in the other one for $\hat{s}_z$, $1251$ and $477$ for $\hat{s}_x$, $1270$ and $484$ for $\hat{s}_y$, $2521$ and $961$ for $\hat{s}_x+\hat{s}_z$.
    }
    \label{fig:LMG_microcanonical_1}
\end{figure}

\begin{figure}[ht!]
    \centering
    \includegraphics[width=.375\textwidth]{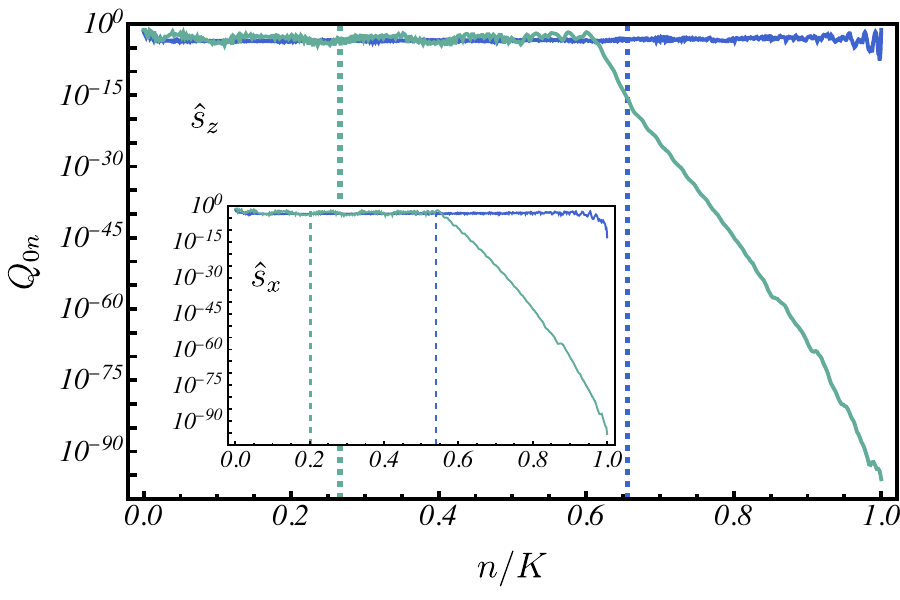}\qquad 
    \includegraphics[width=.375\textwidth]{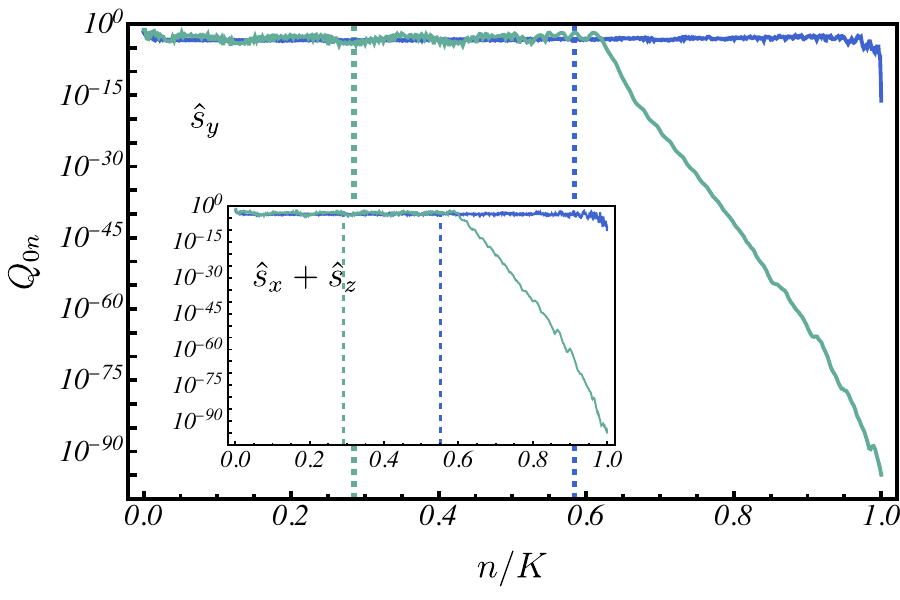}
    \caption{$Q_{0n}$ associated with the data depicted in Figure~\ref{fig:LMG_microcanonical_1}.}
    \label{fig:LMG_microcanonical_Q0n_1}
\end{figure}

\begin{figure}[ht!]
    \centering
    \includegraphics[width=.375\textwidth]{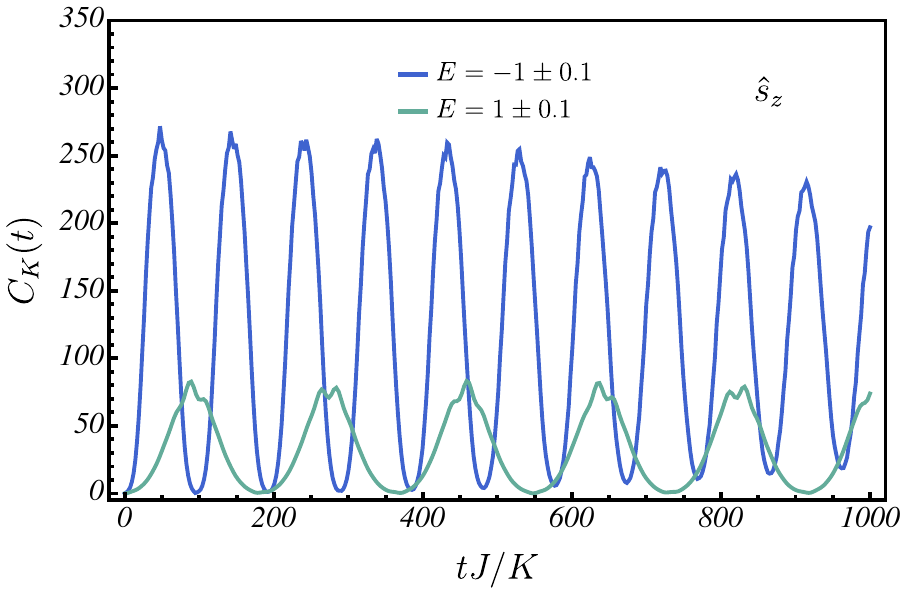}\qquad 
    \includegraphics[width=.375\textwidth]{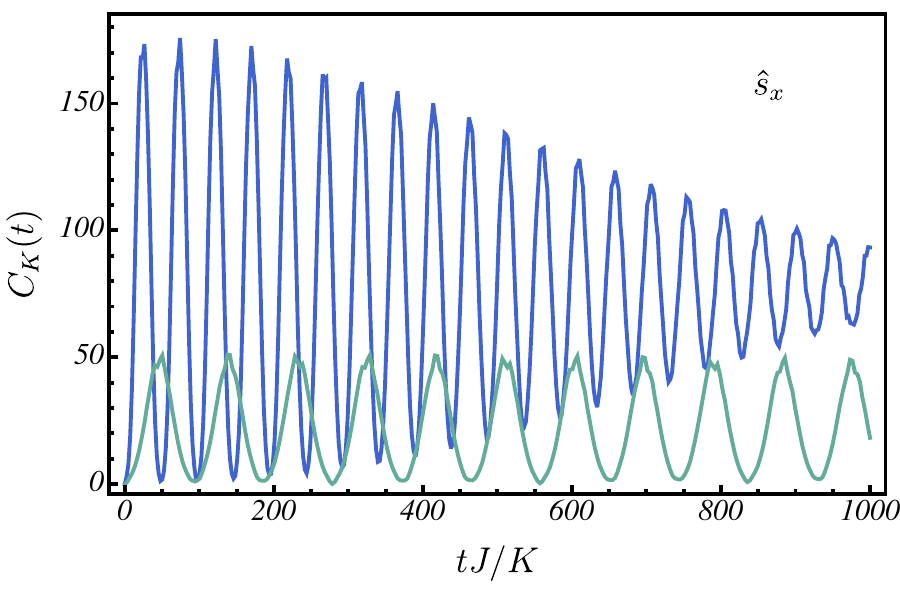}\\
    \includegraphics[width=.375\textwidth]{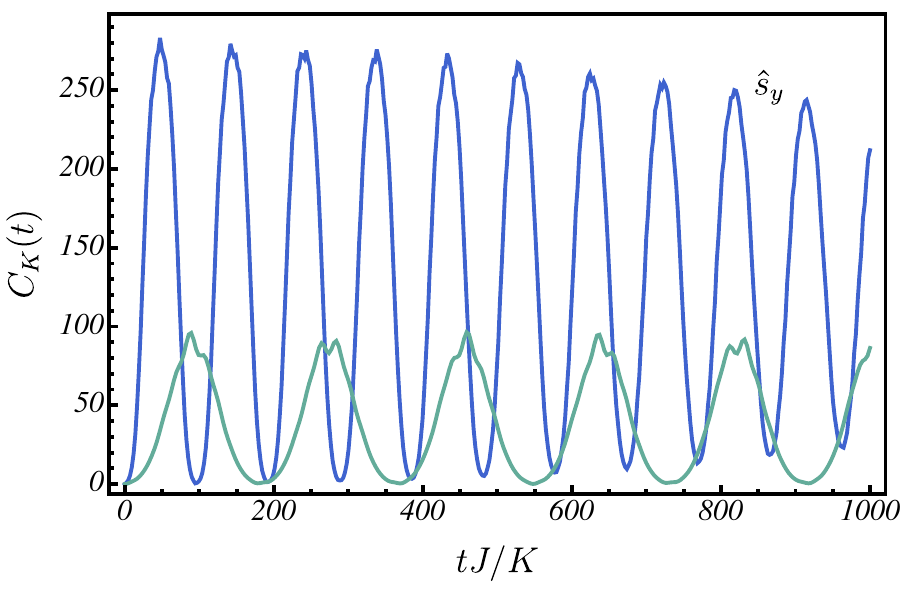}\qquad 
    \includegraphics[width=.375\textwidth]{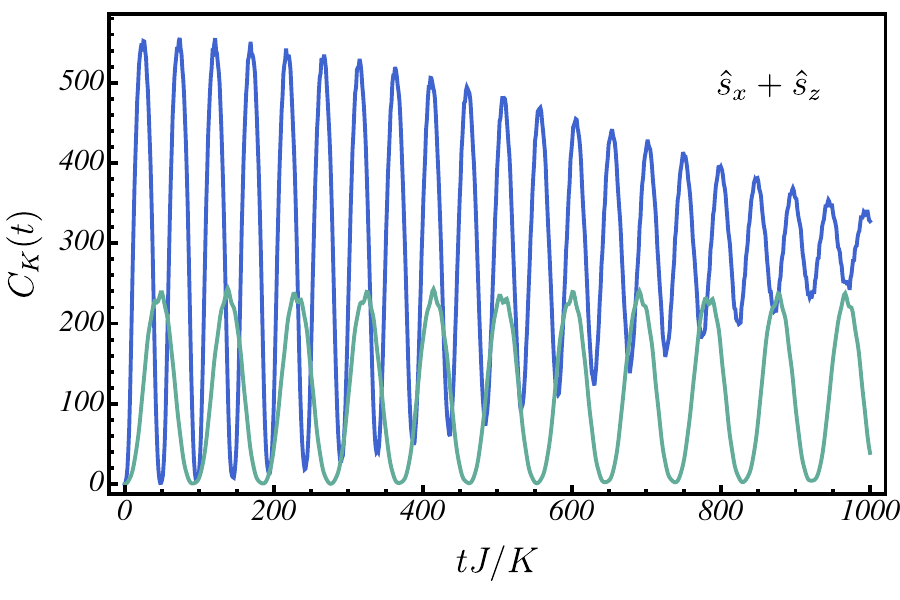}
    \caption{Same as in Figure~\ref{fig:LMG_microcanonical_1} but in the no-saddle regime ($h=2$, $J=1$). The spectrum does not contain a critical energy in this regime. Here, $K=423$ in the shell $E=-1\pm 0.1$ and $K=303$ in the shell $E=1\pm 0.1$ for $\hat{s}_z$, $418$ and $298$ for $\hat{s}_x$, $423$ and $303$ for $\hat{s}_y$, $841$ and $601$ for $\hat{s}_x+\hat{s}_z$.}
    \label{fig:LMG_microcanonical_2}
\end{figure}

\begin{figure}[ht!]
    \centering
    \includegraphics[width=.375\textwidth]{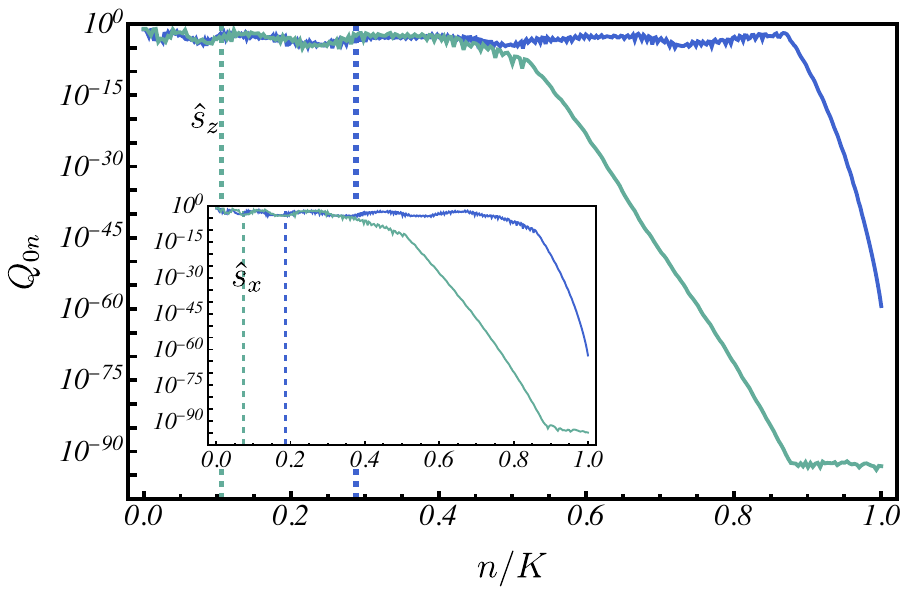}\qquad 
    \includegraphics[width=.375\textwidth]{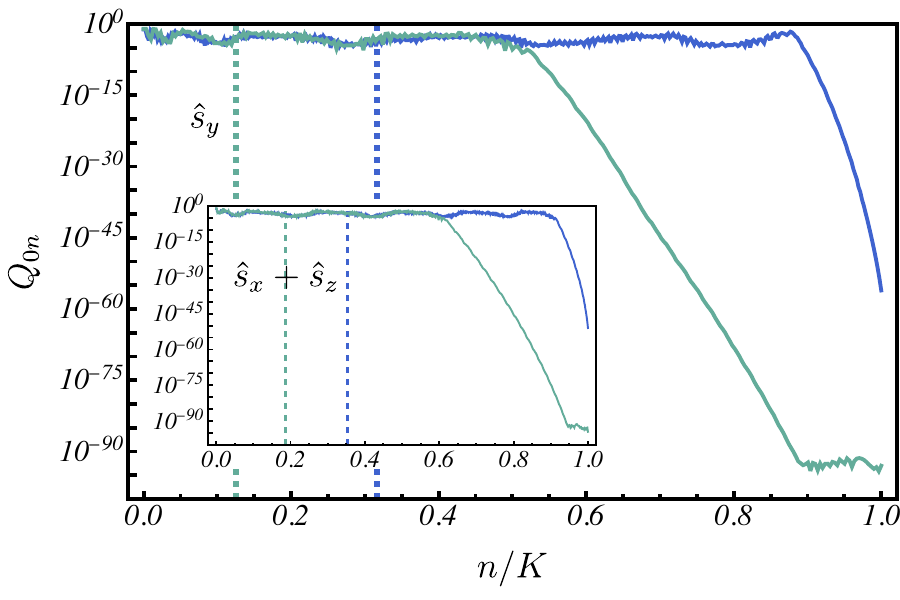}
    \caption{$Q_{0n}$ associated with the data depicted in Figure~\ref{fig:LMG_microcanonical_2}.}
    \label{fig:LMG_microcanonical_Q0n_2}
\end{figure}

\begin{figure}[tp] 
    \centering
    
    \includegraphics[width=.375\textwidth]{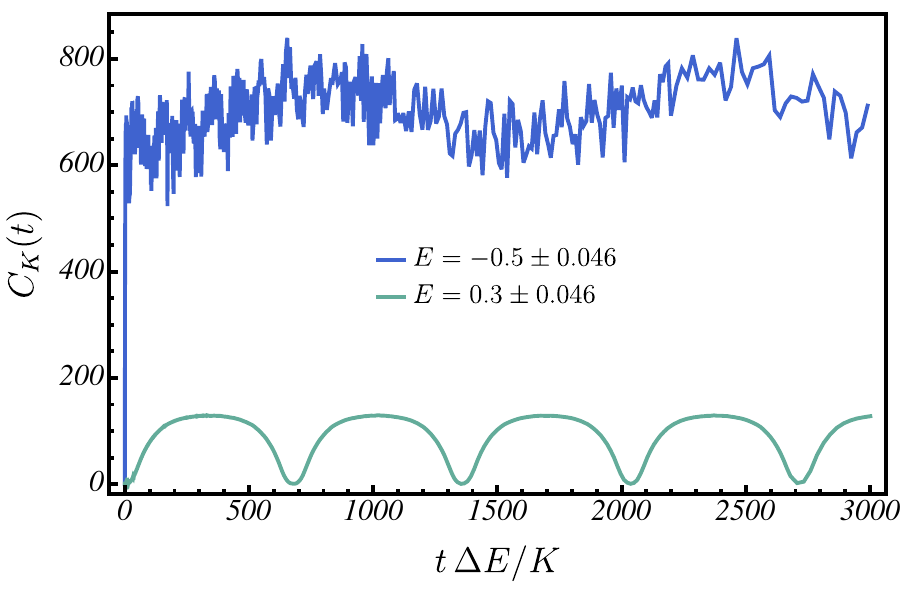}\qquad 
    \includegraphics[width=.375\textwidth]{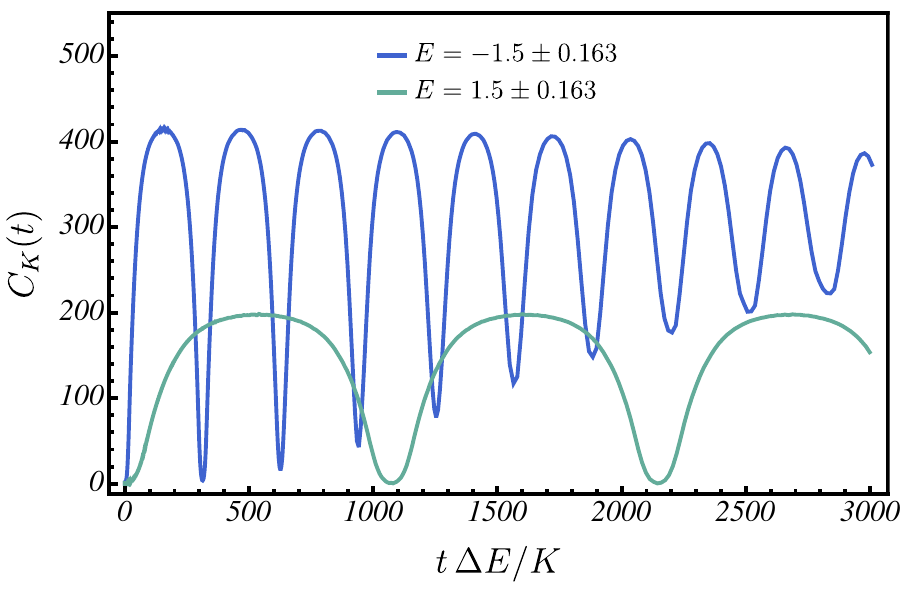}
    \caption{Time evolution of the microcanonical K-complexity (computed using the inner product \eqref{eq:Operator_Inner_product_sym} with the density matrix \eqref{microcanonical_density_matrix}) for the LMG model \eqref{LMG_rescaled_TSS} ($S=300$) with the initial operator $\hat{s}_z$. \textbf{(Left)} Saddle regime with $h=1/2$, $J=1$. The corresponding Krylov dimensions are $K=1440$ in the shell containing the saddle
point and $K=144$ in the other one. \textbf{(Right)} No-saddle regime with $h=2$, $J=1$. The corresponding Krylov dimensions are $K=450$ in the shell $E=-1.5\pm 0.163$ and $K=220$ in the shell $E=1.5\pm 0.163$.}
    \label{fig:LMG_microcanonical_regularized}

    \vspace{1.5cm}

    \includegraphics[width=.375\textwidth]{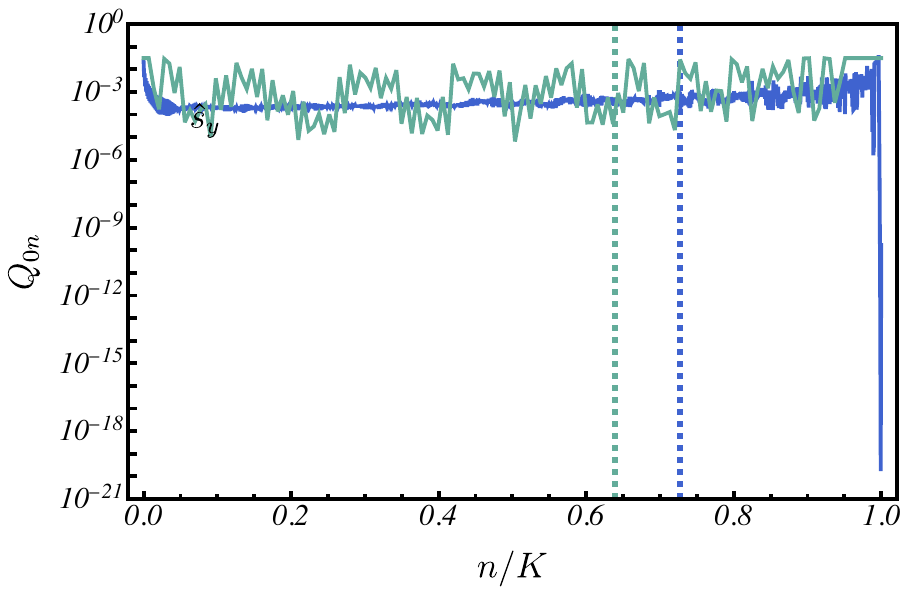}\qquad 
    \includegraphics[width=.375\textwidth]{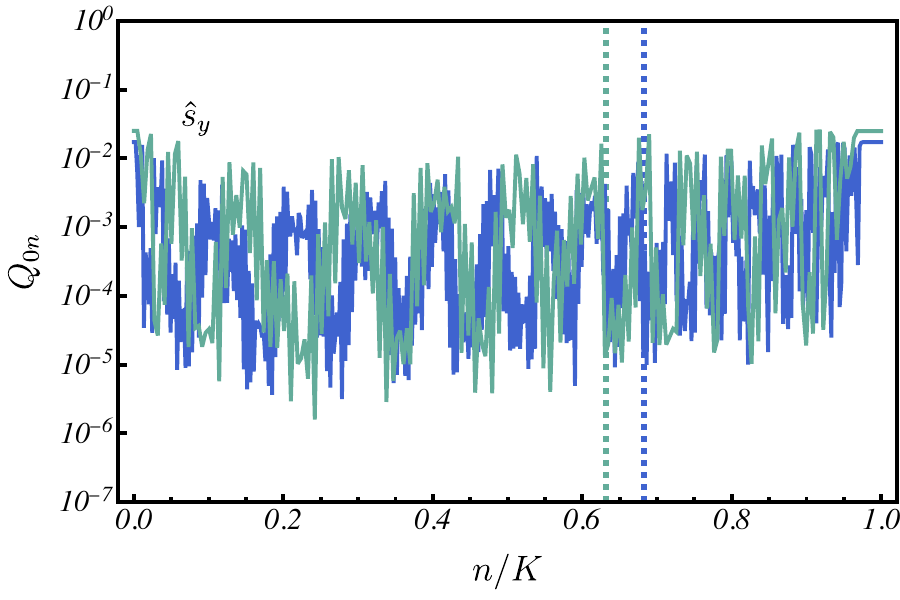}
    \caption{$Q_{0n}$ associated with the data depicted in Figure~\ref{fig:LMG_microcanonical_regularized}.}
    \label{fig:LMG_microcanonical_regularized_Q0n}
\end{figure}

This final appendix contains additional figures relevant to Sections~\ref{sec:late-time_K-complexity} and \ref{sec:quantum_Lanczos_microcanonical}. More specifically, Figures~\ref{fig:Krylov_wavefunctions_LMG}--\ref{fig:Krylov_complexity_FP_appendix} are referenced in Section~\ref{sec:late-time_K-complexity}, while Figures~\ref{fig:LMG_microcanonical_1}--\ref{fig:LMG_microcanonical_regularized_Q0n} are referenced in Section~\ref{sec:quantum_Lanczos_microcanonical}.

Figures \ref{fig:Krylov_wavefunctions_LMG}--\ref{fig:Krylov_complexity_LMG_appendix} present additional material related to the infinite-temperature K-complexity of the LMG model, whereas Figures~\ref{fig:integrability_FP_L6} and \ref{fig:Krylov_complexity_FP_appendix} present the corresponding results for the FP model.

Finally, Figures~\ref{fig:LMG_microcanonical_1}--\ref{fig:LMG_microcanonical_Q0n_2} present additional results for the microcanonical K-complexity of the LMG model obtained using the inner product \eqref{eq:Operator_Inner_product} together with the density matrix defined in \eqref{microcanonical_density_matrix}. Figures~\ref{fig:LMG_microcanonical_regularized} and \ref{fig:LMG_microcanonical_regularized_Q0n} are obtained using the regularized inner product \eqref{eq:Operator_Inner_product_sym} with the same density matrix.


\clearpage

\bibliographystyle{JHEP}
\bibliography{bibli_LMG_FP.bib}

@article{LMG_1,
title = {Validity of many-body approximation methods for a solvable model: {(I)}. {Exact} solutions and perturbation theory},
journal = {Nuclear Physics},
volume = {62},
number = {2},
pages = {188-198},
year = {1965},
issn = {0029-5582},
doi = {https://doi.org/10.1016/0029-5582(65)90862-X},
author = {H.J. Lipkin and N. Meshkov and A.J. Glick}
}

@article{LMG_2,
title = {Validity of many-body approximation methods for a solvable model: {(II)}. {Linearization} procedures},
journal = {Nuclear Physics},
volume = {62},
number = {2},
pages = {199-210},
year = {1965},
issn = {0029-5582},
doi = {https://doi.org/10.1016/0029-5582(65)90863-1},
author = {N. Meshkov and A.J. Glick and H.J. Lipkin}
}

@article{LMG_3,
title = {Validity of many-body approximation methods for a solvable model: {(III)}. {Diagram} summations},
journal = {Nuclear Physics},
volume = {62},
number = {2},
pages = {211-224},
year = {1965},
issn = {0029-5582},
doi = {https://doi.org/10.1016/0029-5582(65)90864-3},
author = {A.J. Glick and H.J. Lipkin and N. Meshkov}
}

@article{LMG_integrable1,
title = {Analytical solutions for the {LMG} model},
author = {Feng Pan and J.P. Draayer},
journal = {Physics Letters B},
volume = {451},
number = {1},
pages = {1-10},
year = {1999},
doi = {https://doi.org/10.1016/S0370-2693(99)00191-4}
}

@article{LMG_integrable2,
title = {Exact solutions for the {LMG} model {Hamiltonian} based on the {Bethe} ansatz},
journal = {Nuclear Physics B},
volume = {737},
number = {3},
pages = {337-350},
year = {2006},
issn = {0550-3213},
doi = {https://doi.org/10.1016/j.nuclphysb.2006.01.015},
url = {https://www.sciencedirect.com/science/article/pii/S0550321306000289},
author = {Hiroyuki Morita and Hiromasa Ohnishi and João {da Providência} and Seiya Nishiyama}
}

@article{LMG_integrable3,
title = {Exactly-solvable models derived from a generalized {Gaudin} algebra},
author = {G. Ortiz and R. Somma and J. Dukelsky and S. Rombouts},
journal = {Nuclear Physics B},
volume = {707},
number = {3},
pages = {421-457},
year = {2005},
doi = {https://doi.org/10.1016/j.nuclphysb.2004.11.008},
}

@article{LMG_integrable4,
doi = {10.1088/1742-6596/492/1/012013},
year = {2014},
month = {3},
publisher = {},
volume = {492},
number = {1},
pages = {012013},
author = {Lerma-H, Sergio and Dukelsky, Jorge},
title = {{The Lipkin-Meshkov-Glick model from the perspective of the SU(1,1) Richardson-Gaudin models}},
journal = {Journal of Physics: Conference Series}
}

@article{FP_1,
title = {Regular and chaotic motion of coupled rotators},
journal = {Physica D: Nonlinear Phenomena},
volume = {9},
number = {3},
pages = {433-438},
year = {1983},
issn = {0167-2789},
doi = {https://doi.org/10.1016/0167-2789(83)90282-8},
author = {Mario Feingold and Asher Peres}
}

@article{FP_2,
  title = {Ergodicity and mixing in quantum theory. {II}},
  author = {Feingold, Mario and Moiseyev, Nimrod and Peres, Asher},
  journal = {Phys. Rev. A},
  volume = {30},
  issue = {1},
  pages = {509--511},
  numpages = {0},
  year = {1984},
  month = {7},
  publisher = {American Physical Society},
  doi = {10.1103/PhysRevA.30.509}
}

@article{FP_modern,
  title = {Quantum chaos for nonstandard symmetry classes in the {Feingold-Peres} model of coupled tops},
  author = {Fan, Yiyun and Gnutzmann, Sven and Liang, Yuqi},
  journal = {Phys. Rev. E},
  volume = {96},
  issue = {6},
  pages = {062207},
  numpages = {12},
  year = {2017},
  month = {12},
  publisher = {American Physical Society},
  doi = {10.1103/PhysRevE.96.062207}
}

@book{Haake,
author = {Haake, Fritz},
title = {Quantum Signatures of Chaos},
year = {2010},
publisher = {Springer Berlin Heidelberg},
address = {Berlin, Heidelberg},
pages = {15--46},
isbn = {978-3-642-05428-0},
doi = {10.1007/978-3-642-05428-0},
}

@article{BT_conjecture,
author = {Berry, Michael Victor  and Tabor, M.  and Ziman, John Michael },
title = {Level clustering in the regular spectrum},
journal = {Proceedings of the Royal Society of London. A. Mathematical and Physical Sciences},
volume = {356},
number = {1686},
pages = {375-394},
year = {1977},
doi = {10.1098/rspa.1977.0140}
}

@article{BGS_conjecture,
  title = {{Characterization of Chaotic Quantum Spectra and Universality of Level Fluctuation Laws}},
  author = {Bohigas, O. and Giannoni, M. J. and Schmit, C.},
  journal = {Phys. Rev. Lett.},
  volume = {52},
  issue = {1},
  pages = {1--4},
  numpages = {0},
  year = {1984},
  month = {1},
  publisher = {American Physical Society},
  doi = {10.1103/PhysRevLett.52.1}
}

@article{r_ratio_Huse,
  title = {Localization of interacting fermions at high temperature},
  author = {Oganesyan, Vadim and Huse, David A.},
  journal = {Phys. Rev. B},
  volume = {75},
  issue = {15},
  pages = {155111},
  numpages = {5},
  year = {2007},
  month = {4},
  publisher = {American Physical Society},
  doi = {10.1103/PhysRevB.75.155111}
}

@article{RMT_math,
  title = {{Distribution of the Ratio of Consecutive Level Spacings in Random Matrix Ensembles}},
  author = {Atas, Y. Y. and Bogomolny, E. and Giraud, O. and Roux, G.},
  journal = {Phys. Rev. Lett.},
  volume = {110},
  issue = {8},
  pages = {084101},
  numpages = {5},
  year = {2013},
  month = {2},
  publisher = {American Physical Society},
  doi = {10.1103/PhysRevLett.110.084101}
}

@article{Universal_Growth_Hypothesis,
  title = {{A Universal Operator Growth Hypothesis}},
  author = {Parker, Daniel E. and Cao, Xiangyu and Avdoshkin, Alexander and Scaffidi, Thomas and Altman, Ehud},
  journal = {Phys. Rev. X},
  volume = {9},
  issue = {4},
  pages = {041017},
  numpages = {29},
  year = {2019},
  month = {10},
  publisher = {American Physical Society},
  doi = {10.1103/PhysRevX.9.041017}
}

@ARTICLE{Adrian_1,
  title = {Operator complexity: a journey to the edge of {Krylov} space},
  author = {Rabinovici, E and S{\'a}nchez-Garrido, A and Shir, R and Sonner, J},
  journal = {Journal of High Energy Physics},
  volume = {2021},
  number = {6},
  pages = {62},
  month = {6},
  year = {2021},
  doi = {10.1007/JHEP06(2021)062}
}

@ARTICLE{Adrian_2,
  title = {{Krylov} localization and suppression of complexity},
  author = {Rabinovici, E and S{\'a}nchez-Garrido, A and Shir, R and Sonner,
              J},
  journal = {Journal of High Energy Physics},
  volume = {2022},
  number = {3},
  pages = {211},
  month = {3},
  year = {2022},
  doi = {10.1007/JHEP03(2022)211}
}

@article{Lanczos_PRO,
  title={{The Lanczos algorithm with partial reorthogonalization}},
  author={Horst D. Simon},
  journal={Mathematics of Computation},
  year={1984},
  volume={42},
  pages={115-142},
  doi={10.1090/S0025-5718-1984-0725988-X}
}

@article{Alessio,
title = {Out-of-equilibrium dynamics of quantum many-body systems with long-range interactions},
journal = {Physics Reports},
volume = {1074},
pages = {1-92},
year = {2024},
note = {Out-of-equilibrium dynamics of quantum many-body systems with long-range interactions},
issn = {0370-1573},
doi = {https://doi.org/10.1016/j.physrep.2024.04.005},
author = {Nicolò Defenu and Alessio Lerose and Silvia Pappalardi}
}

@article{Duminil_Copin_semiclassical,
  title = {Dimerization and {N}{\'e}el {Order in Different Quantum Spin Chains
              Through a Shared Loop Representation}},
  author = {Aizenman, Michael and Duminil-Copin, Hugo and Warzel, Simone},
  journal = {Annales Henri Poincar{\'e}},
  volume =  {21},
  number =  {8},
  pages = {2737--2774},
  month =  {8},
  year =  {2020},
  doi = {10.1007/s00023-020-00924-2}
}

@article{saddle_Krylov,
  title    = "{Krylov} complexity in saddle-dominated scrambling",
  author   = "Bhattacharjee, Budhaditya and Cao, Xiangyu and Nandy, Pratik and
              Pathak, Tanay",
  journal  = "Journal of High Energy Physics",
  volume   =  2022,
  number   =  5,
  pages    = "174",
  month    =  may,
  year     =  2022,
  doi = {10.1007/JHEP05(2022)174}
}

@article{microcanonical_Krylov,
title    = "Random matrix theory for complexity growth and black hole
              interiors",
  author   = "Kar, Arjun and Lamprou, Lampros and Rozali, Moshe and Sully,
              James",
  journal  = "Journal of High Energy Physics",
  volume   =  2022,
  number   =  1,
  pages    = "16",
  month    =  jan,
  year     =  2022,
  doi = {10.1007/JHEP01(2022)016}
}

@article{Adrian_3,
  title = {{Krylov} complexity from integrability to chaos},
  author = {Rabinovici, E and S{\'a}nchez-Garrido, A and Shir, R and Sonner,
              J},
  journal = {Journal of High Energy Physics},
  volume =  {2022},
  number =  {7},
  pages = {151},
  month =  {jul},
  year =  {2022},
  doi = {10.1007/JHEP07(2022)151}
}

@article{Escape_Krylov_space,
  title = {Escaping the {Krylov} space during the finite-precision {Lanczos} algorithm},
  author = {Eckseler, Jannis and Pieper, Max and Schnack, J\"urgen},
  journal = {Phys. Rev. E},
  volume = {112},
  issue = {2},
  pages = {025306},
  numpages = {7},
  year = {2025},
  month = {8},
  publisher = {American Physical Society},
  doi = {10.1103/k94p-vls8}
}

@article{QPT_LMG_1,
  title = {Size {Scaling for Infinitely Coordinated Systems}},
  author = {Botet, R. and Jullien, R. and Pfeuty, P.},
  journal = {Phys. Rev. Lett.},
  volume = {49},
  issue = {7},
  pages = {478--481},
  numpages = {0},
  year = {1982},
  month = {8},
  publisher = {American Physical Society},
  doi = {10.1103/PhysRevLett.49.478}
}

@article{QPT_LMG_2,
  title = {Large-size critical behavior of infinitely coordinated systems},
  author = {Botet, R. and Jullien, R.},
  journal = {Phys. Rev. B},
  volume = {28},
  issue = {7},
  pages = {3955--3967},
  numpages = {0},
  year = {1983},
  month = {10},
  publisher = {American Physical Society},
  doi = {10.1103/PhysRevB.28.3955}
}

@article{QPT_LMG_3,
  title = {{Finite-Size Scaling Exponents of the Lipkin-Meshkov-Glick Model}},
  author = {Dusuel, S\'ebastien and Vidal, Julien},
  journal = {Phys. Rev. Lett.},
  volume = {93},
  issue = {23},
  pages = {237204},
  numpages = {4},
  year = {2004},
  month = {12},
  publisher = {American Physical Society},
  doi = {10.1103/PhysRevLett.93.237204}
}

@article{QPT_LMG_4,
  title = {{Continuous unitary transformations and finite-size scaling exponents in the Lipkin-Meshkov-Glick model}},
  author = {Dusuel, S\'ebastien and Vidal, Julien},
  journal = {Phys. Rev. B},
  volume = {71},
  issue = {22},
  pages = {224420},
  numpages = {18},
  year = {2005},
  month = {6},
  publisher = {American Physical Society},
  doi = {10.1103/PhysRevB.71.224420}
}

@article{ESQPT_LMG,
  title = {Excited-state quantum phase transitions in many-body systems with infinite-range interaction: Localization, dynamics, and bifurcation},
  author = {Santos, Lea F. and T\'avora, Marco and P\'erez-Bernal, Francisco},
  journal = {Phys. Rev. A},
  volume = {94},
  issue = {1},
  pages = {012113},
  numpages = {12},
  year = {2016},
  month = {7},
  publisher = {American Physical Society},
  doi = {10.1103/PhysRevA.94.012113}
}

@article{ESQPT_review,
doi = {10.1088/1751-8121/abdfe8},
url = {https://doi.org/10.1088/1751-8121/abdfe8},
year = {2021},
month = {3},
publisher = {IOP Publishing},
volume = {54},
number = {13},
pages = {133001},
author = {Cejnar, Pavel and Stránský, Pavel and Macek, Michal and Kloc, Michal},
title = {Excited-state quantum phase transitions},
journal = {Journal of Physics A: Mathematical and Theoretical}
}

@article{Coupled_tops_1,
  title = {Quantum entanglement and fixed-point bifurcations},
  author = {Hines, Andrew P. and McKenzie, Ross H. and Milburn, G. J.},
  journal = {Phys. Rev. A},
  volume = {71},
  issue = {4},
  pages = {042303},
  numpages = {9},
  year = {2005},
  month = {4},
  publisher = {American Physical Society},
  doi = {10.1103/PhysRevA.71.042303}
}

@article{Coupled_tops_2,
  title = {Chaos and quantum scars in a coupled top model},
  author = {Mondal, Debabrata and Sinha, Sudip and Sinha, S.},
  journal = {Phys. Rev. E},
  volume = {102},
  issue = {2},
  pages = {020101},
  numpages = {6},
  year = {2020},
  month = {8},
  publisher = {American Physical Society},
  doi = {10.1103/PhysRevE.102.020101}
}

@Article{Coupled_tops_QPT,
author = {Mao, Wen-Jian and Ye, Tian and Duan, Liwei and Wang, Yan-Zhi},
title = {{Quantum Phase Transition in the Coupled-Top Model: From $Z_2$ to U(1) Symmetry Breaking}},
journal = {Entropy},
volume = {27},
year = {2025},
number = {5},
article-number = {474},
PubMedID = {40422429},
ISSN = {1099-4300},
doi = {10.3390/e27050474}
}

@article{Coupled_tops_classical_1,
  title = {Chaos in a two-spin system with applied magnetic field},
  author = {Robb, Daniel T. and Reichl, L. E.},
  journal = {Phys. Rev. E},
  volume = {57},
  issue = {2},
  pages = {2458--2459},
  numpages = {0},
  year = {1998},
  month = {2},
  publisher = {American Physical Society},
  doi = {10.1103/PhysRevE.57.2458}
}

@article{Coupled_tops_classical_2,
  author  = {B. Skellett and C. A. Holmes},
  title   = {InterJournal Complex Systems},
  journal = {InterJournal, Complex Systems},
  volume  = {518},
  year    = {2002}
}

@article{ESQPT_FP,
  title = {Signatures of excited-state quantum phase transitions in quantum many-body systems: Phase space analysis},
  author = {Wang, Qian and P\'erez-Bernal, Francisco},
  journal = {Phys. Rev. E},
  volume = {104},
  issue = {3},
  pages = {034119},
  numpages = {15},
  year = {2021},
  month = {9},
  publisher = {American Physical Society},
  doi = {10.1103/PhysRevE.104.034119}
}

@article{LMG_entaglement,
  title = {Entanglement dynamics in the {Lipkin-Meshkov-Glick} model},
  author = {Vidal, Julien and Palacios, Guillaume and Aslangul, Claude},
  journal = {Phys. Rev. A},
  volume = {70},
  issue = {6},
  pages = {062304},
  numpages = {10},
  year = {2004},
  month = {12},
  publisher = {American Physical Society},
  doi = {10.1103/PhysRevA.70.062304},
  url = {https://link.aps.org/doi/10.1103/PhysRevA.70.062304}
}

@article{frequency_Heiss,
  title = {{Large-$N$ Scaling Behavior of the Lipkin-Meshkov-Glick Model}},
  author = {Leyvraz, F. and Heiss, W. D.},
  journal = {Phys. Rev. Lett.},
  volume = {95},
  issue = {5},
  pages = {050402},
  numpages = {4},
  year = {2005},
  month = {7},
  publisher = {American Physical Society},
  doi = {10.1103/PhysRevLett.95.050402}
}

@article{Richardson_Gaudin,
  title = {{Colloquium: Exactly solvable Richardson-Gaudin models for many-body quantum systems}},
  author = {Dukelsky, J. and Pittel, S. and Sierra, G.},
  journal = {Rev. Mod. Phys.},
  volume = {76},
  issue = {3},
  pages = {643--662},
  numpages = {0},
  year = {2004},
  month = {8},
  publisher = {American Physical Society},
  doi = {10.1103/RevModPhys.76.643}
}

@article{Integrability_Husimi,
doi = {10.1088/0305-4470/23/10/017},
url = {https://doi.org/10.1088/0305-4470/23/10/017},
year = {1990},
month = {5},
publisher = {},
volume = {23},
number = {10},
pages = {1765},
author = {P Leboeuf and A Voros},
title = {Chaos-revealing multiplicative representation of quantum eigenstates},
journal = {Journal of Physics A: Mathematical and General}
}

@article{LMG_deformed_polynomials,
doi = {10.1088/0305-4470/34/15/305},
url = {https://doi.org/10.1088/0305-4470/34/15/305},
year = {2001},
month = {4},
publisher = {},
volume = {34},
number = {15},
pages = {3265},
author = {N Debergh and Fl Stancu},
title = {{On the exact solutions of the Lipkin-Meshkov-Glick model}},
journal = {Journal of Physics A: Mathematical and General}
}

@article{Level_spacings_harmonic,
  title = {Level spacings for harmonic-oscillator systems},
  author = {Pandey, Akhilesh and Ramaswamy, Ramakrishna},
  journal = {Phys. Rev. A},
  volume = {43},
  issue = {8},
  pages = {4237--4243},
  numpages = {0},
  year = {1991},
  month = {4},
  publisher = {American Physical Society},
  doi = {10.1103/PhysRevA.43.4237},
  url = {https://link.aps.org/doi/10.1103/PhysRevA.43.4237}
}

@article{Holstein_Primakoff,
  title = {{Field Dependence of the Intrinsic Domain Magnetization of a Ferromagnet}},
  author = {Holstein, T. and Primakoff, H.},
  journal = {Phys. Rev.},
  volume = {58},
  issue = {12},
  pages = {1098--1113},
  numpages = {0},
  year = {1940},
  month = {12},
  publisher = {American Physical Society},
  doi = {10.1103/PhysRev.58.1098}
}

@Book{Recursion_method_book,
  author = {V. S. Viswanath and Gerhard M{\"u}ller},
  title = {The Recursion Method: Application to Many-Body Dynamics},
  publisher = {Springer},
  address = {Berlin, Heidelberg},
  year = {1994},
  series = {Lecture Notes in Physics Monographs},
  volume = {23},
  isbn = {3-540-58319-X},
  doi = {10.1007/978-3-540-48651-0}
}

@article{Lanczos_algorithm_historical,
    author = {Lanczos, Cornelius},
    title = {An iteration method for the solution of the eigenvalue problem of linear differential and integral operators},
    doi = {10.6028/jres.045.026},
    journal = {J. Res. Natl. Bur. Stand. B},
    volume = {45},
    pages = {255--282},
    year = {1950}
}

@ARTICLE{Krylov_SU2,
  title = {Quasinormal modes and complexity in saddle-dominated {SU(N}) spin
              systems},
  author = {Aguilar-Gutierrez, Sergio E and Fu, Yichao and Pal, Kuntal and
              Parmentier, Klaas},
  journal = {Journal of High Energy Physics},
  volume =  {2025},
  number =  {9},
  pages = {39},
  month =  {9},
  year =  {2025},
  doi = {10.1007/JHEP09(2025)039}
}

@article{Surprise_LMG,
  title = {Surprises in the deep {Hilbert} space of all-to-all systems: From superexponential scrambling to slow entanglement growth},
  author = {Qi, Zihao and Scaffidi, Thomas and Cao, Xiangyu},
  journal = {Phys. Rev. B},
  volume = {108},
  issue = {5},
  pages = {054301},
  numpages = {22},
  year = {2023},
  month = {Aug},
  publisher = {American Physical Society},
  doi = {10.1103/PhysRevB.108.054301},
  url = {https://link.aps.org/doi/10.1103/PhysRevB.108.054301}
}

@article{spherical_harmonics_Poisson,
    author = {{Freidel}, Laurent and {Krasnov}, Kirill},
    title = "{The fuzzy sphere {\ensuremath{\star}}-product and spin networks}",
    journal = {Journal of Mathematical Physics},
    year = {2002},
    month = {4},
    volume = {43},
    number = {4},
    pages = {1737-1754},
    doi = {10.1063/1.1456255}
}

@book{Arnold,
author = {Arnold, V. I.},
title = {Mathematical Methods of Classical Mechanics},
year = {1989},
publisher = {Springer New York},
address = {New York, NY},
doi = {10.1007/978-1-4757-2063-1}
}

@article{Silvia_LMG,
  title = {Scrambling and entanglement spreading in long-range spin chains},
  author = {Pappalardi, Silvia and Russomanno, Angelo and \ifmmode \check{Z}\else \v{Z}\fi{}unkovi\ifmmode \check{c}\else \v{c}\fi{}, Bojan and Iemini, Fernando and Silva, Alessandro and Fazio, Rosario},
  journal = {Phys. Rev. B},
  volume = {98},
  issue = {13},
  pages = {134303},
  numpages = {11},
  year = {2018},
  month = {Oct},
  publisher = {American Physical Society},
  doi = {10.1103/PhysRevB.98.134303},
  url = {https://link.aps.org/doi/10.1103/PhysRevB.98.134303}
}

@article{Scrambling_chaos,
  title = {{Does Scrambling Equal Chaos?}},
  author = {Xu, Tianrui and Scaffidi, Thomas and Cao, Xiangyu},
  journal = {Phys. Rev. Lett.},
  volume = {124},
  issue = {14},
  pages = {140602},
  numpages = {7},
  year = {2020},
  month = {Apr},
  publisher = {American Physical Society},
  doi = {10.1103/PhysRevLett.124.140602}
}

@article{PhysRevLett.70.3852,
  title = {Spectral density of the {QCD Dirac} operator near zero virtuality},
  author = {Verbaarschot, J. J. M. and Zahed, I.},
  journal = {Phys. Rev. Lett.},
  volume = {70},
  issue = {25},
  pages = {3852--3855},
  numpages = {0},
  year = {1993},
  month = {Jun},
  publisher = {American Physical Society},
  doi = {10.1103/PhysRevLett.70.3852},
  url = {https://link.aps.org/doi/10.1103/PhysRevLett.70.3852}
}

@article{PhysRevLett.72.2531,
  title = {Spectrum of the {QCD Dirac} operator and chiral random matrix theory},
  author = {Verbaarschot, Jacobus},
  journal = {Phys. Rev. Lett.},
  volume = {72},
  issue = {16},
  pages = {2531--2533},
  numpages = {0},
  year = {1994},
  month = {Apr},
  publisher = {American Physical Society},
  doi = {10.1103/PhysRevLett.72.2531},
  url = {https://link.aps.org/doi/10.1103/PhysRevLett.72.2531}
}

@article{PhysRevLett.76.3420,
  title = {{Random Matrix Theory of a Chaotic Andreev Quantum Dot}},
  author = {Altland, Alexander and Zirnbauer, Martin R.},
  journal = {Phys. Rev. Lett.},
  volume = {76},
  issue = {18},
  pages = {3420--3423},
  numpages = {0},
  year = {1996},
  month = {Apr},
  publisher = {American Physical Society},
  doi = {10.1103/PhysRevLett.76.3420},
  url = {https://link.aps.org/doi/10.1103/PhysRevLett.76.3420}
}

@article{Altland_nonstandard_symmetry,
  title = {Nonstandard symmetry classes in mesoscopic normal-superconducting hybrid structures},
  author = {Altland, Alexander and Zirnbauer, Martin R.},
  journal = {Phys. Rev. B},
  volume = {55},
  issue = {2},
  pages = {1142--1161},
  numpages = {0},
  year = {1997},
  month = {Jan},
  publisher = {American Physical Society},
  doi = {10.1103/PhysRevB.55.1142},
  url = {https://link.aps.org/doi/10.1103/PhysRevB.55.1142}
}

@article{Zirnbauer:1996zz,
    title = {Quantum complexity in gravity, quantum field theory, and quantum information science},
    journal = {Physics Reports},
    volume = {1159},
    pages = {1-77},
    year = {2026},
    note = {Quantum complexity in gravity, quantum field theory, and quantum information science},
    issn = {0370-1573},
    doi = {10.1016/j.physrep.2025.11.001},
    author = {Stefano Baiguera and Vijay Balasubramanian and Pawel Caputa and Shira Chapman and Jonas Haferkamp and Michal P. Heller and Nicole Yunger Halpern}
}

@book{spherical_harmonics_relations,
  author = {D. A. Varshalovich and A. N. Moskalev and V. K. Khersonskii},
  title = {Quantum Theory of Angular Momentum: Irreducible Tensors, Spherical Harmonics, Vector Coupling Coefficients, 3nj Symbols},
  publisher = {World Scientific},
  year = {1988},
  address = {Singapore},
  pages = {524},
  doi = {10.1142/0270},
  isbn = {978-9971-5-0107-5}
}

@article{Rabinovici:2025otw,
    author = "Rabinovici, Eliezer and S{\'a}nchez-Garrido, Adri{\'a}n and Shir, Ruth and Sonner, Julian",
    title = "{{Krylov} Complexity}",
    eprint = "2507.06286",
    archivePrefix = "arXiv",
    primaryClass = "hep-th",
    reportNumber = "CERN-TH-2025-128",
    month = "7",
    year = "2025",
    url = "https://arxiv.org/abs/2507.06286"
}

@article{Baiguera:2025dkc,
    title = {Quantum complexity in gravity, quantum field theory, and quantum information science},
    journal = {Physics Reports},
    volume = {1159},
    pages = {1-77},
    year = {2026},
    note = {Quantum complexity in gravity, quantum field theory, and quantum information science},
    issn = {0370-1573},
    doi = {https://doi.org/10.1016/j.physrep.2025.11.001},
    url = {https://www.sciencedirect.com/science/article/pii/S0370157325003023},
    author = {Stefano Baiguera and Vijay Balasubramanian and Pawel Caputa and Shira Chapman and Jonas Haferkamp and Michal P. Heller and Nicole Yunger Halpern},
}

@article{Nandy:2024evd,
title = {Quantum dynamics in {Krylov} space: Methods and applications},
journal = {Physics Reports},
volume = {1125-1128},
pages = {1-82},
year = {2025},
note = {Quantum dynamics in {Krylov} space: Methods and applications},
issn = {0370-1573},
doi = {https://doi.org/10.1016/j.physrep.2025.05.001},
url = {https://www.sciencedirect.com/science/article/pii/S0370157325001462},
author = {Pratik Nandy and Apollonas S. Matsoukas-Roubeas and Pablo Martínez-Azcona and Anatoly Dymarsky and Adolfo {del Campo}},
}

@article{Balasubramanian:2022tpr,
title = {Quantum chaos and the complexity of spread of states},
  author = {Balasubramanian, Vijay and Caputa, Pawel and Magan, Javier M. and Wu, Qingyue},
  journal = {Phys. Rev. D},
  volume = {106},
  issue = {4},
  pages = {046007},
  numpages = {28},
  year = {2022},
  month = {Aug},
  publisher = {American Physical Society},
  doi = {10.1103/PhysRevD.106.046007}
}

@article{Caputa:2024vrn,
title    = "Krylov complexity of density matrix operators",
  author   = "Caputa, Pawel and Jeong, Hyun-Sik and Liu, Sinong and Pedraza,
              Juan F and Qu, Le-Chen",
  journal  = "Journal of High Energy Physics",
  volume   =  2024,
  number   =  5,
  pages    = "337",
  month    =  may,
  year     =  2024,
  doi = {10.1007/JHEP05(2024)337}
}

@article{Krylov:1931,
    author = "A. N. {Krylov}",
    title = "{On the numerical solution of the equation determining the frequencies of small oscillations of material systems in applied mechanics questions.}",
    journal = "Proceedings of the Academy of Sciences of the USSR. VII series. Department of Mathematical and Natural Sciences",
    volume = "4",
    pages = "491--539",
    year = "1931",
    note = "(In Russian. Zentralblatt MATH number: 2561215, JFM: 57.1454.02)",
    zbMATH = {2561215},
    JFM = {57.1454.02}
}

@book{full_ortho_algorith,
author = {Parlett, Beresford N.},
title = {The Symmetric Eigenvalue Problem},
publisher = {Society for Industrial and Applied Mathematics},
year = {1998},
doi = {10.1137/1.9781611971163},
address = {},
edition   = {}
}

@article{Polkovnikov_phase_space_QM,
title = {Phase space representation of quantum dynamics},
journal = {Annals of Physics},
volume = {325},
number = {8},
pages = {1790-1852},
year = {2010},
issn = {0003-4916},
doi = {https://doi.org/10.1016/j.aop.2010.02.006},
author = {Anatoli Polkovnikov}
}

@article{ESQPT_general,
title = {Excited state quantum phase transitions in many-body systems},
journal = {Annals of Physics},
volume = {323},
number = {5},
pages = {1106-1135},
year = {2008},
issn = {0003-4916},
doi = {https://doi.org/10.1016/j.aop.2007.06.011},
url = {https://www.sciencedirect.com/science/article/pii/S0003491607001042},
author = {M.A. Caprio and P. Cejnar and F. Iachello},
}

@article{sigma_localization,
doi = {10.1088/0022-3719/10/6/003},
year = {1977},
month = {mar},
publisher = {},
volume = {10},
number = {6},
pages = {L125},
author = {L Fleishman and D C Licciardello},
title = {Fluctuations and localization in one dimension},
journal = {Journal of Physics C: Solid State Physics}
}

@article{barbon2019,
  title = {On the evolution of operator complexity beyond scrambling},
  author = {J.L.F. Barbón and E. Rabinovici and R. Shir and R. Sinha},
  journal = {Journal of High Energy Physics},
  year = {2019},
  volume = {2019},
  number = {10},
  pages = {264},
  isbn = {1029-8479},
  doi = {10.1007/JHEP10(2019)264},
  url = {https://doi.org/10.1007/JHEP10(2019)264}
}

@article{PhysRevA.30.504,
  title = {Ergodicity and mixing in quantum theory. {I}},
  author = {Peres, Asher},
  journal = {Phys. Rev. A},
  volume = {30},
  issue = {1},
  pages = {504--508},
  numpages = {0},
  year = {1984},
  month = {Jul},
  publisher = {American Physical Society},
  doi = {10.1103/PhysRevA.30.504},
  url = {https://link.aps.org/doi/10.1103/PhysRevA.30.504}
}

@article{PhysRevA.34.591,
  title = {Distribution of matrix elements of chaotic systems},
  author = {Feingold, Mario and Peres, Asher},
  journal = {Phys. Rev. A},
  volume = {34},
  issue = {1},
  pages = {591--595},
  numpages = {0},
  year = {1986},
  month = {Jul},
  publisher = {American Physical Society},
  doi = {10.1103/PhysRevA.34.591},
  url = {https://link.aps.org/doi/10.1103/PhysRevA.34.591}
}

@article{Basu:2024tgg,
  title    = "Complexity growth and the {Krylov-Wigner} function",
  author   = "Basu, Ritam and Ganguly, Anirban and Nath, Souparna and Parrikar,
              Onkar",
  journal  = "Journal of High Energy Physics",
  volume   =  2024,
  number   =  5,
  pages    = "264",
  month    =  may,
  year     =  2024,
  doi = {10.1007/JHEP05(2024)264}
}

@article{Basu:2025mmm,
  title    = "Wigner negativity, random matrices and gravity",
  author   = "Basu, Ritam and Chowdhury, Pratyusha and Ganguly, Anirban and
              Nath, Souparna and Parrikar, Onkar and Paul, Suprakash",
  journal  = "Journal of High Energy Physics",
  volume   =  2026,
  number   =  1,
  pages    = "106",
  month    =  jan,
  year     =  2026,
  doi = {10.1007/JHEP01(2026)106}
}

@article{Maldacena:2015waa,
  title    = "A bound on chaos",
  author   = "Maldacena, Juan and Shenker, Stephen H and Stanford, Douglas",
  journal  = "Journal of High Energy Physics",
  volume   =  2016,
  number   =  8,
  pages    = "106",
  month    =  aug,
  year     =  2016,
  doi = {10.1007/JHEP08(2016)106}
}

@article{Dymarsky:2021bjq,
  title = {Krylov complexity in conformal field theory},
  author = {Dymarsky, Anatoly and Smolkin, Michael},
  journal = {Phys. Rev. D},
  volume = {104},
  issue = {8},
  pages = {L081702},
  numpages = {5},
  year = {2021},
  month = {Oct},
  publisher = {American Physical Society},
  doi = {10.1103/PhysRevD.104.L081702},
}

@article{Avdoshkin:2022xuw,
  title    = "Krylov complexity in quantum field theory, and beyond",
  author   = "Avdoshkin, Alexander and Dymarsky, Anatoly and Smolkin, Michael",
  journal  = "Journal of High Energy Physics",
  volume   =  2024,
  number   =  6,
  pages    = "66",
  month    =  jun,
  year     =  2024,
  doi = {10.1007/JHEP06(2024)066}
}

@article{Camargo:2022rnt,
    title    = "{Krylov} complexity in free and interacting scalar field theories
              with bounded power spectrum",
  author   = "Camargo, Hugo A and Jahnke, Viktor and Kim, Keun-Young and
              Nishida, Mitsuhiro",
  journal  = "Journal of High Energy Physics",
  volume   =  2023,
  number   =  5,
  pages    = "226",
  month    =  may,
  year     =  2023,
  doi = {10.1007/JHEP05(2023)226}
}

@article{Kundu:2023hbk,
 title    = "{State dependence of Krylov complexity in 2d CFTs}",
  author   = "Kundu, Arnab and Malvimat, Vinay and Sinha, Ritam",
  journal  = "Journal of High Energy Physics",
  volume   =  2023,
  number   =  9,
  pages    = "11",
  month    =  sep,
  year     =  2023,
  doi = {10.1007/JHEP09(2023)011}
}

@article{Altland:2020ccq,
	title = {Late time physics of holographic quantum chaos},
	pages = {034},
	author = {Altland, Alexander and Sonner, Julian},
	journal = {SciPost Phys.},
	volume = {11},
	year = {2021},
	publisher = {SciPost},
	doi = {10.21468/SciPostPhys.11.2.034}
}

@article{Altland:2021rqn,
 title = {From operator statistics to wormholes},
  author = {Altland, Alexander and Bagrets, Dmitry and Nayak, Pranjal and Sonner, Julian and Vielma, Manuel},
  journal = {Phys. Rev. Res.},
  volume = {3},
  issue = {3},
  pages = {033259},
  numpages = {20},
  year = {2021},
  month = {Sep},
  publisher = {American Physical Society},
  doi = {10.1103/PhysRevResearch.3.033259}
}

@book{Marino:2021lne,
    author = "Mari{\~n}o, Marcos",
    title = "{Advanced Topics in Quantum Mechanics}",
    doi = "10.1017/9781108863384",
    isbn = "978-1-108-86338-4, 978-1-108-49587-5",
    publisher = "Cambridge University Press",
    month = "12",
    year = "2021"
}

@article{Rundle:2021sku,
author = {Rundle, Russell P. and Everitt, Mark J.},
title = {Overview of the Phase Space Formulation of Quantum Mechanics with Application to Quantum Technologies},
journal = {Advanced Quantum Technologies},
volume = {4},
number = {6},
pages = {2100016},
doi = {10.1002/qute.202100016},
year = {2021}
}

@article{Stratonovich1957,
  author  = {Stratonovich, R. L.},
  title   = {{On Distributions in Representation Space}},
  journal = {Sov. Phys. JETP},
  year    = {1957},
  volume  = {4},
  pages   = {891--898},
  note    = {[Zh. Eksp. Teor. Fiz. \textbf{31}, 1012 (1956)]},
  url     = {https://jetp.ras.ru/cgi-bin/e/index/e/4/6/p891?a=list}
}

@article{VARILLY1989107,
title = {The moyal representation for spin},
journal = {Annals of Physics},
volume = {190},
number = {1},
pages = {107-148},
year = {1989},
issn = {0003-4916},
doi = {https://doi.org/10.1016/0003-4916(89)90262-5},
url = {https://www.sciencedirect.com/science/article/pii/0003491689902625},
author = {Joseph C Várilly and JoséM Gracia-Bondía}
}

@article{Amiet_1991,
doi = {10.1088/0305-4470/24/7/023},
url = {https://doi.org/10.1088/0305-4470/24/7/023},
year = {1991},
month = {apr},
publisher = {},
volume = {24},
number = {7},
pages = {1515},
author = {J -P Amiet and M B Cibils},
title = {Description of quantum spin using functions on the sphere ${S}^2$},
journal = {Journal of Physics A: Mathematical and General}
}

@article{Amiet:2000rju,
    title = {{Contracting the Wigner kernel of a spin to the Wigner kernel of a particle}},
  author = {Amiet, Jean-Pierre and Weigert, Stefan},
  journal = {Phys. Rev. A},
  volume = {63},
  issue = {1},
  pages = {012102},
  numpages = {5},
  year = {2000},
  month = {Dec},
  publisher = {American Physical Society},
  doi = {10.1103/PhysRevA.63.012102}
}

@article{PhysRevA.63.012105,
  title = {{Discrete Moyal-type representations for a spin}},
  author = {Heiss, Stephan and Weigert, Stefan},
  journal = {Phys. Rev. A},
  volume = {63},
  issue = {1},
  pages = {012105},
  numpages = {11},
  year = {2000},
  month = {Dec},
  publisher = {American Physical Society},
  doi = {10.1103/PhysRevA.63.012105},
  url = {https://link.aps.org/doi/10.1103/PhysRevA.63.012105}
}

@article{Klimov_2002,
doi = {10.1088/0305-4470/35/40/305},
url = {https://doi.org/10.1088/0305-4470/35/40/305},
year = {2002},
month = {sep},
publisher = {},
volume = {35},
number = {40},
pages = {8435},
author = {A B Klimov and P Espinoza},
title = {{Moyal-like form of the
star product for generalized SU(2) Stratonovich-Weyl symbols}},
journal = {Journal of Physics A: Mathematical and General}
}

@ARTICLE{Shepelyansky:2020,
AUTHOR = {Shepelyansky, D. },
TITLE   = {{E}hrenfest time and chaos},
YEAR    = {2020},
JOURNAL = {Scholarpedia},
VOLUME  = {15},
NUMBER  = {9},
PAGES   = {55031},
DOI     = {10.4249/scholarpedia.55031},
NOTE    = {revision \#197901}
}

@article{PhysRevE.66.016217,
  title = {Universal relationship between a quantum phase transition and instability points of classical systems},
  author = {Heiss, W. D. and M\"uller, M.},
  journal = {Phys. Rev. E},
  volume = {66},
  issue = {1},
  pages = {016217},
  numpages = {5},
  year = {2002},
  month = {Jul},
  publisher = {American Physical Society},
  doi = {10.1103/PhysRevE.66.016217},
  url = {https://link.aps.org/doi/10.1103/PhysRevE.66.016217}
}

@article{Craps:2024suj,
 title = {{Multiseed Krylov Complexity}},
  author = {Craps, Ben and Evnin, Oleg and Pascuzzi, Gabriele},
  journal = {Phys. Rev. Lett.},
  volume = {134},
  issue = {5},
  pages = {050402},
  numpages = {7},
  year = {2025},
  month = {Feb},
  publisher = {American Physical Society},
  doi = {10.1103/PhysRevLett.134.050402}
}

@article{PhysRevB.102.075127,
  title = {{Low-frequency behavior of off-diagonal matrix elements in the integrable XXZ chain and in a locally perturbed quantum-chaotic XXZ chain}},
  author = {Brenes, Marlon and Goold, John and Rigol, Marcos},
  journal = {Phys. Rev. B},
  volume = {102},
  issue = {7},
  pages = {075127},
  numpages = {7},
  year = {2020},
  month = {Aug},
  publisher = {American Physical Society},
  doi = {10.1103/PhysRevB.102.075127},
  url = {https://link.aps.org/doi/10.1103/PhysRevB.102.075127}
}

@article{Camargo:2026szl,
title={{Towards a Refinement of Krylov Complexity: Scrambling, Classical Operator Growth and Replicas}}, 
      author={Hugo A. Camargo and Yichao Fu and Keun-Young Kim and Yeong Han Park},
      year={2026},
      eprint={2603.19359},
      archivePrefix={arXiv},
      primaryClass={hep-th},
      url={https://arxiv.org/abs/2603.19359}, 
}

@article{Pal:2026otf,
      title={{A phase space approach to the wavefunction spreading and operator growth in the Krylov basis}}, 
      author={Kunal Pal and Kuntal Pal and Keun-Young Kim},
      year={2026},
      eprint={2601.13872},
      archivePrefix={arXiv},
      primaryClass={quant-ph},
      url={https://arxiv.org/abs/2601.13872}, 
}

@article{Das:2026hbw,
	author={Das, Rathindra Nath and Demulder, Saskia},
	title={{Integrability breaking in semiclassical strings in Koopman-Krylov space}},
	journal={Journal of Physics A: Mathematical and Theoretical},
	url={http://iopscience.iop.org/article/10.1088/1751-8121/ae895d},
	year={2026},
    doi={10.1088/1751-8121/ae895d}
}

@misc{KrylovQuantumClassical,
  author = {Nicolas {De Ro}},
  title  = {\href{https://doi.org/10.5281/zenodo.21337092}{``\text{{KrylovQuantumClassical}}''}},
  year   = {(2026)},

}
\end{document}